\documentclass[aps,physrev,ppreprint,groupedaddress,nofootinbib]{revtex4-2}
\usepackage[utf8]{inputenc}
\usepackage{physics}
\usepackage{amsmath,latexsym,amsfonts,amssymb,amsthm}
\usepackage{enumitem}
\usepackage{graphicx}
\usepackage{wrapfig}
\usepackage{bbold}
\usepackage[margin=1in]{geometry}
\usepackage{comment}
\usepackage{blkarray}
\usepackage{listings}
\usepackage{xcolor}
\usepackage{bm}
\usepackage[colorlinks=true,citecolor=black,linkcolor=black,urlcolor=black]{hyperref}

\let\oldvec\vec
\def\vec#1{\oldvec{#1}\,}

\newcommand{\dT}{\Delta}
\newcommand{\dTwoT}{B}

\newcommand{\LaplacianMatrix}{\mathcal{L}}

\newcommand{\VertexSpace}{\mathcal{V}}
\newcommand{\EdgeSpace}{\mathcal{E}}
\newcommand{\FaceSpace}{\mathcal{F}}
\newcommand{\potential}{\psi}
\newcommand{\Current}{I}
\newcommand{\current}{i}
\newcommand{\Adjacency}{W}
\newcommand{\Conductance}{\kappa}
\newcommand{\Graph}{G}

\newcommand{\OperatorAccent}{\hat}
\newcommand{\Op}[1]{\OperatorAccent{#1}}

\newcommand{\EdgeVecAccent}{\vec}
\newcommand{\EdgeVec}[1]{\EdgeVecAccent{#1}}
\newcommand{\NodeVecAccent}{\vec}
\newcommand{\NodeVec}[1]{\NodeVecAccent{#1}}

\graphicspath{{Images/}}

\begin{document}

\title{A Beginner's Guide to Physical Networks}
\author{Aaron Winn}
\email{winna@sas.upenn.edu}
\affiliation{Department of Physics and Astronomy, University of Pennsylvania, Philadelphia, Pennsylvania 19104, USA}

\author{Eleni Katifori}
\email{katifori@sas.upenn.edu}
\affiliation{Department of Physics and Astronomy, University of Pennsylvania, Philadelphia, Pennsylvania 19104, USA}
\affiliation{Center for Computational Biology, Flatiron Institute, New York, NY 10010, USA}


\maketitle


Network theory studies entities (nodes) and the relationships between them (edges). While these entities and their relationships may be abstract in general, this tutorial focuses on the special case of \textbf{physical networks}: spatially embedded networks that carry physical quantities such as electric charge, fluid volume, or momentum. The underlying structure is described mathematically by a graph (Section I), with the incidence matrix playing a central role in the analysis. Viewing a graph as a special case of a cell complex reveals the topological structure of the incidence matrix and its relation to familiar operators from calculus (Section II). Because these networks are physical, flows are conserved at each node and often derive from the gradient of a potential. Models that satisfy these principles behave analogously to circuits; accordingly, we first review circuit theory (Section III), then examine related fluid flow (Section IV) and mechanical (Section V) networks. Briefly, the topics of random walkers (Section VI) and nonlinear networks (Section VII) will be discussed, followed by concluding remarks (Section VIII). Analytical examples and Python implementations, in lieu of pseudocode, are included throughout. Exercises are included as footnotes. This tutorial is based on a short course taught at the Boulder School for Condensed Matter in 2019 and a course taught at the University of Pennsylvania in Spring 2024. It should be a useful resource for students, teachers, and researchers applying graph theory to physical networks. Knowledge of linear algebra is assumed, but no prior knowledge of graph theory or physics is necessary.

\section{Graph Theory}

\subsection{Definitions}
A network is represented mathematically as a graph. A \textbf{graph} $\Graph$ is a collection of nodes (also known as vertices) $\VertexSpace$ and edges $\EdgeSpace$. In particular, we will be interested in graphs whose edges have an orientation; such graphs are called \textbf{oriented graphs}. Edge orientations establish an arbitrary sign convention for direction, analogous to choosing a coordinate system.\footnote{An oriented graph should not be confused with a \textbf{directed graph}. In a directed graph, the edges have a particular (not arbitrary) direction, as will be explained in a later section. As a physical example, an oriented graph can be used to study fluid flow through pipes where the fluid is allowed to flow in either direction. We need a means of specifying when the fluid is transported forward versus backward through the pipe. If the flow is along the edge orientation, we say the flow is positive, and if the flow is against the edge orientation, we say the flow is negative. If, however, the pipes contain valves, then the fluid is only permitted to flow in one direction, so the graph is directed.} As a convenient notation, a node will be identified by a Latin index, $v_i \in \VertexSpace$. An edge will be identified by a Greek index $e_{\mu} \in \EdgeSpace$. We may also identify an oriented edge with an ordered pair of nodes: If an edge connects node $v_i$ to node $v_j$, we may write the edge as $e_{ij} \equiv (v_i,v_j)$,\footnote{This notation becomes problematic for \textbf{multigraphs}, graphs with multiple edges between the same two nodes. But, aside from sometimes making the notation cumbersome, the analysis in our review applies perfectly well to multigraphs.} in which case $v_i$ is called the tail of edge $e_{ij}$, and $v_j$ is called the head. We will not allow for the possibility of self-loops where an edge leaves and enters the same node. The number of nodes in $\VertexSpace$ and edges in $\EdgeSpace$ will be denoted $|\VertexSpace|$ and $|\EdgeSpace|$, respectively.

To each abstract node $v_i$ in the set $\VertexSpace$, we associate a vector $\vec{v}_i$ where $\vec{v}_i$ has a $1$ in the $i$-th entry and a $0$ in all other $|\VertexSpace|-1$ entries. The vectors $\{\vec{v}_i\}$ form a basis for a vector space isomorphic to $\mathbb{R}^{|\VertexSpace|}$ called the \textbf{node space}.\footnote{Of course, the node space could instead be defined over complex numbers or any other field, but we will typically denote the node space $\mathbb{R}^{|\VertexSpace|}$.} Similarly, the vectors $\{\vec{e}_\mu \}$ form a basis for the \textbf{edge space} isomorphic to $\mathbb{R}^{|\EdgeSpace|}$. This vector space structure allows for efficient calculation of physical quantities on graphs using linear algebra. Throughout the paper, vectors in the node or edge space will be distinguished using an arrow (ex: $\vec{\psi} \in \mathbb{R}^{|\VertexSpace|}$), while spatial vectors will be distinguished using boldface (ex: $\boldsymbol{x} \in \mathbb{R}^3$). Similarly, linear maps between node and edge spaces will be distinguished using a hat (ex: $\hat{\LaplacianMatrix}: \mathbb{R}^{|\VertexSpace|} \rightarrow \mathbb{R}^{|\VertexSpace|}$) while linear maps between spatial vectors will be distinguished using boldface (ex: $\boldsymbol{\sigma}: \mathbb{R}^{3} \rightarrow \mathbb{R}^{3}$).

\begin{figure}
    \centering
    \includegraphics[width=\textwidth]{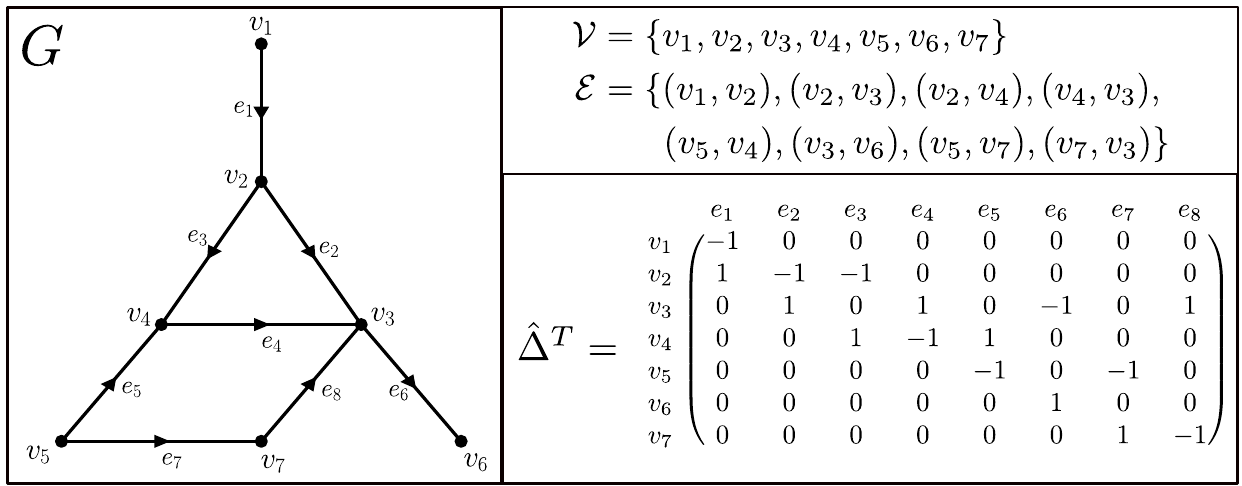}
    \caption{Three equivalent representations of the oriented graph $\Graph$ are shown: pictorially, as a set of vertices $\mathcal{V}$ and oriented edges $\mathcal{E}$, and with its incidence matrix $\Op{\dT}^T$.}
    \label{fig:flow_network_example}
\end{figure}

We can encode the oriented connectivity of the network using the $|\VertexSpace| \cross |\EdgeSpace|$ \textbf{incidence matrix}\footnote{More specifically, this is the \textbf{node-edge incidence matrix}. Once we consider more general cell complexes, we will encounter other incidence matrices, but for the case of graphs, the node-edge incidence matrix is the only incidence matrix. So, throughout most of the paper, we will simply refer to the node-edge incidence matrix as \textit{the} incidence matrix.} $\Op{\dT}^T: \mathbb{R}^{|\EdgeSpace|} \rightarrow \mathbb{R}^{|\VertexSpace|}$ given by
\begin{equation}
    \dT^T_{i\nu} = \begin{cases}
        +1, & \text{if edge $e_\nu$ enters node $v_i$}\\
        -1, & \text{if edge $e_\nu$ leaves node $v_i$}\\
        0, & \text{otherwise}\\
    \end{cases}. \label{eq:incidence}
\end{equation}
An example of an oriented graph with its incidence matrix is shown in figure \ref{fig:flow_network_example}. 

If the graph is weighted, then $G$ is also endowed with a vector of positive \textbf{edge weights} $\vec{\Conductance} \in \mathbb{R}^{|\EdgeSpace|}$. The notation $\hat{\Conductance} = \text{diag}(\Vec{\Conductance})$ will also be used to denote the $|\EdgeSpace| \cross |\EdgeSpace|$ matrix whose diagonal contains the weights of each edge and is otherwise zero. Finally, $\Conductance_{ij}$ will be understood as the weight along edge $e_{ij}$. For an unweighted graph, $\Conductance_\mu = 1$. For a physical network, the edge weights often correspond to an edge conductance.

The $|\VertexSpace| \cross |\VertexSpace|$ \textbf{adjacency matrix} $\Op{\Adjacency}: \mathbb{R}^{|\VertexSpace|} \rightarrow \mathbb{R}^{|\VertexSpace|}$ whose elements are
\begin{equation}
    \Adjacency_{ij} = \begin{cases}
        \Conductance_{ij}, & \text{if } (v_i,v_j) \in \mathcal{E}\\
        \Conductance_{ji}, & \text{if } (v_j,v_i) \in \mathcal{E}\\
        0, & \text{otherwise}\\
    \end{cases} \label{eq:adjacency}
\end{equation} 
is a symmetric matrix which contains information on the connectivity and weights of the network, but not on the orientation. The \textbf{weighted degree} of node $v_i$ is given by summing the weights of the edges adjacent to a node: $d_i = \sum_k W_{ik}$. The $|\VertexSpace| \cross |\VertexSpace|$ diagonal matrix of weighted degrees is denoted $\Op{D} \equiv \text{diag}(\vec{d})$.

The $|\VertexSpace| \cross |\VertexSpace|$ \textbf{Laplacian matrix} $\hat{\LaplacianMatrix}: \mathbb{R}^{|\VertexSpace|} \rightarrow \mathbb{R}^{|\VertexSpace|}$ is a symmetric matrix that can be expressed in two different ways\footnote{\textbf{EXERCISE}: Prove that these two forms of the Laplacian matrix are equivalent.}:
\begin{equation}
    \Op{\LaplacianMatrix} = \Op{\dT}^T \hat{\Conductance} \Op{\dT} = \hat{D} - \Op{\Adjacency} .\label{eq:laplacian}
\end{equation}
From the first representation of the Laplacian, we can see that $\Op{\LaplacianMatrix}$ is positive semi-definite. However, $\Op{\LaplacianMatrix}$ always has a zero eigenvalue corresponding to an eigenvector with entries all equal to 1. The number of zero eigenvalues is the number of connected components of the graph, so for a connected graph, this is the only zero eigenvalue. The Perron-Frobenius Theorem discussed in the appendix is a useful tool for studying the spectrum of the Laplacian and adjacency matrices. Although $\Op{\LaplacianMatrix}$ is not invertible, its pseudo-inverse takes a simple form. Note that for a real, symmetric, positive semi-definite matrix, the singular value decomposition is equivalent to the eigendecomposition. If the eigenvectors and eigenvalues of $\Op{\LaplacianMatrix}$ are $\Vec{\xi}_i$ and $\lambda_i$, then we can write $\Op{\LaplacianMatrix}$ and its pseudo-inverse $\Op{\LaplacianMatrix}^+$ as 
\begin{eqnarray}
    \Op{\LaplacianMatrix} = \sum_i \lambda_i \Vec{\xi}_i \Vec{\xi}_i^T ,
    \hspace{2cm}\Op{\LaplacianMatrix}^+ = \sum_{\{i|\lambda_i \neq 0\}}  \lambda_i^{-1} \Vec{\xi}_i \Vec{\xi}_i^T \label{eq:pseudoinverse}.
\end{eqnarray}
The pseudo-inverse will be useful when solving linear systems involving the Laplacian matrix. 

\subsection{Paths, cycles, and trees}
Some intuitive terminology is used to describe traversals in a network. These definitions do not require the graph to be oriented. For more definitions and related theorems, see chapter 3 of \cite{Wilson1996IntroductionEdition}. 

A \textbf{path} is a sequence of non-repeating edges where any two consecutive edges are connected by a node. Two nodes are \textbf{connected} if there exists a path between them; a graph is connected if all nodes are connected. A disconnecting set is any set of edges whose removal increases the number of connected components of the graph. In particular, a \textbf{cutset} is a disconnecting set with no proper subset that is a disconnecting set. For example, the set of edges adjacent to a node form a cutset. A \textbf{cycle} is a closed path for which the starting and ending edge are also connected by a node. A \textbf{tree} is a graph in which any two nodes are connected by exactly one path. In other words, a tree is a connected graph which contains no cycles. We can use this property of trees to find a convenient basis for the cycles of a graph. A \textbf{spanning tree} of a connected graph $G$ is a subgraph which is a tree and which includes all nodes of $G$. Using spanning trees, one can show that the total number of independent cycles in a graph is $|\EdgeSpace|-|\VertexSpace|+k$, where $k$ is the number of connected components of the graph.\footnote{\textbf{EXERCISE}: Draw a spanning tree for the graph in figure \ref{fig:flow_network_example}. Show that by adding an edge from $G$ not in the spanning tree, you obtain a graph with one cycle. From this construction, you can form a cycle basis of the graph. Generalize this idea to any connected graph. How many edges are in a spanning tree for a graph with $|\VertexSpace|$ vertices? How many edges are left in $G$? How many independent cycles must $G$ have? How would the counting change if we had a graph with $k$ connected components?}

\subsection{Properties of the incidence matrix}
Although it doesn't have a special name, the transpose of the incidence matrix $\hat{\Delta}$ is in some ways more intuitive to work with.\footnote{Warning: We use the symbol $\hat{\Delta}^T$ for the incidence matrix. Its transpose is $\hat{\Delta}$. It's easy to get $\hat{\Delta}$ and $\hat{\Delta}^T$ mixed up, but the easiest way to remember which is which is to remember formula \eqref{eq:Delta_psi} since it is common to use the symbol $\Delta$ to denote a difference in some quantity. Indeed, that is why our group has adopted this notation!} Consider a quantity (such as the electric potential) with values on each node, that is, a vector $\vec{\psi} \in \mathbb{R}^{|\VertexSpace|}$. If edge $e_\mu = (v_i,v_j)$, then 
\begin{eqnarray}
    (\hat{\Delta} \vec{\psi})_{\mu} = \psi_j - \psi_i \label{eq:Delta_psi}
\end{eqnarray}
measures the voltage rise, along edge $e_\mu$. The operator $\hat{\Delta}$ takes the difference of the quantity at the terminal node and the quantity at the starting node. Its image can be thought of as all edge vectors which derive from a difference in node potentials. Note that if $\psi_i = \psi_j$, then $(\hat{\Delta} \vec{\psi})_\mu = 0$. Thus, the vector with all $|\VertexSpace|$ entries equal to 1 is in the null space of the incidence matrix:
\begin{align}
    \hat{\Delta} \Vec{1} &= \vec{0} \nonumber \\ \sum_{j} \Delta_{\mu j} &= 0 . \label{eq:incidence_property}
\end{align}
In fact, for a connected graph, this is the only vector in the null space of $\hat{\Delta}$ \cite{Van_Mieghem_2011}. To see this, note that if the graph is connected and $\psi_i = \psi_j$ for each $e_{ij} \in \mathcal{E}$, then each $\psi_i$ in the entire network must be the same. For a graph which is not connected, we only require that the $\psi_i$ be the same on each connected component. Therefore, the null space of $\hat{\Delta}$ is spanned by vectors which are equal to 1 on a particular connected component and 0 elsewhere. The nullity of $\hat{\Delta}$ is the number of connected components $k$. By the rank-nullity theorem,\footnote{Recall, the rank-nullity theorem states that for any matrix $A$, the rank (dimension of the image) of $\hat{A}$ plus the nullity (dimension of the kernel, or null space) of $\hat{A}$ is equal to the number of columns of $\hat{A}$.} the rank of $\hat{\Delta}$ is equal to $|\VertexSpace|-k$. When a graph is disconnected, $\hat{\Delta}$ can be put in block diagonal form, decoupling whatever problem we wish to solve into $k$ distinct problems. 

Now consider an element of the incidence matrix acting on a quantity $\Vec{I}$ (for example, the electric current) defined on the edges:
\begin{equation}
    (\hat{\Delta}^T \vec{I})_i = \sum_{\mu \rightarrow i} I_\mu - \sum_{\nu \leftarrow i} I_\nu .\label{eq:DeltaT_I}
\end{equation}
The first sum runs over all edges oriented into node $v_i$ and the second sum runs over all edges oriented out of node $v_i$. The result can be thought of as the net flow into node $v_i$. For many cases of physical interest, this quantity will be zero, corresponding to a conservation law (for example, electric charge). 

As is the case with any matrix, the rank of $\hat{\Delta}^T$ is equal to the rank of its transpose, in this case $|\VertexSpace|-k$. By the rank-nullity theorem, the nullity of $\hat{\Delta}^T$ is $|\EdgeSpace| - (|\VertexSpace|-k)$. It is no coincidence that this is the same number of cycles we found in the previous section. To see this, it will be useful to recognize the incidence matrix as a boundary operator. The boundary of an edge (or more generally, a path) is the starting and ending node. A cycle is a path which has no boundary. When traversing an edge in a path, the step direction may or may not agree with the orientation of the network. Consider an indicator vector $\vec{e} \in \mathbb{R}^{|\EdgeSpace|}$ whose entries are $1$ if a step is taken along the edge orientation, $-1$ if a step is taken against the edge orientation, and $0$ if the edge is not in the path. Then, the path is a cycle if and only if $\hat{\Delta}^T \Vec{e} = 0$. The cycle space is the null space of $\hat{\Delta}^T$. 

\begin{figure}
    \centering
    \includegraphics[width=0.8\linewidth]{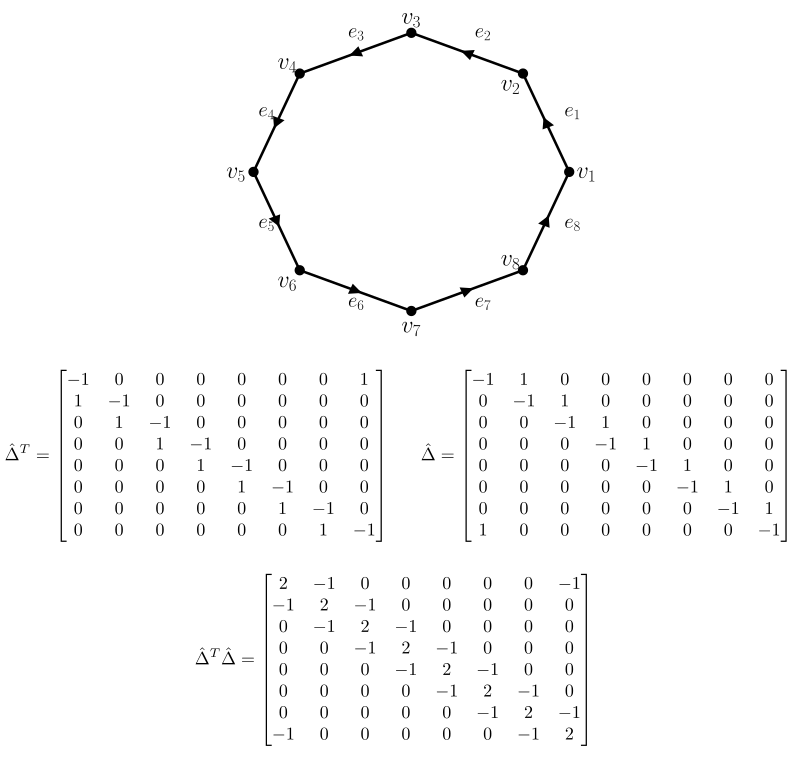}
    \caption{The incidence matrix $\hat{\Delta}^T$, the transpose of the incidence matrix $\hat{\Delta}$, and the Laplacian $\hat{\Delta}^T \hat{\Delta}$ for an unweighted, oriented circle graph with 8 nodes and 8 edges. Notice that $\hat{\Delta}$ and $-\hat{\Delta}^T$ resemble forward and backward finite difference operators in a 1D discretized system with periodic boundary conditions, while $-\hat{\Delta}^T \hat{\Delta}$ resembles a discrete Laplacian operator. }
    \label{fig:circle_graph}
\end{figure}

It is helpful to think of node quantities as analogous to continuous scalar functions and edge quantities as analogous to continuous vector functions with direction given by the edge orientation. Equation \eqref{eq:Delta_psi} suggests that $\hat{\Delta}$ acts like a discrete derivative (or more precisely, gradient) operator, acting on a node-valued function to give an oriented edge-valued function.\footnote{Unfortunately, in calculus, the symbol $\Delta$ usually refers to the Laplacian. Again, our notation should make you think of $\hat{\Delta}$ as a difference operator, equation \eqref{eq:Delta_psi}, not a Laplacian.} Equation \eqref{eq:DeltaT_I} suggests that $\hat{\Delta}^T$ acts like the negative of a divergence operator, acting on an edge-valued function to produce a node-valued function.\footnote{Negative because when $(\hat{\Delta}^T \Vec{I})_i$ is positive, there is a convergence of flows at node $v_i$.} For line and circle graphs ordered and oriented in the obvious way, the operators $\hat{\Delta}$, $-\hat{\Delta}^T$, and $-\hat{\Delta}^T \hat{\Delta}$ are the forward difference, backward difference, and Laplacian operators used in a typical finite difference scheme. See figure \ref{fig:circle_graph}. 

It's quite remarkable that $\hat{\Delta}^T$ acts both as a boundary operator and a derivative operator depending on whether it acts on an indicator vector or a node-valued function. We will explore these ideas further in section II.

Even though our graphs are oriented, it is occasionally useful to consider the \textbf{unoriented incidence matrix} which keeps track of which edges and nodes are connected but not the orientation; its entries are the absolute value of the entries of $\Op{\dT}^T$, so it will be denoted $|\Op{\dT}|^T$. If $\Op{\dT}$ acts as a difference operator, then $|\Op{\dT}|$ acts as a summation operator associating to an edge the sum of its two adjacent node-valued quantities. For example, if the potential $\vec{\psi}$ is defined on the nodes, then the quantity $\frac{1}{2}|\Op{\dT}| \vec{\psi} $ associates to each edge the average value of the potential on the two adjacent nodes. Similarly, $|\Op{\dT}|^T $ associates to a node the sum of the adjacent edge-valued quantities. For example, the weighted degree vector can be written $\vec{d} = |\Op{\dT}|^T \vec{\kappa}$. 

\subsection{Directed graphs}
A \textbf{directed graph} consists of nodes and directed edges, differing from an oriented graph in that the orientation is not arbitrary, but rather it has some additional physical meaning. In the context of flow networks, a directed edge $e_{ij}$ only permits flow from $v_i$ to $v_j$. If flow is permitted both from $v_i$ to $v_j$ and from $v_j$ to $v_i$, then we will need to include both $e_{ij}$ and $e_{ji}$ in our edge list, though $\Conductance_{ij}$ need not equal $\Conductance_{ji}$. The adjacency matrix takes a rather intuitive form for a directed graph: 
\begin{equation}
    \Adjacency_{ij} = \begin{cases}
        \Conductance_{ij}, & \text{if } (v_i,v_j) \in \mathcal{E}\\
        0, & \text{otherwise}\\
    \end{cases} \label{eq:adjacencyDirected}.
\end{equation} 
This encodes the weights across all ordered pairs of vertices, but it will not necessarily be symmetric. We can define two different weighted degrees for each node: the \textbf{indegree} $d_{+,i} = \sum_k W_{ki}$ and the \textbf{outdegree} $d_{-,i} = \sum_k W_{ik}$. 

The incidence matrix can still be defined using \eqref{eq:incidence} with the only caveat being that what we mean by entering versus leaving is not arbitrary anymore; it comes from the direction of the edge. The incidence matrix is not the most fundamental operator needed to describe directed networks. Define the matrices $\Op{\dT}_+^T$ and $\Op{\dT}_-^T$ using the following: 
\begin{equation}
    \dT^T_{+,i\nu} = \begin{cases}
        +1, & \text{if edge $e_\nu$ enters node $v_i$}\\
        0, & \text{otherwise}\\
    \end{cases}. \label{eq:incidencePlus}
\end{equation}
\begin{equation}
    \dT^T_{-,i\nu} = \begin{cases}
        +1, & \text{if edge $e_\nu$ leaves node $v_i$}\\
        0, & \text{otherwise}\\
    \end{cases}. \label{eq:incidenceMinus}
\end{equation}
Clearly, $\Op{\dT}^T = \Op{\dT}_+^T - \Op{\dT}_-^T$, but in some cases, we will only want to consider the incoming or outgoing edges from a node, in which case we'd use $\Op{\dT}_+^T$ or $\Op{\dT}_-^T$, respectively. 

Certain physics applications require the use of directed graphs (ex: advection of nutrients in a fluid flow), though we will mostly be working with undirected, oriented graphs, until sections VI and VII. 

\subsection{NetworkX}
The small network examples in this tutorial can be studied by hand, but we recommend reproducing the calculations in Python. The Python package NetworkX allows for the creation, manipulation, and plotting of graphs \cite{SciPyProceedings_11}. For installation instructions and documentation, visit \url{https://networkx.org/documentation/stable/}. We also recommend importing Numpy, SciPy, and Matplotlib. In this tutorial, we will only review the essentials for getting started using NetworkX to solve physical network problems. 

Graphs can be constructed either by using predefined functions or by explicitly adding nodes and edges. Note that NetworkX does not understand the difference between an oriented, undirected graph and a directed graph, so to construct an oriented graph, we actually use \texttt{nx.DiGraph()}. In the snippet below, we construct an oriented square graph by explicitly adding nodes and edges. 
\begin{lstlisting}[language=Python, caption={Drawing a simple graph}] 
import networkx as nx
G = nx.DiGraph() #Creates an oriented graph
G.add_node(1, pos=(0,0)) #Adds a node named '1'.
G.add_node(2, pos=(0,1)) #Each node has attribute 'pos' specifying its position.
G.add_node(3, pos=(1,1)) 
G.add_node(4, pos=(1,0))
G.add_edge(1,2, weight=2) #Adds an edge from node 1 to node 2.
G.add_edge(2,3, weight=4) #Each edge has attribute 'weight' specifying its width.
G.add_edge(3,4, weight=6)
G.add_edge(4,1, weight=10)
positions = nx.get_node_attributes(G,'pos') #Save node positions in a dictionary
weights = [G[i][j]['weight'] for (i, j) in G.edges()] #Save edge weights in a list
nx.draw(G, pos=positions, width=weights, with_labels=True, node_color='orange')
\end{lstlisting}
Nodes are each given a label when added. It could be anything, but here we use the integers 1 through 4. Edges are added by specifying the starting and ending node label. Nodes and edges can be assigned attributes. Here, we assign the node attribute \texttt{pos} for defining the node positions and the edge attribute \texttt{weight} for defining the edge weights. Note that \texttt{positions} is a dictionary while \texttt{weights} is a list.\footnote{\textbf{EXERCISE}: Using NetworkX, generate an unweighted, oriented cycle graph with 10 nodes. Study the incidence matrix and its transpose. How are they related to finite difference operators? How is the Laplacian matrix related to the discretized Laplacian operator from calculus? What about when we include edge weights?} 

Once we have defined a graph, the next thing we want to do is obtain its incidence matrix and the Laplacian matrix.
\begin{lstlisting}[language=Python, caption={Constructing the incidence matrix and Laplacian matrix}] 
import numpy as np
import scipy.sparse as sp
DeltaT = nx.incidence_matrix(G, oriented = True) #Saved as a SciPy sparse matrix
Delta = DeltaT.T
print(DeltaT.toarray()) #Convert to an array only to visualize the matrix.
Conductance = sp.diags(weights)
Laplacian = DeltaT @ Conductance @ Delta
print(Laplacian)
\end{lstlisting}
By default, \texttt{nx.incidence\_matrix(G, oriented = True)}
is saved as a sparse matrix. It is a good habit to perform matrix computations using sparse matrices since most of the elements of $\hat{\Delta}^T$ and $\hat{\LaplacianMatrix}$ are zero. For the tiny graphs in this review, it won't matter, but if your network has thousands of nodes, sparse operations will save you time and memory.

\newpage
\section{Discrete Calculus}
The type of graphs we are considering (oriented graphs with no self loops) are a special case of a more general mathematical object called a cell complex. By understanding the topology of cell complexes, we can better understand the incidence matrix and its extension to higher dimensions. We will also learn how to study calculus and homology of graphs, as well as how to formally construct a dual graph. We hope to not only give some intuition for common mathematical operations on graphs, but also provide an intuitive introduction to algebraic topology. This section is largely based on the book \textit{Discrete Calculus} by Grady and Polimeni \cite{Grady_Polimeni_2010} and a short review by Desbrun, et al \cite{DesbrunDiscreteModeling}. A more complete resource for the application of topology to physical problems can be found in \cite{Nakahara2003}.

\subsection{Graphs...with faces}
A graph is only concerned with 0-dimensional nodes and 1-dimensional edges. Before considering a completely general cell complex, let's first look in just one higher dimension and consider faces. We will denote the set of faces $\mathcal{F}$. Note that a face is a 2-dimensional object, not to be confused with a cycle which is a closed 1-dimensional path. A face will be identified by a capital Latin index $f_I \in \FaceSpace$. Much like how we had to assign an orientation to the edges by specifying an order of two nodes, we can assign an orientation to a face by specifying an order to at least three edges (ex: clockwise or counterclockwise). For a given face, the orientation also induces an orientation on the edges bounding the face which may or may not match the actual orientation of the edge. We describe how oriented faces are made up of oriented edges using the \textbf{edge-face incidence matrix} $\hat{B}^T: \mathbb{R}^{|\mathcal{F}|} \rightarrow \mathbb{R}^{|\mathcal{E}|} $ whose elements are given by 
\begin{equation}
    B_{\nu I}^T = \begin{cases}
        0, & \text{if edge $e_\nu$ is not on the boundary of face $f_I$}\\
        +1, & \text{if the orientation of edge $e_\nu$ is the orientation induced by face $f_I$}\\
        -1, & \text{if the orientation of edge $e_\nu$ is opposite to the orientation induced by face $f_I$}\\
    \end{cases}. \label{eq:B_def}
\end{equation}
It is often convenient whenever possible to assign the orientation ``counterclockwise" to all faces. Such a notion only makes sense when the graph is planar. 

\begin{figure}
    \centering
    \includegraphics[width=0.75\linewidth]{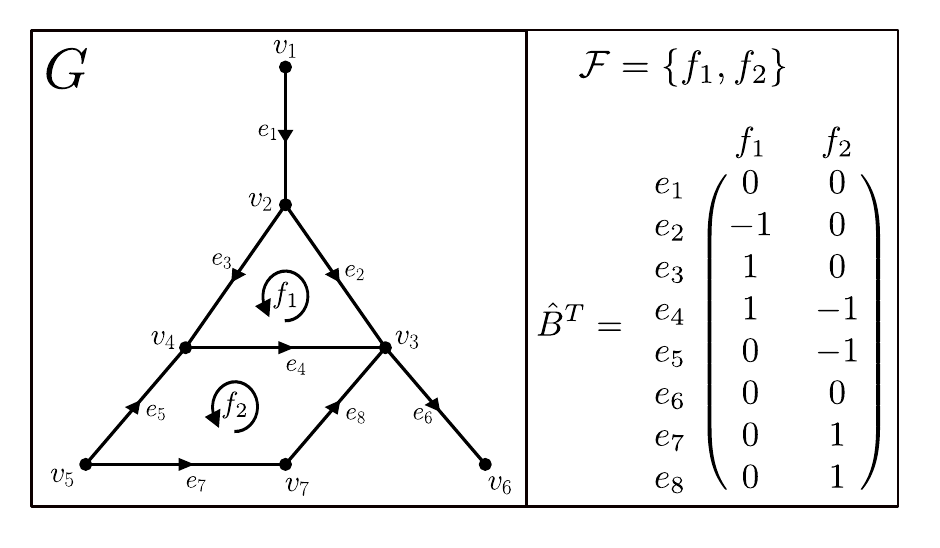}
    \caption{The oriented graph $G$ from figure \ref{fig:flow_network_example}, now with a set of faces $\mathcal{F}$. The edge-face incidence matrix $\hat{B}^T$ is shown.}
    \label{fig:GraphWithFaces}
\end{figure}

Although cycles and faces are distinct, when we want to measure a physical flux through a loop, we will often fill in each independent cycle with a face. Assigning faces to each such cycle in the graph shown in figure \ref{fig:flow_network_example} gives the graph-with-faces shown in figure \ref{fig:GraphWithFaces}. The edge-face incidence matrix of that graph-with-faces is calculated. For example, face $f_1$ is composed of edges $e_2$, $e_3$, and $e_4$. The orientations of edges $e_3$ and $e_4$ agree with the orientation of face $f_1$ (both the edges and the face are oriented counterclockwise), but the orientation of edge $e_2$ disagrees with the orientation of face $f_1$ (the edge is oriented clockwise but the face is oriented counterclockwise). Thus, in the $f_1$ column, we put a $-1$ in the $e_2$ row, a $1$ in the $e_3$ and $e_4$ rows, and a $0$ in all other rows. 

Unlike the node-edge incidence matrix $\hat{\Delta}^T$, the columns of the edge-face incidence matrix need not sum to zero. The fundamental difference is that while each edge must be composed of one tail node and one head node, faces are composed of many edges.

\begin{figure}
    \centering
    \includegraphics[width=\linewidth]{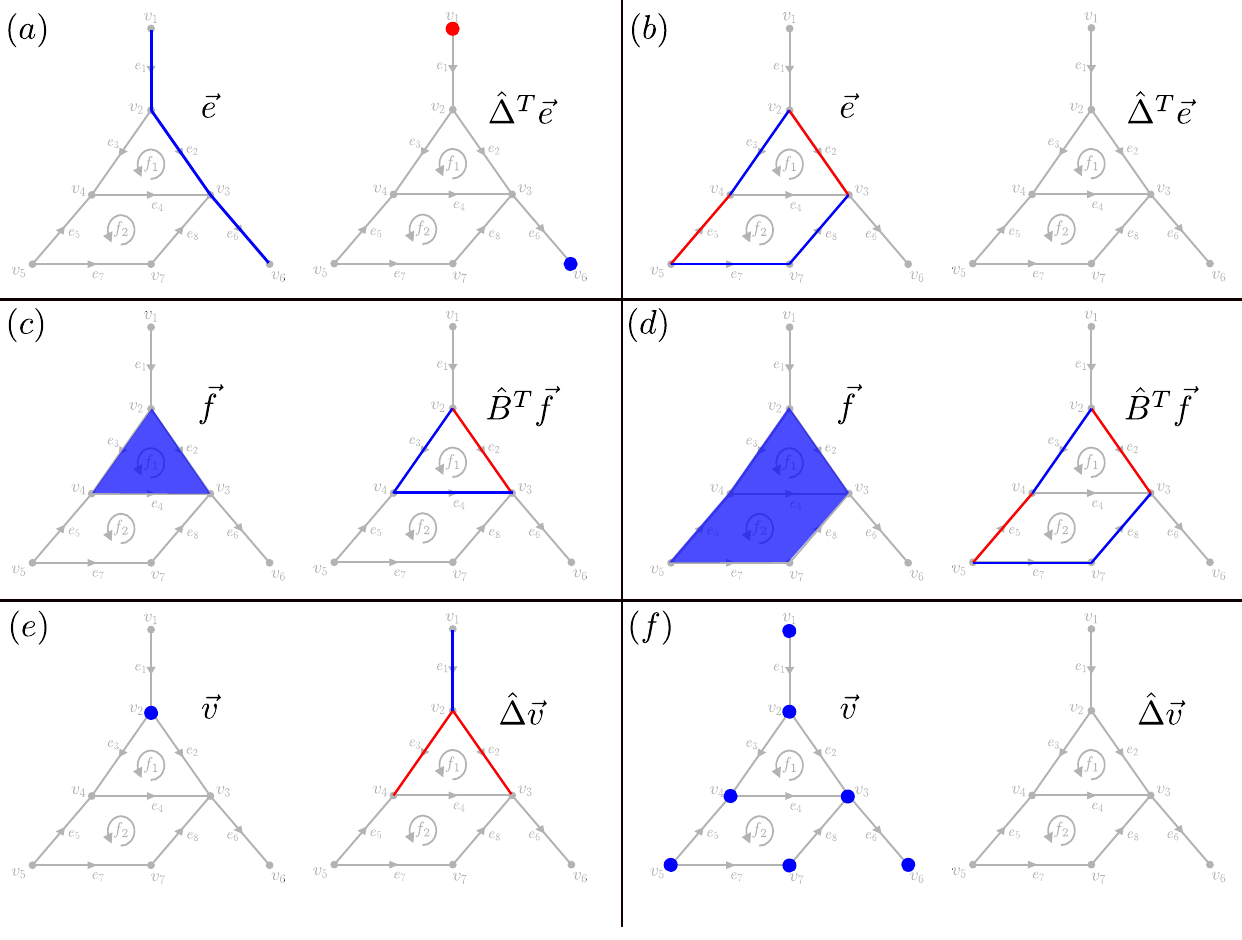}
    \caption{Examples of the node-edge incidence matrix, edge-face incidence matrix, and the transpose of the node-edge incidence matrix acting on chains. In all cases, rather than a vector representation, a pictorial representation is shown where blue indicates a $+1$ entry and red a $-1$ entry. For example, in $(c)$, $\vec{f} = (1, 0)^T$, and $\hat{B}^T \vec{f} = (0, -1, 1, 1, 0, 0, 0, 0)^T$.  $(a)$ The boundary of the oriented sequence of edges is one source and one sink. $(b)$ The boundary of a cycle of edges is empty. Note that a red edge in $\vec{e}$ is traversed opposite the edge orientation. $(c)$ The boundary of one face is three edges. $(d)$ The boundary of two adjacent faces is the cycle from $(b)$. $(e)$ The coboundary of a single node describes which edges are entering (blue) or leaving (red) the node. $(f)$ The coboundary of the nodes in the entire graph is empty since each edge enters one node and leaves one node. }
    \label{fig:DiscreteCalculus}
\end{figure}

We can also interpret $\hat{B}^T$ as a boundary operator mapping a face to its boundary consisting of edges. Analogous to the calculus result that the boundary of a boundary is zero, we might expect
\begin{equation}
    \hat{\Delta}^T \hat{B}^T  = 0. \label{eq:AB}
\end{equation}
To see that this is true, note that the image of $\hat{B}^T$ consists of cycles of the graph, which when acted on by $\Op{\dT}^T$ produces the zero vector in the space of vertices.\footnote{\textbf{EXERCISE}: Check that equation \eqref{eq:AB} is true for the graph-with-faces shown in figure \ref{fig:GraphWithFaces}.} The transposed expression is also true:
\begin{eqnarray}
    \hat{B} \hat{\Delta} = 0 \label{eq:BA}.
\end{eqnarray}
To visualize the incidence matrices $\hat{\Delta}^T$ and $\hat{B}^T$ as boundary operators, consider how these matrices act on indicator vectors on the edges and faces. See figure \ref{fig:DiscreteCalculus}.

When acting on face-valued functions and edge-valued functions, $\Op{\dTwoT}^T$ and $\Op{\dTwoT}$ act like discrete versions of the curl operator. With this in mind, equations \eqref{eq:AB} and \eqref{eq:BA} are the familiar statements $\vec{\nabla}\cdot (\vec{\nabla} \cross \vec{F}) = 0$ and $\vec{\nabla} \cross (\vec{\nabla} f )= \vec{0}$. 

One can also define a face Laplacian matrix $\hat{\LaplacianMatrix}_2: \mathbb{R}^{|\mathcal{F}|}  \rightarrow \mathbb{R}^{|\mathcal{F}|} $
\begin{equation}
    \hat{\LaplacianMatrix}_2 = \hat{B} \hat{\Conductance}^{-1} \hat{B}^T
\end{equation}
analogous to the (node) Laplacian matrix we have already seen. It will become clear later why we are using $\hat{\Conductance}^{-1}$ as a weight instead of $\hat{\Conductance}$. In many cases, the face Laplacian matrix is invertible.

We now have discrete analogues for common differential operators of vector calculus. Integration on graphs is even more straightforward. If we want to ``integrate" a quantity $\vec{\psi}$ defined on the nodes, we need only sum over $\psi_i$ where $v_i$ is in our region of integration. Letting the indicator vector $\vec{v} \in \mathbb{R}^{|\VertexSpace|}$ equal 1 for any nodes within the domain of integration and 0 for nodes outside the domain of integration, an integral over nodes is simply the quantity $\langle \vec{\psi}, \vec{v}\rangle = \vec{\psi}^T \vec{v}$. If we pick only a single node $v_i$, then the indicator vector $\vec{v}_i$ is the $i$-th basis vector of the node space, and $\vec{\psi}^T \vec{v}_i = \psi_i$ can be thought of evaluating the scalar function $\vec{\psi}$ on node $v_i$. Integration over edges is similar, only one will need to be more careful keeping track of the direction along which the edge is integrated. To integrate over a path, we define a vector $\vec{e} \in \mathbb{R}^{|\EdgeSpace|}$ which is equal to $+1$ if the edge is traversed along the edge orientation, $-1$ if the edge is traversed against the edge orientation, and 0 if the edge is not in the path. Then integration over a quantity $\vec{I}$ defined on the edges is $\vec{I}^T \vec{e}$. There's nothing stopping us from assigning other numbers to the components of $\vec{e}$. For example, a $2$ would correspond to integrating over the edge twice along the direction of the edge orientation. A summary of the discrete analogues of derivative and integral operators are shown in table \ref{tab:operators}.\footnote{In this section, we are focusing on the topology of graphs, but to fully complete the analogy to calculus, we occasionally need to consider the geometry. The role of the metric tensor is played by the edge weights (and node weights and face weights, though we usually weight all nodes and faces equally). The metric allows us to define a Hodge star operator on complexes and explains why $\hat{\LaplacianMatrix}$ has a factor of $\hat{\Conductance}$ while $\hat{\LaplacianMatrix}_2$ has a factor of $\hat{\Conductance}^{-1}$. We direct the mathematically curious reader to \cite{Grady_Polimeni_2010}. The form of the node Laplacian and face Laplacian arise naturally when studying physical networks, so we will avoid the unnecessarily technical discussion here.} 

In vector calculus, the relationship between derivatives and integrals are summarized in the fundamental theorem of calculus and its relatives. Using the relations in table \ref{tab:operators}, we can write down the discrete versions of these theorems, as summarized in table \ref{tab:calculus_theorems}. The statements on graphs amount to changing the order of operation of matrix multiplications. When an incidence matrix acts on a physical quantity (think the electric potential $\vec{\psi}$, the voltage $\vec{\Psi}$, and the current $\vec{I}$), it acts like a derivative operator, but when it acts on an indicator vector it acts like a boundary operator. To understand the discrete Fundamental Theorem of Calculus, return to figure \ref{fig:DiscreteCalculus}$(a)$, and suppose we wish to calculate $\langle \hat{\Delta} \psi, \vec{e} \rangle$ for some physical quantity $\vec{\psi}$ defined on the nodes. Physically, this would mean summing up the gradients of $\vec{\psi}$ along the blue edges. By the Fundamental Theorem of Calculus, we can instead calculate $\langle \vec{\psi}, \hat{\Delta}^T \vec{e} \rangle$ which amounts to evaluating $\vec{\psi}$ at only two points. When the integral of a gradient is taken over a closed loop, the result is zero regardless of the function. Stokes' theorem can be understood in panel $(d)$ as the curl of a function defined on the two faces canceling out in the middle edge, so the curl picks out only the boundary edges. Finally, Gauss's Theorem can be understood as current conservation. For panel $(e)$, if a current exits node $v_2$, it must have entered from the adjacent edges (signed to account for direction with respect to edge orientation). These theorems are closely related to Kirchhoff's Laws in circuits.\footnote{\textbf{EXERCISE}. Show that, for any vector $\vec{I} \in \mathbb{R}^{|\EdgeSpace|}$, if $\vec{v} \in \mathbb{R}^{|\VertexSpace|}$ is the vector with all node-valued entries equal to 1, then $\langle \hat{\dT}^T\vec{I}, \vec{v} \rangle = 0 $. Give a physical interpretation of this equation.}

There is also a discrete analogue to the Helmholtz Decomposition so that one can decompose an edge-valued function into a gradient term and a curl term. The details are left as an exercise.\footnote{\textbf{EXERCISE}: In continuous space, a Helmholtz decomposition can be performed for any sufficiently smooth and rapidly decaying function $\vec{F}$ by writing $\vec{F}$ as the sum of an irrotational and solenoidal piece: $\Vec{F}=\Vec{F}_{irrot} + \Vec{F}_{sol} = \grad \phi + \curl \Vec{A}$. Notably, the second term is divergence-less. For an edge-valued vector $\vec{I}$, there exists a $\vec{\psi}\in \mathbb{R}^n$ such that one can perform a Helmholtz decomposition with $ \Vec{I}_{irrot} = -\hat{\Delta} \Vec{\psi}$ and $\Vec{I}_{sol} = \hat{B}^T \vec{i}_{\ell}$. Show that $\Vec{I}_{irrot}$ and $\Vec{I}_{sol}$ are orthogonal and that such a construction is always possible. Then, demonstrate how to calculate $\vec{\psi}$ when $\hat{\Delta}^T \vec{I}$ is known and $\vec{i}_{\ell}$ when $\hat{B} \vec{I}$ is known. Is $\vec{\psi}$ unique? [Hint: If you get stuck, read the next section on circuit theory, and return. How does this relate to the node and loop analyses in a circuit with unit conductances?]}

\renewcommand{\arraystretch}{1.5}
\begin{table}[]
    \centering
    \begin{tabular}{|c|c|}
\hline
 Continuous Operator  & Discrete Operator \\ 
 \hline \hline
 Gradient $\boldsymbol{\nabla} $ & Transposed node-edge incidence matrix $\hat{\dT}: \mathbb{R}^{|\mathcal{V}|} \rightarrow \mathbb{R}^{|\mathcal{E}|}$  \\ \hline
 Divergence $\boldsymbol{\nabla} \cdot $ & Negative of node-edge incidence matrix $-\hat{\dT}^T: \mathbb{R}^{|\mathcal{E}|} \rightarrow \mathbb{R}^{|\mathcal{V}|}$  \\ \hline
 Curl 1 $\boldsymbol{\nabla} \cross $ & Transposed edge-face incidence matrix $\hat{\dTwoT}: \mathbb{R}^{|\mathcal{E}|} \rightarrow \mathbb{R}^{|\mathcal{F}|}$  \\ \hline
 Curl 2 $\boldsymbol{\nabla} \cross $ &  Edge-face incidence matrix $\hat{\dTwoT}^T: \mathbb{R}^{|\mathcal{F}|} \rightarrow \mathbb{R}^{|\mathcal{E}|}$  \\ \hline
 Laplacian ${\nabla}^2  $ &  Negative of Laplacian matrix $-\hat{\LaplacianMatrix}: \mathbb{R}^{|\mathcal{V}|} \rightarrow \mathbb{R}^{|\mathcal{V}|}$  \\ \hline
 Line integral $\int_l  d\boldsymbol{l} \cdot$ & $\langle \hspace{.1cm} \cdot \hspace{.1cm}, \vec{e} \rangle: \mathbb{R}^{|\mathcal{E}|} \rightarrow \mathbb{R}$ \\ \hline
 Surface integral $\int_A  d\boldsymbol{A} \cdot$ & $\langle \hspace{.1cm} \cdot \hspace{.1cm}, \vec{f} \rangle: \mathbb{R}^{|\mathcal{F}|}\rightarrow \mathbb{R}$ \\ \hline
 Volume integral $\int_V  dV \hspace{.1cm}$ & $\langle \hspace{.1cm} \cdot \hspace{.1cm}, \vec{v} \rangle: \mathbb{R}^{|\mathcal{V}|} \rightarrow \mathbb{R}$ \\ \hline
    \end{tabular}
    \caption{Operators from continuous vector calculus and their discrete analogues. }
    \label{tab:operators}
\end{table}

\begin{table}[]
    \centering
    \begin{tabular}{ |{c}|c|c| } 
 \hline
 Theorem & Continuous & Discrete \\ 
 \hline \hline
 Fundamental Theorem of Calculus & $\int_{l} \boldsymbol{\nabla} f \cdot d\boldsymbol{l} = f(b) - f(a) $ & $\langle \hat{\dT}\vec{\psi}, \vec{e} \rangle = \langle \vec{\psi}, \hat{\dT}^T \vec{e} \rangle $\\ \hline
 Stokes' Theorem I & $\int_A \boldsymbol{\nabla} \cross \boldsymbol{F} \cdot d\boldsymbol{A} = \oint_{\partial A} \boldsymbol{F} \cdot d\boldsymbol{l}$ & $\langle \hat{B}\vec{\Psi}, \vec{f} \rangle = \langle \vec{\Psi}, \hat{B}^T \vec{f} \rangle $ \\ \hline Stokes' Theorem II & $\int_A \boldsymbol{\nabla} \cross \boldsymbol{F} \cdot d\boldsymbol{A} = \oint_{\partial A} \boldsymbol{F} \cdot d\boldsymbol{l}$ & $\langle \hat{B}^T \vec{i}_{\ell}, \vec{e} \rangle = \langle \vec{i}_{\ell}, \hat{B} \vec{e} \rangle $ \\ \hline 
 Gauss's Theorem & $\int_V \boldsymbol{\nabla}\vdot \boldsymbol{F} \hspace{.1cm} dV = \oint_{\partial V} \boldsymbol{F} \cdot d\boldsymbol{A}$ & $\langle \hat{\dT}^T\vec{I}, \vec{v} \rangle = \langle \vec{I}, \hat{\dT} \vec{v} \rangle $ \\ \hline
\end{tabular}
    \caption{Theorems from calculus and their discrete analogues. Think of $\vec{\psi}$ as the electric potential, $\vec{\Psi}$ as the voltage, $\vec{i}_\ell$ as the loop current, and $\vec{I}$ as the edge current in an electric circuit. }
    \label{tab:calculus_theorems}
\end{table}

\subsection{A primer on algebraic topology}

At this point, you may be thinking...why stop at faces? We considered 0-dimensional nodes, 1-dimensional edges, and 2-dimensional faces, all of which are oriented and fit together nicely. But, you could imagine incorporating 3-dimensional volumes, 4-dimensional hyper-volumes, and so on. These $p$-dimensional oriented objects are called $p$-\textbf{cells}. When glued together in a nice way, the cells form a \textbf{cell complex}.\footnote{Formally, a $p$-cell is defined as a set of points homeomorphic to the unit $p$-ball. The boundary of each $p$-cell comprises the union of lower-order $p$-cells, and the intersection of any two cells is either empty or a boundary element of both cells. The orientation is specified by an ordering of the nodes, with all even permutations of this ordering being equivalent. We will mainly need to consider 0-cells (nodes), 1-cells (edges), and 2-cells (faces), where the phrase ``glued together in a nice way" is intuitive. Note that a cell complex, also known as a CW complex, is more general than a simplicial complex which is only built out of $p$-simplices which consist of $p+1$ nodes (triangles, tetrahedra, etc.). } The cell complex is the natural generalization of the oriented graph, and it is the object of study in algebraic topology. Here, we will give a brief and intuitive introduction to the topology of cell complexes.  A more thorough study can be found in \cite{Grady_Polimeni_2010}. 

From the abstract objects called nodes we built a vector space called the node space. Actually, there are two related vector spaces that up until now we have not been careful to distinguish. There are ``indicator vectors" $\vec{v}$ which specify particular nodes to evaluate or perform integration, often by placing a $-1$, $0$, or $1$ in the entry for that node to indicate whether a node is included and its orientation. These indicator vectors are more formally called $0$-chains, and the vector space of $0$-chains is denoted $\mathcal{C}_0$. Dual to this vector space is the space of linear functionals from $\mathbb{R}^{|\VertexSpace|}$ to $\mathbb{R}$. These are the physical functions such as the electric potential $\vec{\psi}$ defined on the nodes. These vectors are called $0$-cochains, and the vector space of $0$-cochains is denoted $\mathcal{C}^0$. Both vector spaces are isomorphic to $\mathbb{R}^{|\VertexSpace|}$, so for most of the review we will just write $\mathbb{R}^{|\VertexSpace|}$ for either. Edges and faces possess this same structure. In general, indicator vectors on $p$-cells are called $p$-\textbf{chains} and the physical functions defined on $p$-cells are called $p$-\textbf{cochains}. The vector space of $p$-chains is denoted $\mathcal{C}_p$, and the vector space of $p$-cochains is denoted $\mathcal{C}^p$.

A $p$-cell induces an orientation on its boundary which may or may not agree with the orientation imposed on the $(p-1)$-cells which compose the boundary. The topological structure of the $p$-complex is completely described by the \textbf{incidence matrices} $\{\hat{\partial}_p\}$ of the cell complex, where $\hat{\partial}_1 = \hat{\Delta}^T$ and $\hat{\partial}_2 = \hat{B}^T$. Higher incidence matrices are the generalization of equation \eqref{eq:B_def} where one compares the induced orientation from the higher-dimensional entities to the orientation of the lower-dimensional entities and assigns $+1$'s and $-1$'s according to whether these orientations agree.\footnote{For completeness, $0$-cells are all given the orientation of ``sourceness" such that $\hat{\partial}_1$ reduces to the node-edge incidence matrix $\Op{\dT}^T$ we have already encountered. Using the opposite sign convention has the advantage of making the operator $\Op{\dT}^T$ map to the continuous divergence operator, and $\Op{\LaplacianMatrix}$ map to the continuous Laplacian operator without the annoying minus signs. We opted for the conventions most commonly used in the literature.} Note that $0$-cells do not have a boundary, so $\hat{\partial}_0$ maps all nodes to 0. The incidence matrix is a representation of the \textbf{boundary operator} $\hat{\partial}_{p}: \mathcal{C}_p \rightarrow \mathcal{C}_{p-1}$ while its transpose is a representation of the \textbf{coboundary operator} $\hat{\partial}_{p}^T: \mathcal{C}^{p-1} \rightarrow \mathcal{C}^{p}$.\footnote{We use the rather unconventional notation $\hat{\partial}_p^T$ for the coboundary operator because we want to ground ourselves in linear algebra. If we understand the boundary operator as an incidence matrix, then the coboundary operator is just its transpose. Mathematicians would write the coboundary operator as $d_{p-1}: \mathcal{C}^{p-1} \rightarrow \mathcal{C}^p$. This is the differential that appears in exterior algebra. } 
The incidence matrices satisfy 
\begin{eqnarray}
    \hat{\partial}_p \hat{\partial}_{p+1}= \hat{0}.
\end{eqnarray}
The adjointness relation
\begin{equation}
    \langle \hat{\partial}_p^T \vec{\omega}, \vec{\Omega} \rangle = \langle  \vec{\omega}, \hat{\partial}_p \vec{\Omega} \rangle
\end{equation}
for $p$-chain $\vec{\Omega}$ and $(p-1)$-cochain $\omega$ is just the Generalized Stokes' Theorem $\int_{\partial \Omega} \omega = \int_\Omega d\omega $ on discrete spaces! The integral operation is achieved by the inner product $\langle \cdot \rangle$, the cell complex represents the entire domain, the chain represents the domain of integration, and the incidence matrix represents the boundary operator. Rather than evaluating a derivative operator on a cochain which is a physical function like voltage, we can evaluate a boundary or coboundary operator on a chain which represents the space itself.

\subsection{Homology on cell complexes}
The homology groups of a cell complex $G$, denoted $H_p(G)$, describe the cycles which are not boundaries. We already have representations for the boundary operators, so we can write this in a form which will allow us to explicitly calculate the homology groups. The \textbf{homology groups} are defined as the quotient space:\footnote{Mathematicians usually study homology groups with coefficients in $\mathbb{Z}$. For our purposes, the coefficients are over $\mathbb{R}$, and the homology groups can be thought of as vector spaces. Perhaps it would be more appropriate for us to use the term ``homology spaces". }
\begin{equation}
    H_p(G) = \text{ker}(\hat{\partial}_p) / \text{im}(\hat{\partial}_{p+1}).
\end{equation}
The dimension of this space is called the Betti number $\beta_p$:
\begin{equation}
    \beta_p = \text{nullity}(\hat{\partial}_p) - \text{rank}(\hat{\partial}_{p+1}).
\end{equation}
Because our vector space is over the real numbers, $H_p(G)$ is isomorphic to $\mathbb{R}^{\beta_p}$. Two useful facts from linear algebra to recall are the rank-nullity theorem which implies that 
\begin{equation}
    \text{rank}(\hat{\partial}_p) + \text{nullity}(\hat{\partial}_p) = |\mathcal{C}_p|
\end{equation}
and the fact that the rank of a linear operator and its transpose are the same:
\begin{equation}
    \text{rank}(\hat{\partial}_p) = \text{rank}(\hat{\partial}_p^T).
\end{equation}
Let's work out the first couple homology groups, most relevant for graphs. 

Recall that $\hat{\partial}_0$ is the operator which maps all $0$-chains to 0, so its nullity is $|\mathcal{V}|$. The rank of $\hat{\partial}_1 = \hat{\Delta}^T$ is also the rank of $\hat{\partial}_1^T = \hat{\Delta}$, which as stated in the introduction, is $|\mathcal{V}| - k$ where $k$ is the number of connected components. Thus, $\beta_0$ counts the number of connected components. The quotient space $H_0$ is all vertices modulo connectedness and is spanned by node representatives of each disconnected component. 

For a graph, with no faces filled in (only nodes and edges), $\text{rank}(\hat{B}^T) = 0$. The nullity of $\hat{\Delta}^T$ is $|\mathcal{E}| - |\mathcal{V}| + k$, the number of cycles in the graph, as argued in the introduction. Indeed, $\beta_1$ counts the number of cycles in the graph, and $H_1$ is spanned by a cycle basis. More generally, if we allow for faces, $H_1$ is spanned by all cycles which are not the boundary of faces. 

Generalizing these ideas, the Betti number $\beta_p$ counts the number of independent topological $p$-spheres in the cell complex, and $H_p$ gives these $p$-spheres algebraic structure.

Homology is a useful tool for characterizing topological features in a graph. Persistent homology tells us which of these features is important. \textbf{Persistent homology} considers how homological features persist when constructing a sequence of cell complexes by varying a parameter. This sequence of cell complexes is called a filtration. See \cite{Aktas_Akbas_Fatmaoui_2019} for a detailed review.\footnote{The most famous filtration is the Rips Complex used to characterize high-dimensional data clouds, though this is less relevant for the physical networks considered here.} The features (connected components, cycles, etc.) which persist throughout the filtration are considered important features of the graph, though the jump from persistent to important requires a carefully chosen filtration based on the physical system of interest. For example, suppose the magnitude of the pressure drops on the edges of a flow network are used as a filtration to determine the important edges in a flow network \cite{Rocks_Liu_Katifori_2020}. Edges are added to an initially empty graph starting with the edge that has the smallest pressure drop. A new feature is born when an edge is added that is disconnected from all other edges in the graph. A feature dies when an edge is added which connects two existing clusters. Features with a long lifetime are considered important for the function of the network. 

\subsection{Dual complexes}
An embedding space for a graph is a space in which the graph can be embedded without any edge crossings. All graphs can be embedded in $\mathbb{R}^3$, but not all graphs can be embedded in $\mathbb{R}^2$. Those graphs which can be embedded in $\mathbb{R}^2$ are called \textbf{planar graphs}. Every unoriented, planar graph $G$ has an unoriented \textbf{dual graph} $G^*$ which can be constructed by replacing each face of the planar graph with a node in the dual graph. The edges of $G^*$ transversely intersect the edges in $G$. The cycles in $G$ correspond to cutsets in $G^*$ and vice versa \cite{Wilson1996IntroductionEdition}. It is necessary to add an exterior face to a graph before taking its dual so that the exterior face gets mapped to a single node. With this extra exterior face / node, the graph is not embedded on the plane, but rather on the 2-sphere.

\begin{figure}
    \centering
    \includegraphics[width=\linewidth]{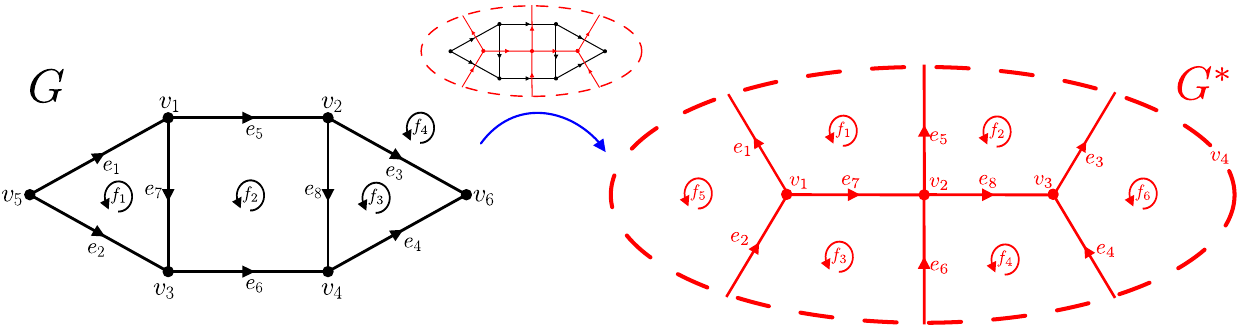}
    \caption{The dual graph $G^*$ of a graph $G$ embedded on the 2-sphere is constructed by replacing nodes with faces, faces with nodes, and edges with edges perpendicular to the original edges. Note that because the graph lives on a sphere, we must include an exterior face $f_4$ which gets mapped to a single node $v_4$ represented by a dashed line (think of it as living on the opposite side of the sphere as the rest of the graph). If the orientations of each element are carefully chosen (edges in the dual graph rotated by 90 degrees and all nodes and faces have the same orientation), then the incidence matrices satisfy $\hat{\Delta}^* = \pm \hat{B}^T$ and $\hat{B}^* = \pm \hat{\Delta}^T$. }
    \label{fig:dual_graph}
\end{figure}

In order to preserve the topological structure and define an incidence matrix, we need to be able to orient the dual graph. It is not true that every oriented planar graph has an oriented dual graph. Indeed the graph encountered in the previous figures does not possess an oriented dual graph.\footnote{\textbf{EXERCISE}: For the graph with faces shown in figure \ref{fig:GraphWithFaces}, add an extra exterior face, and construct an unoriented dual graph pictorially. Then attempt to construct an oriented dual graph. What's the problem?} In order to define duality properly, we will need a special type of cell complex which is itself a closed manifold. Details can be found in \cite{Grady_Polimeni_2010}, but we will explain how this notion relates to the dual of planar graphs. When taking the dual of a planar graph, we had to introduce an exterior face. This creates a cell complex with the topology of a sphere. Asserting that all of the faces have the same face orientation, say all counterclockwise, induces a consistent orientation of nodes, say all sources. Furthermore, this construction makes it so that each edge $e_\mu$ in $G$ is incident to two faces, inducing opposite orientations on $e_\mu$. This will allow us to define an orientation on the edge $e^*_{\mu} \in G^*$ which cuts through $e_\mu \in G$ by drawing $e^*_\mu$ from the face in $G$ where the induced orientation disagrees with the edge orientation to the face where the induced orientation agrees with the edge orientation. This construction not only allows us to pictorially construct a dual graph, but also defines an incidence matrix on the dual graph. The incidence matrices of $G$ are related to the incidence matrices of $G^*$ by  $\hat{\partial}_p = \pm(\hat{\partial}_{3-p}^*)^T$, for $p=1$ or $2$. The overall $+$ or $-$ is related to the global orientation of the space (whether we rotate the edge vectors clockwise or counterclockwise and how we orient the faces of $G^*$ relative to $G$). This is most easily demonstrated with an example. See figure \ref{fig:dual_graph}.\footnote{\textbf{EXERCISE}: Show that for the orientations chosen in this figure, $\hat{\Delta}^{*} = \hat{B}^T$ and $\hat{B}^* = -\hat{\Delta}^T$.} This idea can be generalized to cell complexes which have the topology of the $n$-sphere:
\begin{eqnarray}
    \hat{\partial}_p = \pm (\hat{\partial}_{n-p+1}^*)^T.
\end{eqnarray}
Thus, the homology of a cell complex is related to that of its dual complex, a relation known as Poincaré duality.

\subsection{Relationship to the Exterior Calculus of Differential Forms}
Cochains provide a discrete counterpart to \textbf{differential forms}, the continuum objects used to describe quantities that can be integrated over points, curves, surfaces, and volumes. A differential $p$-form is integrated over a $p$-dimensional region of a cell complex to produce a $p$-cochain. For example, the electric field is a 1-form whose integral along an edge gives the voltage drop: a 1-cochain. Essentially, cochains are just discrete versions of differential forms, and the branch of math that studies differential forms is called exterior calculus. This connection makes discrete calculus a practical framework for translating continuum equations into numerical schemes that retain their topological structure, with applications ranging from continuum physics simulations to 3D image processing \cite{Crane2013DEC}. 

\newpage
\section{Circuit Theory}
Now that we are familiar with oriented graphs, we are ready to study physical networks. In the next few sections, we will study different examples of linear flow networks. The flows on these networks are driven by a linear response from a potential field and satisfy a conservation law at the nodes. Many examples of physical networks fit into this framework, but we will begin by motivating with perhaps the most familiar example: electronic circuits. The connection between circuits and graph theory can be found in the classic engineering textbook by Strang \cite{Strang_1986}.

A \textbf{linear flow network}\footnote{A nonlinear flow network also satisfies three laws, except Ohm's Law is replaced by a nonlinear current-voltage relationship.} is a weighted graph with edge currents $\EdgeVec{I} \in \mathbb{R}^{|\EdgeSpace|}$ and edge voltages $\EdgeVec{\Psi} \in \mathbb{R}^{|\EdgeSpace|}$ satisfying the following laws at steady state:
\begin{itemize}
    \item \textbf{Ohm's Law}
    \begin{equation}
        \EdgeVec{\Current} = \Op{\Conductance} \vec{\Psi}
        \label{eq:Ohm}
    \end{equation}
    \item \textbf{Kirchhoff's Current Law (KCL)}
    \begin{equation}
        \Op{\dT}^T \EdgeVec{\Current} = \NodeVec{\current}\label{eq:KCL}
    \end{equation} 
    \item \textbf{Kirchhoff's Voltage Law (KVL)}
    \begin{equation}
        \hat{B}\vec{\Psi} = \vec{\psi}_{\ell}  .\label{eq:KVL}
    \end{equation}
\end{itemize}
The edge weights $\EdgeVec{\Conductance}$ represent the conductances of the edges (how easily the quantities on the edges flow). Each edge is assumed to have a finite conductance. In circuit theory, you could say that $\mathcal{E}$ is the set of resistors. It is most convenient not to consider voltage sources and current sources as living on the edges, but rather, these are quantities that live either on the nodes or the faces, depending on whether a node analysis or a loop analysis is performed. In a node analysis, the vector $\NodeVec{\current}\in \mathbb{R}^{|\VertexSpace|}$ is used to account for excess outflow ($\current_j>0$) or inflow ($\current_j<0$) at each node, and the edge voltage (``voltage drop") is written as the gradient of a node voltage (``electric potential"): $\vec{\Psi} = - \hat{\Delta}\vec{\psi}$. In a loop analysis, the vector $\vec{\psi}_\ell \in \mathbb{R}^{|\mathcal{F}|}$ is used to account for excess voltage accumulated when traversing a closed loop (``EMF"), and the current is written as the curl of a loop current: $\vec{I} = \hat{B}^T \vec{i}_\ell$. We will demonstrate how to solve linear flow networks using both the node analysis and the loop analysis.

Away from steady state, there are additional linear relationships we can include (ex: capacitors and inductors) that involve the time derivatives of the currents and voltages. We will demonstrate techniques for solving dynamic flow networks using the node analysis.

Of course, in contexts other than circuits, $\vec{I}$ and $\vec{\Psi}$ may be called something else, but the terms ``Ohm's Law", ``KCL", and ``KVL" are still often used for any equations of this form on graphs. Ohm's Law encodes the physical linear response of the network. For most networks, the right-hand-side of equation KCL (KVL) is zero at all but a small number of nodes (faces) where boundary conditions are applied. Aside from these special nodes (faces), KCL (KVL) is a purely topological property of the graph. 

\begin{figure}
    \centering
    \includegraphics[width=\linewidth]{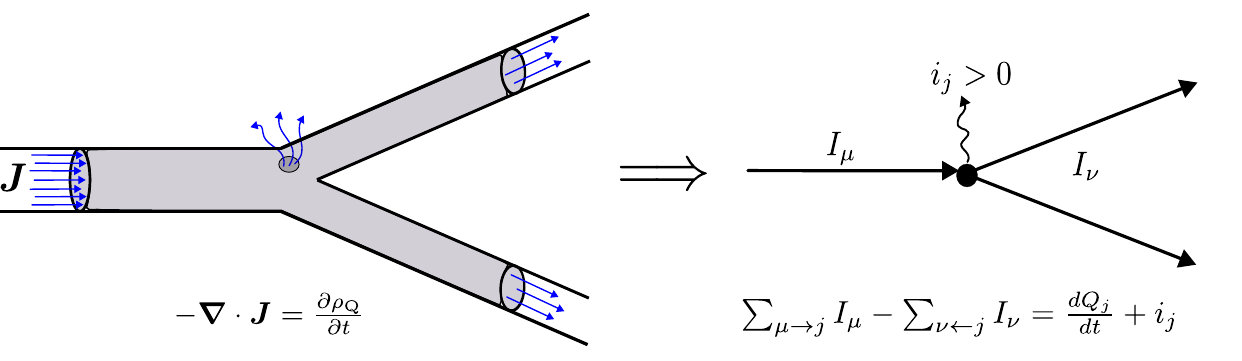}
    \caption{Visualization of how local charge conservation gives rise to Kirchhoff's Current Law (KCL). A Gaussian surface is drawn around the wires at a junction (node $v_j$). Gauss's theorem converts the integral over the volume to a sum of surface integrals over the wires, thus reproducing KCL.}
    \label{fig:flux}
\end{figure}

\subsection{KCL as a Continuity Equation}
It is useful to see how KCL derives from physics in continuous space. This calculation demonstrates a uniting feature of all physical networks. Throughout the paper, boldface will be used to denote vectors in three-dimensional space, while arrows always denote elements in the vector space of nodes or edges. 

We begin by identifying a conserved quantity which is being transported in our network. For circuits, this is the electric charge Q.\footnote{The transported quantities that appear throughout the paper will be denoted with Roman font. This notation will also be used in the subscript for the associated density.} Local charge conservation is expressed in terms of the continuity equation:
\begin{eqnarray}
    - \boldsymbol{\nabla}\cdot \boldsymbol{J} = \frac{\partial \rho_{\text{Q}}}{\partial t} \label{eq:charge_conservation}
\end{eqnarray}
where $\boldsymbol{J}$ is the current density, and $\rho_{\text{Q}}$ is the charge density. 

We can derive KCL directly from this local statement of charge conservation. We only need to assume that within any edge, the current is spatially uniform. For circuits, this is generally accepted to be the case. The current piercing a cross-section of the wire in edge $e_\mu$ is defined by
\begin{eqnarray}
    I_\mu \equiv \pm \int_{S_\mu} \boldsymbol{J}\cdot d\boldsymbol{A},
\end{eqnarray}
where the $-$ sign is needed whenever the edge orientation and $d\boldsymbol{A}$ orientation disagree.
Now, consider the gray volume $\Omega$ shown in figure \ref{fig:flux} enclosing node $v_j$ and bounded by the surface of the wires $S_w$ and the disks $\{S_\mu\}$ which intersect the wires. The current in the wires pierces each of the disks $\{S_\mu\}$, but there may also be some leakage through $S_w$. Let the wire / edge orientation agree with the direction of $\boldsymbol{J}$. Integrating equation \eqref{eq:charge_conservation} and applying Gauss's Law gives:
\begin{align}
    -\int_\Omega \boldsymbol{\nabla}\cdot \boldsymbol{J} dV &= -\sum_\mu \int_{S_\mu} \boldsymbol{J} \cdot d\boldsymbol{A} - \int_{S_w} \boldsymbol{J} \cdot d\boldsymbol{A} \nonumber\\
    &= \Big(\sum_{\mu \rightarrow j} I_\mu - \sum_{\nu \leftarrow j} I_\nu \Big) - i_j \nonumber\\
    &= (\hat{\Delta}^T \vec{I})_j- i_j, \label{eq:KCL01}
\end{align}
where we defined the node current
\begin{eqnarray}
    i_j \equiv \int_{S_w} \boldsymbol{J} \cdot d\boldsymbol{A} 
\end{eqnarray}
to be the total current leaking out of the walls of the integration region (due to, for example, a current sink). In the second line, we simply applied the definition of current and the assumption that the current is constant along the length of one wire. Note that the orientation of $\boldsymbol{J}$ relative to the boundary normal gives rise to an orientation of an edge relative to the node, and thus an element of the incidence matrix. Similarly, integrating the right side of the continuity equation gives
\begin{eqnarray}
    \int_{\Omega} \frac{\partial \rho_{\text{Q}}}{\partial t} dV = \frac{d\text{Q}_j}{dt} . \label{eq:KCL02}
\end{eqnarray}
The difference between the current entering the node and the current leaving the node can be explained by an accumulation of charge $\text{Q}_j$ at the node. Equating equations \eqref{eq:KCL01} and \eqref{eq:KCL02} recovers \eqref{eq:KCL}, generalized to the non-steady case. Note that we didn't actually apply any physics in this derivation! We only required that there is some conserved scalar quantity being transported. The physics enters the calculation when we say, for example, that the stored charge is related to the node potential by a capacitance.

Given any physical network, we will have a quantity whose continuity equation gives rise to KCL. A similar, though perhaps not as intuitive, calculation can be performed for KVL.\footnote{\textbf{EXERCISE}: Try to find an equivalent derivation of KVL from continuum equations. A starting point may be to claim that a change in flux through a surface must be driven by a circulation around the surface's boundary.}  Ohm's Law, on the other hand, is entirely a statement about the physics of a linear response. We could derive a discrete Ohm's Law from the continuous Ohm's Law (where $\boldsymbol{J}$ is related to the electric field), but for resistive circuits, the voltage drop occurs not in the wire but only over the discrete resistor element. In this case, \eqref{eq:Ohm} can be considered an empirical law. For other physical networks, like fluid flow networks, deriving the linear response equation from conservation laws and established empirical laws in three-dimensional space is an essential step in determining the appropriate network equations. Very often, one will find that no linear response equation exists.

\subsection{Solving linear flow networks I: Node analysis}
The node analysis and loop analysis are two equally valid ways of solving linear flow networks. Depending on how boundary conditions are interpreted, one can use either. In the node analysis, boundary conditions (such as battery voltages) are applied at the nodes. As a consequence, there are no loop voltages: KVL is satisfied at every face with $\vec{\psi}_\ell = 0$. This allows us to write the voltage as the gradient of a potential using \eqref{eq:BA}:
\begin{eqnarray}
    \hat{B}\vec{\Psi} = \vec{0} \implies \vec{\Psi} = -\hat{\Delta} \vec{\psi}. \label{eq:Vpsi}
\end{eqnarray}
Decoupling \eqref{eq:Vpsi}, \eqref{eq:Ohm}, and \eqref{eq:KCL} gives a single equation on the nodes written in terms of the graph Laplacian $\Op{\LaplacianMatrix}=\Op{\dT}^T \Op{\Conductance} \Op{\dT}$: 
\begin{equation}
    \Op{\LaplacianMatrix}\NodeVec{\potential} = - \NodeVec{\current}. \label{eq:poisson}
\end{equation}
Before solving this equation, let's address the question of why $\Op{\LaplacianMatrix}$ is called a Laplacian with an example. Let $G$ be a square grid with uniform conductances $\Conductance_k = 1$. The nodes are represented with the coordinates on the grid, $(x,y)\in \mathcal{V}$. Equation \eqref{eq:poisson} for the node $(x,y)$ reads
\begin{equation}  
 [\Op{\LaplacianMatrix}\potential]_{(x,y)} =  -(\potential_{(x,y+\Delta y)} + \vec{\potential}_{(x+\Delta x,y)} + \potential_{(x,y-\Delta y)}+ \potential_{(x-\Delta x,y)})+4  \potential_{(x,y)}. \nonumber
\end{equation}
If this graph is embedded in $\mathbb{R}^2$, and the potential changes smoothly over a length scale much larger than the lattice spacing, then we can approximate this expression with a Taylor expansion:
\begin{equation*}
 [\Op{\LaplacianMatrix}\vec{\potential}]_{(x,y)}\approx  - \Delta x^2\left(\frac{\partial^2 \potential}{\partial y^2} +  \frac{\partial^2 \potential}{\partial x^2}\right)_{(x,y)} = - \Delta x^2 \nabla^2 \potential_{(x,y)}.
\end{equation*}
Thus, for a grid, $-\Op{\LaplacianMatrix}$ approximates the familiar Laplacian operator $\nabla^2$ in continuous space. In general, the topology of $\hat{\LaplacianMatrix}$ is encoded in the incidence matrix, while the geometry is encoded in the conductance. Equation \eqref{eq:poisson} is analogous to a Poisson equation in a curved space. Where the node currents are zero, we can take advantage of properties of harmonic functions, as described in the appendix. 

Returning to the problem at hand, we wish to solve equation \eqref{eq:poisson}. The complication comes from the fact that $\Op{\LaplacianMatrix}$ is not invertible. If all of the node currents are known, then there are a couple simple computational ways of addressing this problem, such as using the pseudo-inverse described in the first section,\footnote{
\textbf{EXERCISE}: Prove that $\Vec{\psi} = - \hat{\LaplacianMatrix}^+ \Vec{i}$ solves $\hat{\LaplacianMatrix} \Vec{\psi} = -\Vec{i}$, where $\hat{\LaplacianMatrix}^+$ is the pseudo-inverse defined in equation \eqref{eq:pseudoinverse}. Then, show that $\Vec{\psi} = - \hat{\LaplacianMatrix}^+ \Vec{i} + [\hat{\mathbb{1}}- \hat{\LaplacianMatrix}^+ \hat{\LaplacianMatrix}] \Vec{w}$ also solves $\hat{\LaplacianMatrix} \Vec{\psi} = -\Vec{i}$ for any choice of $\Vec{w} \in \mathbb{R}^{|\VertexSpace|}$.} but this does not generalize easily when some of the node flows are unknown. The following technique is a general method of inverting the Laplacian for any node boundary conditions. An equivalent description of this method is outlined in Chapter 12 of \cite{Spielman_2019} and in \cite{Dorfler2013KronNetworks}, but we will make use of projection operators which are more computationally explicit. 

Denote the set of boundary node voltages $\mathcal{B} \subseteq \mathcal{V}$ and the set of interior nodes $\mathcal{I} = \mathcal{V} \backslash \mathcal{B}$. Suppose the vectors $\{\vec{b} \}$ form a basis for the vector space spanned by boundary nodes. Note that these basis vectors only have $|\mathcal{B}|$ elements. We will define the boundary restriction operator $\hat{\mathbb{P}}_\mathcal{B}: \mathbb{R}^{|\mathcal{V}|} \rightarrow \mathbb{R}^{|\mathcal{B}|}$ as
\begin{eqnarray}
    \hat{\mathbb{P}}_{\mathcal{B}} = \sum_{i\in \mathcal{B}} \sum_{j\in \mathcal{V}} \vec{b}_i \vec{v}_j^T.
\end{eqnarray}
Its matrix elements are 1 if the $i$th node in $\mathcal{B}$ is the $j$th node in $\mathcal{V}$ and 0 otherwise. When acting on a vector defined over all of $\mathcal{V}$, it removes the entries of elements in $\mathcal{I}$. The transposed operator $\hat{\mathbb{P}}_\mathcal{B}^T$ embeds $\mathcal{B}$ in $\mathcal{V}$. The interior restriction operator $\hat{\mathbb{P}}_\mathcal{I}$ can be defined similarly, and the two matrices are orthogonal: $\hat{\mathbb{P}}_{\mathcal{B}}\hat{\mathbb{P}}_{\mathcal{I}}^T = \hat{0}$ and $\hat{\mathbb{P}}_{\mathcal{I}}\hat{\mathbb{P}}_{\mathcal{B}}^T = \hat{0}$. Note that $\hat{\mathbb{P}}_\mathcal{B}^T \hat{\mathbb{P}}_\mathcal{B}: \mathbb{R}^{|\mathcal{V}|}\rightarrow \mathbb{R}^{|\mathcal{V}|}$ is a projection operator which sets all components of interior vertices to zero whereas $\hat{\mathbb{P}}_{\mathcal{B}}$ completely removes those entries and will be more useful for our purposes. 

Let $\vec{\psi}_{\mathcal{B}} = \hat{\mathbb{P}}_{\mathcal{B}} \vec{\psi} \in \mathbb{R}^{|\mathcal{B}|}$ and $\vec{i}_{\mathcal{I}} = \hat{\mathbb{P}}_{\mathcal{I}} \vec{i} \in \mathbb{R}^{|\mathcal{I}|}$ be known quantities, while $\vec{\psi}_{\mathcal{I}} = \hat{\mathbb{P}}_{\mathcal{I}} \vec{\psi}\in \mathbb{R}^{|\mathcal{I}|}$ and $\vec{i}_{\mathcal{B}} = \hat{\mathbb{P}}_{\mathcal{B}} \vec{i} \in \mathbb{R}^{|\mathcal{B}|}$ are unknown. In order to obtain a unique solution for the potential, it is necessary that $\mathcal{B}$ be non-empty. When $\vec{i}_{\mathcal{I}} = 0$, current is conserved at all interior nodes, and one is left with \textbf{Dirichlet boundary conditions}; in this case,  $\mathcal{B}$ must contain at least two elements to get a non-trivial solution. When only one potential is fixed to zero (the ground node) and the remaining boundary conditions are set on the current (typically zero for all but a small number of sources and sinks), one is left with \textbf{Neumann boundary conditions}. Note that these current source and sink nodes are, by our definition, elements of $\mathcal{I}$; only the potential boundary nodes are elements of $\mathcal{B}$.

The advantage to introducing the previous notation for describing boundary conditions is that the operator $\hat{\LaplacianMatrix}_{\mathcal{I}\mathcal{I}} \equiv \hat{\mathbb{P}}_{\mathcal{I}}\hat{\LaplacianMatrix}\hat{\mathbb{P}}_{\mathcal{I}}^T: \mathbb{R}^{|\mathcal{I}|} \rightarrow \mathbb{R}^{|\mathcal{I}|}$, called the \textbf{reduced Laplacian}, is invertible. The reduced Laplacian is simply the Laplacian with the potential boundary rows and columns removed. We can also define $\hat{\LaplacianMatrix}_{\mathcal{I}\mathcal{B}} \equiv \hat{\mathbb{P}}_{\mathcal{I}}\hat{\LaplacianMatrix}\hat{\mathbb{P}}_{\mathcal{B}}^T$, etc. Equation \eqref{eq:poisson} can be decomposed as follows\footnote{\textbf{EXERCISE}: This equation is most apparent when the nodes in $\mathcal{V}$ are already ordered such that all the interior nodes come first and all the boundary nodes come after, but this form holds true regardless of the ordering in $\mathcal{V}$. Prove this decomposition holds in general. The completeness relation $\hat{\mathbb{P}}_{\mathcal{B}}^T \hat{\mathbb{P}}_{\mathcal{B}} + \hat{\mathbb{P}}_{\mathcal{I}}^T \hat{\mathbb{P}}_{\mathcal{I}} = \hat{\mathbb{1}}$ may be useful. }:
\begin{eqnarray}
    \begin{pmatrix}
        \hat{\LaplacianMatrix}_{\mathcal{I}\mathcal{I}} & \hat{\LaplacianMatrix}_{\mathcal{I}\mathcal{B}} \\
        \hat{\LaplacianMatrix}_{\mathcal{B}\mathcal{I}} & \hat{\LaplacianMatrix}_{\mathcal{B}\mathcal{B}}  
    \end{pmatrix}
    \begin{pmatrix}
        \vec{\psi}_\mathcal{I} \\\vec{\psi}_\mathcal{B} 
    \end{pmatrix} = - \begin{pmatrix}
        \vec{i}_\mathcal{I} \\\vec{i}_\mathcal{B} 
    \end{pmatrix}.
\end{eqnarray}
Solving for the unknowns,

\begin{eqnarray}
    \vec{\psi}_{\mathcal{I}} = \hat{\LaplacianMatrix}_{\mathcal{II}}^{-1} \left[-\vec{i}_{\mathcal{I}} - \hat{\LaplacianMatrix}_{\mathcal{IB}}\vec{\psi}_{\mathcal{B}} \right] \label{eq:pI}
\end{eqnarray}
\begin{equation}
    \vec{i}_\mathcal{B} = (\hat{\LaplacianMatrix}_{\mathcal{BI}})(\hat{\LaplacianMatrix}_{\mathcal{II}})^{-1} \vec{i}_\mathcal{I} -\left[ (\hat{\LaplacianMatrix}_{\mathcal{BB}})-(\hat{\LaplacianMatrix}_{\mathcal{BI}}) (\hat{\LaplacianMatrix}_{\mathcal{II}})^{-1} (\hat{\LaplacianMatrix}_{\mathcal{IB}})\right]\vec{\psi}_\mathcal{B}. \label{eq:qB}
\end{equation}
These expressions can often be simplified. For the special case of Neumann boundary conditions, $\vec{\psi}_\mathcal{B} = \vec{0}$. For the special case of Dirichlet boundary conditions, $\vec{i}_\mathcal{I} = \vec{0}$, and equation \eqref{eq:qB} looks a lot like equation \eqref{eq:poisson} with the quantity in brackets playing the role of a Laplacian on the boundary nodes.\footnote{Mathematicians would call $(\hat{\LaplacianMatrix}_{\mathcal{BB}})-(\hat{\LaplacianMatrix}_{\mathcal{BI}}) (\hat{\LaplacianMatrix}_{\mathcal{II}})^{-1} (\hat{\LaplacianMatrix}_{\mathcal{IB}})$ the Schur complement of $\hat{\LaplacianMatrix}$ with respect to $\hat{\LaplacianMatrix}_{\mathcal{II}}$. This is a new Laplacian defined only on the boundary obtained by performing Gaussian elimination on the interior nodes \cite{Spielman_2019}. The operator $(\hat{\LaplacianMatrix}_{\mathcal{BI}})(\hat{\LaplacianMatrix}_{\mathcal{II}})^{-1}$ maps interior flows to boundary flows, and one can interpret equation \eqref{eq:qB} as an equation on a reduced network consisting only of the boundary nodes \cite{Dorfler2013KronNetworks}.} Once the interior potentials have been solved for, we can repackage the potentials back into our original basis: $\vec{\psi} = \hat{\mathbb{P}}_{\mathcal{B}}^T \vec{\psi}_{\mathcal{B}} + \hat{\mathbb{P}}_{\mathcal{I}}^T \vec{\psi}_{\mathcal{I}} $. Then, the edge currents can be obtained using \eqref{eq:Ohm}. 

A step-by-step demonstration of the node analysis in a system with both a current and voltage source is shown in the left column of figure \ref{fig:circuit}. A minimal Python code to solve that network is shown below. In step 1, the graph is constructed. It is important to ensure that a consistent basis is used for nodes and edges. One simple way to do this is to save the nodes and edges in lists (\texttt{ordered\_nodes} and \texttt{ordered\_edges}), and pass these lists into \texttt{nx.incidence\_matrix}. In step 2, we define the physics on the graph by specifying the conductance matrix $\hat{\Conductance}$, the boundary potential vector $\vec{\psi}_\mathcal{B}$, and the interior current vector $\vec{i}_\mathcal{I}$. Note that we have, at this point, fixed a basis for the boundary and interior spaces, and that the sum of the dimensions of these spaces must equal the total number of nodes. In step 3, the reduced Laplacian is inverted to solve for $\vec{\psi}_\mathcal{I}$. Note that \texttt{PP\_I} and \texttt{PP\_B} are the interior restriction and boundary restriction operators $\hat{\mathbb{P}}_\mathcal{I}$ and $\hat{\mathbb{P}}_\mathcal{B}$,  constructed by taking the first $|\mathcal{I}|$ rows and last $|\mathcal{B}|$ rows of a $|\mathcal{V}| \cross |\mathcal{V}|$ identity matrix, respectively. Even if we had not put all of the boundary nodes at the end of \texttt{ordered\_nodes}, we could select the rows of the identity matrix based on which nodes are in the interior versus boundary. At the end of step 3, the node potentials $\vec{\psi}$ and edge currents $\vec{I}$ are completely solved for in the bases specified by \texttt{ordered\_nodes} and \texttt{ordered\_edges}. This algorithm can be adapted to any linear flow network. 

\begin{lstlisting}[language=Python, caption={Solving a linear flow network computationally}] 
import networkx as nx
import numpy as np
import scipy.sparse as sp

#Step 1: Label graph, calculate incidence matrix
G = nx.DiGraph()
ordered_nodes = [1, 2, 3, 4, 5, 6]
ordered_edges = [(5,1),(5,3),(2,6),(4,6),(1,2),(3,4),(1,3),(2,4)]
G.add_nodes_from(ordered_nodes)
G.add_edges_from(ordered_edges)
nx.set_node_attributes(G, {1: (0,1), 2: (1,1), 3: (0,0), 4: (1,0), 
                           5: (-np.sqrt(3)/2,1/2), 6: (1+np.sqrt(3)/2,1/2)}, 
                       "pos")
DeltaT = nx.incidence_matrix(G, oriented=True, 
                             nodelist=ordered_nodes, edgelist=ordered_edges)
Delta = DeltaT.T

#Step 2: Identify Knowns and Unknowns
conductance = np.array([1, 1, 1, 1, 1, 1, 1, 1])
psi_B = np.array([1, 0])
i_I = np.array([-1, 1, 0, 0])
N_B = len(psi_B)
N_I = len(i_I)
N = nx.number_of_nodes(G)
M = nx.number_of_edges(G)
assert N_I + N_B == N and N_B > 0, "Invalid boundary conditions"
assert len(conductance) == M, "Invalid conductance assignment"
    
#Step 3: Solve for unknowns by inverting reduced Laplacian
Conductance = sp.diags(conductance, format="csr")
Laplacian = DeltaT @ Conductance @ Delta
PP_I = sp.eye(N, N, format="csr")[:N_I, :]
PP_B = sp.eye(N, N, format="csr")[-N_B:, :]
L_II = PP_I @ Laplacian @ PP_I.T
L_IB = PP_I @ Laplacian @ PP_B.T
psi_I = sp.linalg.spsolve(L_II, -i_I - L_IB @ psi_B)
psi = PP_I.T @ psi_I + PP_B.T @ psi_B
I = -Conductance @ Delta @ psi

print(psi)
print(I)
\end{lstlisting}

\begin{figure}
    \centering
    \includegraphics[width=.94\linewidth]{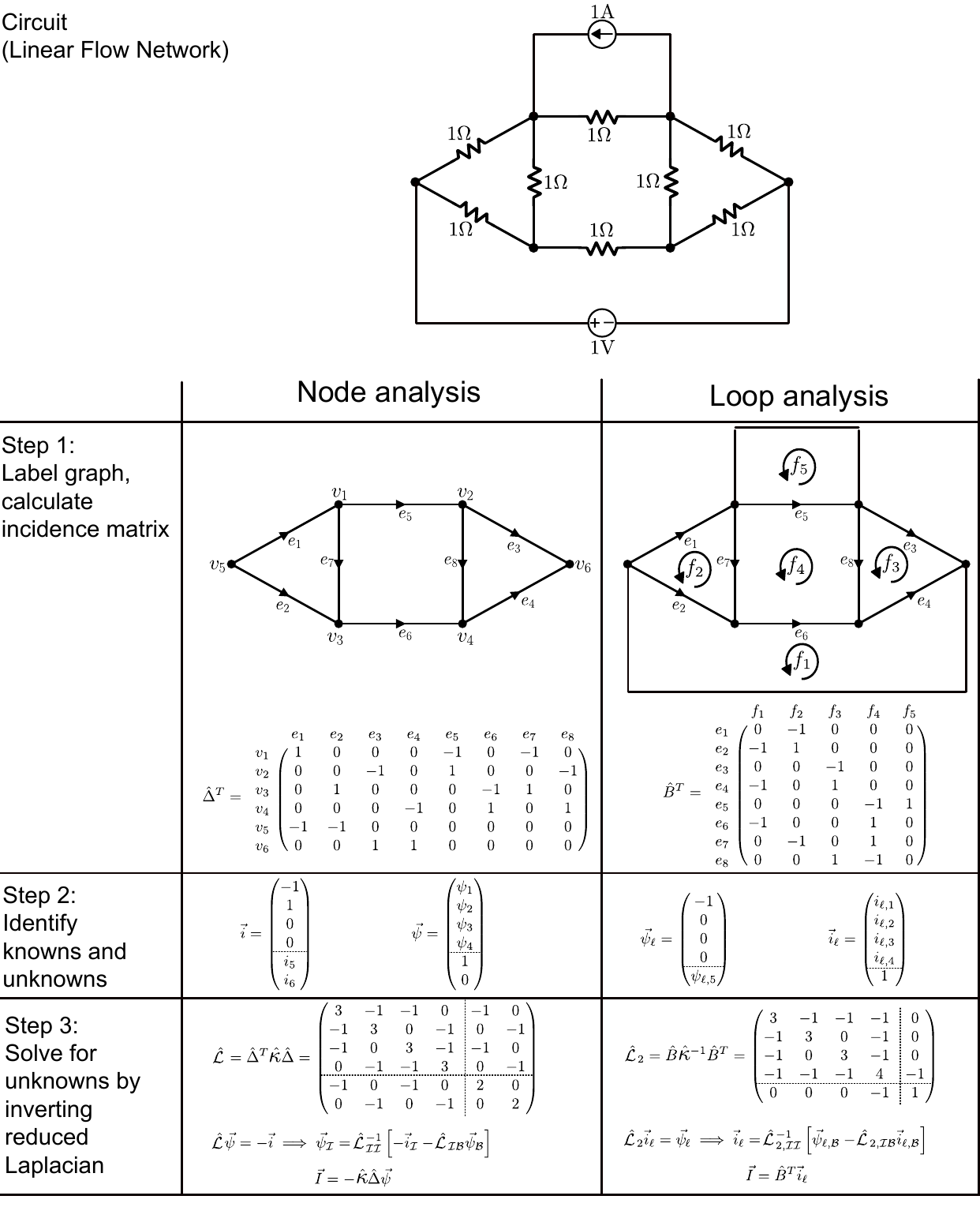}
    \caption{Review of circuit theory. An example circuit is shown on the top with a 1 Amp current source, 1 Volt DC battery, and eight 1 Ohm resistors. The edge currents $\vec{I}$ can be found by using either the node analysis or loop analysis. In the node analysis, the node-edge incidence matrix $\hat{\Delta}^T$ is used, the interior node currents $\vec{i}_{\mathcal{I}}$ and boundary potentials $\vec{\psi}_{\mathcal{B}}$ are known, the unknown potentials $\vec{\psi}_I$ are found by inverting the reduced node Laplacian $\hat{\LaplacianMatrix}_{\mathcal{II}}$, and the edge currents $\vec{I}$ are found using Ohm's Law. In the loop analysis, the edge-face incidence matrix $\hat{B}^T$ is used, the interior loop voltages $\vec{\psi}_{\ell, \mathcal{I}}$ and boundary loop currents $\vec{i}_{\ell, \mathcal{B}}$ are known, the unknown loop currents $\vec{i}_{\ell, \mathcal{I}}$ are found by inverting the reduced face Laplacian $\hat{\LaplacianMatrix}_{2,\mathcal{II}}$, and the edge currents $\vec{I}$ are found by taking the curl of the loop currents. The dashed lines in the matrices are used to separate interior and boundary elements.}
    \label{fig:circuit}
\end{figure}

\subsection{Solving Linear Flow Networks II: Loop analysis}
The loop analysis is more appropriately named the face analysis since its equations are solved on the faces of the graph. In the loop analysis \cite{Strang_1986,Irwin_2020}, boundary conditions are imposed on the faces, and KCL is everywhere satisfied with $\vec{i}=0$, which allows us to write the current as the curl of a loop current using \eqref{eq:AB}: 
\begin{equation}
    \hat{\Delta}^T \Vec{I} = \Vec{0} \implies \Vec{I} = \hat{B}^T \vec{i}_{\ell} \label{eq:loop_currents}.
\end{equation}
Notably, if an edge is incident to only a single face, then the loop current is equal to the edge current (if the orientation of the cycle and the edge agree, otherwise the loop current is the negative edge current). Decoupling equations \eqref{eq:loop_currents}, \eqref{eq:Ohm}, and \eqref{eq:KVL} gives an equation in terms of the face Laplacian $\hat{\LaplacianMatrix}_2 = \hat{B} \hat{\Conductance}^{-1} \hat{B}^T$:
\begin{equation}
    \hat{\LaplacianMatrix}_2 \vec{i}_\ell \equiv \hat{B} \hat{\Conductance}^{-1} \hat{B}^T \Vec{i}_{\ell} = \vec{\psi}_{\ell}. \label{eq:poisson_dual}
\end{equation}
At this point, the computation parallels that of the node analysis (it's even easier in fact), but constructing an appropriate basis and defining the loop currents $\vec{i}_\ell$ and loop voltages $\vec{\psi}_\ell$ can be confusing. Typically, the cycle basis for a planar graph is chosen to form a mesh: cycles in a planar graph which do not contain other cycles, and all cycles are taken to be faces. When a battery supplies a voltage, one can draw a face on the loop containing the battery, and if the orientation of the battery and face agree, then a positive loop voltage equal to the battery voltage is fixed on that face (if the orientations disagree, the loop voltage is equal to minus the battery voltage). The same can be said for loop currents. In the circuit in figure \ref{fig:circuit}, the loop voltage and loop current can be identified intuitively. Carefully choosing a cycle basis can prevent confusion regarding these quantities.\footnote{\textbf{EXERCISE}: Make sure you know how to go from the circuit at the top of figure \ref{fig:circuit} to the vectors in step 2. Find the edge currents $\vec{I}$ using both the node and loop analyses. Did you get the same answer?} 

The loop current $\vec{i}_\ell$ is unknown everywhere except along the cycles with current sources, and the loop voltage, or EMF, $\Vec{\psi}_\ell$ is known everywhere except along the cycles with current sources. Just as with our node analysis where we had to decompose our space into boundary nodes (where the potential was known) and interior nodes (all other nodes, including node current sources), here we will have to decompose our space into boundary faces (where the loop current is known) and interior faces (all other faces, including voltage sources). Then, the loop currents can be found by inverting $\hat{B} \hat{\Conductance}^{-1} \hat{B}^T$ restricted to the interior in equation \eqref{eq:poisson_dual}. Unlike the node analysis, the boundary here can be empty since $\hat{B} \hat{\Conductance}^{-1} \hat{B}^T$ does not necessarily have a zero eigenvalue. The physical currents can then be found using \eqref{eq:loop_currents} and the voltages can be found using \eqref{eq:Ohm}.

When should one use the loop analysis instead of the node analysis? Recall that there are $|\EdgeSpace|-|\VertexSpace|+1$ loops and $|\VertexSpace|$ nodes in a connected graph, so if there are many more edges than nodes (many parallel connections), then a node analysis is more efficient whereas if the number of edges is comparable to the number of nodes (many series connections), a loop analysis is more efficient.\footnote{\textbf{EXERCISE}: Prove using a loop analysis that for a network with resistors connected in series to a single voltage source, the loop current is related to the source voltage by $\psi_{\ell} = (\sum_\mu \Conductance_\mu^{-1} ) i_\ell $. Then, prove using a node analysis that for a network containing two nodes with many resistors in parallel, the potential difference between the two nodes is related to the node current by $\sum_\mu \Conductance_\mu (\psi_1 - \psi_2) = i_2$. Comment on the difficulty on trying to prove these relations using the opposite analyses. } However, if computational efficiency isn't a concern, then it is best to use whichever method is easiest and most intuitive, which for many of our applications is the node analysis. The loop analysis is also harder to implement in Python. Note that our code for analyzing the node analysis was simple once the graph was constructed because we could use the function \texttt{nx.incidence\_matrix} to calculate the node-edge incidence matrix. There is no function in NetworkX for calculating the edge-face incidence matrix. It is good to know that the loop analysis exists as it helps to clarify the structure of Kirchhoff's Laws and may greatly improve efficiency of solving certain networks, but we (the authors) can honestly say we never use it. For the remainder of the paper, we will only be using the node analysis.

\subsection{Power minimization}
Physical laws are often the result of energy minimization principles, or in the case of circuits, power. For an electric circuit (in the node analysis where KVL is everywhere satisfied), the solution $\vec{\psi}$ to an equation of the form \eqref{eq:poisson} minimizes the functional\footnote{\textbf{EXERCISE}: Show that if you take the gradient of equation \eqref{eq:Power} with respect to $\vec{\psi}$, you recover equation \eqref{eq:poisson}.}
\begin{equation}
    E[\vec{\psi}] = \frac{1}{2} \vec{\psi}^T \hat{\LaplacianMatrix} \vec{\psi} + \vec{\psi}^T \vec{i} = \frac{1}{2} (\hat{\Delta }\vec{\psi})^T \hat{\Conductance} \hat{\Delta }\vec{\psi} + \vec{\psi}^T \vec{i} . \label{eq:Power}
\end{equation}
The physical potentials are the $\vec{\psi}$ which minimize $E$ subject to the constraints given by the boundary conditions. This is just the familiar formula for the power dissipated in an electric circuit: The power is the sum of the squared voltage drops over all edges, weighted by the conductance of each edge. Note also that 
\begin{equation}
    \frac{\partial^2 E}{\partial \psi_i \partial \psi_j} = \LaplacianMatrix_{ij}.
\end{equation}
That is, the Laplacian we have been working with is the Hessian of the power. It tells us about the curvature in the space of node voltages. The zero eigenvalue of the Laplacian suggests (quite intuitively in this case) that uniform shifts in the potential cost no energy. Changes in the potential aligned with a low curvature (small eigenvalue) direction are less energetically costly than those aligned with a high curvature (large eigenvalue) direction. 

The loop analysis (assuming KCL is satisfied everywhere) suggests that we should minimize a power function of the form
\begin{equation}
    E[\vec{i}_\ell] = \frac{1}{2} \vec{i}_\ell^T \hat{\LaplacianMatrix}_2 \vec{i}_\ell - \vec{i}_\ell^T \vec{\psi}_\ell = \frac{1}{2} (\hat{B}^T \vec{i}_\ell)^T\hat{\Conductance}^{-1} \hat{B}^T \vec{i}_\ell - \vec{i}_\ell^T \vec{\psi}_\ell
\end{equation}
in order to recover equation \eqref{eq:poisson_dual}. Recalling that $\hat{B}^T \vec{i}_\ell = \vec{I}$, we see that this expression is another familiar form of the power: a sum of the currents over all edges, weighted by the resistance of each edge. The Laplacian $\hat{\LaplacianMatrix}_2$ acts as a Hessian in the power landscape of loop currents. 

\subsection{Quantifying networks with the effective resistance}
For any connected graph, the \textbf{effective resistance between nodes $v_j$ and $v_k$}, denoted $R_{jk}$ is defined as the ratio of the voltage to the current when a current $i$ is injected at node $v_j$ and extracted at node $v_k$:
\begin{eqnarray}
    R_{jk} \equiv \frac{\psi_j - \psi_k}{i}.
\end{eqnarray}
That is, we solve the Neumann problem with $i_j = -i$, $i_k=+i$, and all other node flows are zero, and calculate the above ratio. The result does not depend on the value of $i$ and is always non-negative (zero for $j=k$). The effective graph resistance $R_G$ is defined as 
\begin{equation}
    R_G \equiv \frac{1}{2} \sum_{j=1}^{|\VertexSpace|} \sum_{k=1}^{|\VertexSpace|} R_{jk}.
\end{equation}
This characterizes the overall difficulty for flows in the graph. At first glance, this seems like a rather complicated quantity to calculate, since each $R_{ij}$ requires us to solve a different flow problem. Remarkably, it has been shown \cite{Klein_Randic_1993, Van_Mieghem_2011} that the effective graph resistance satisfies 
\begin{eqnarray}
    R_G = |\VertexSpace| \sum_{\{k|\lambda_k \neq 0\} } \lambda_k^{-1},
\end{eqnarray}
where $\lambda_k$ is an eigenvalue of $\hat{\LaplacianMatrix}$. Thus, the Laplacian not only informs the local energetics of the graph, it can also be used to globally characterize the resistance of the graph.\footnote{\textbf{EXERCISE}: Calculate $R_{56}$ and $R_{G}$ for the graph shown in figure \ref{fig:circuit}.}

\subsection{Solving Dynamic Linear Flow Networks (RLC Circuits)}
The techniques presented so far can be applied to any network with steady potential and current sources. In this section, we will demonstrate how to generalize the node analysis to linear networks with dynamics. 

Consider a circuit consisting of edge resistors $R$, edge inductors $L$, node capacitors $C$, and dynamic sources.\footnote{A node capacitor has one of its terminals connected to ground such that the voltage drop across the capacitor is the node potential (minus zero). One could also consider edge capacitors where the edge current is proportional to the time derivative of the edge voltage, though it is the node capacitors which have a clear analogy to other problems studied in this tutorial, so we will introduce them here. } The responses of individual circuit elements are:
\begin{align}
    \Psi &= R I \\
    \Psi &= L \frac{dI}{dt} \\
    \frac{d\text{Q}}{dt} &= C \frac{d\psi}{dt}.
\end{align}
Thus, many dynamic linear circuits with voltage sources can be solved using the node analysis and the following equations:
\begin{align}
    -\hat{\Delta} \vec{\psi} &= \hat{R} \vec{I} + \hat{L} \frac{d\vec{I}}{dt} \label{Ohm_complex}\\
    \hat{\Delta}^T \vec{I} &= \hat{C}\frac{d\vec{\psi}}{dt} + \vec{i} \label{eq:KCL_complex}.
\end{align}
This suggests that each edge has an inductor and resistor in series, and each interior node has a capacitor connected to ground plus an exterior current source, though any of these quantities may be zero. 

\begin{figure}
    \centering
    \includegraphics[width=0.9\linewidth]{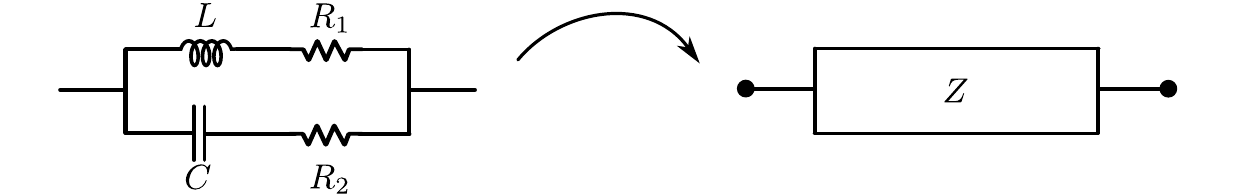}
    \caption{An example lumped element with complex impedance $Z$ is obtained by grouping four circuit elements together.}
    \label{fig:impedance}
\end{figure}

Of course, there are other linear circuits that one could solve. Each edge may consist of lumped elements of resistors, capacitors, and inductors arranged in a particular way, and identifying an appropriate lumped element is often the first step to solving a linear flow network. An example lumped element consisting of two resistors, an inductor, and a capacitor is shown in figure \ref{fig:impedance}. We could have let each of the four circuit elements be its own edge, but that could lead to complications while solving. When the capacitor is fully charged, the impedance of that branch diverges (the conductance goes to zero), but the impedance of the entire lumped element $Z$ remains finite as long as $R_1$ is finite.

\subsubsection{Method 1: Solving dynamic networks in the frequency domain}
We seek the steady state solution to a circuit driven by periodic (AC) current and voltage sources. There may still be a constant (DC) component driving a constant current. Let $\omega_0$ denote the fundamental frequency such that the boundary conditions have period $2\pi/\omega_0$. Then, we can expand our functions in a Fourier series:
\begin{equation}
    \vec{\psi}(t) = \sum_{m=-\infty}^{\infty} \vec{\tilde{\psi}}^{(m)} e^{i m \omega_0 t}, \hspace{1cm} \vec{I}(t) = \sum_{m=-\infty}^{\infty} \vec{\tilde{I}}^{(m)} e^{i m \omega_0 t}, \label{eq:FT}
\end{equation}
where $\tilde{\psi}^{(m)}_i$ denotes the $m$-th Fourier coefficient of the potential on node $v_i$, and $\tilde{I}^{(m)}_\mu$ denotes the $m$-th Fourier coefficient of the current on edge $e_\mu$. Defining a complex impedance on edge $e_\mu$ as $Z_\mu (\omega) = R_\mu + i \omega L_{\mu}$, Fourier transforming \eqref{Ohm_complex} and \eqref{eq:KCL_complex} gives
\begin{align}
    \vec{\tilde{I}}^{(m)} &= - \hat{Z}(m \omega_0)^{-1} \hat{\Delta} \vec{\tilde{\psi}}^{(m)}, \label{eq:complex_Ohm} \\
    \hat{\Delta}^T \vec{\tilde{I}}^{(m)} &= im\omega_0 \hat{C}\vec{\tilde{\psi}}^{(m)} + \vec{\tilde{i}}^{(m)}
\end{align}
Decoupling these equations gives a complex Laplacian:
\begin{eqnarray}
    \left[im\omega_0 \hat{C} + \hat{\Delta}^T \hat{Z}^{-1}(m\omega_0) \hat{\Delta} \right] \vec{\tilde{\psi}}^{(m)} = -\vec{\tilde{i}}^{(m)}. \label{eq:complex_Laplacian}
\end{eqnarray}
Solving for $\vec{\tilde{\psi}}^{(m)}$ proceeds exactly as outlined in section III.B. One needs to invert a complex-valued Laplacian for each value of $m$. Note that for $m \neq 0$ and nonzero node capacitance, the operator in brackets is invertible since time-varying potentials lead to measurable changes in current, while the $m=0$ (static) component reduces to the usual Laplacian for a resistive circuit which is not invertible. In practice, the driving currents and voltages typically take a simple form, and one only needs to compute a small number of $m$. For example, if the circuit is driven by an AC voltage at node 1, $\psi_1(t) = \sin (\omega_0 t) = (\frac{1}{2i} e^{i\omega_0 t} - \frac{1}{2i} e^{-i\omega_0 t})$, then only the $m=1$ and $m=-1$ equations will need to be considered. Also, since $\psi_i(t)$ is real, $\tilde{\psi}_i^{(m)} = \big(\tilde{\psi}_i^{(-m)}\big)^*$, so one does not need to separately solve the $m>0$ and $m<0$ cases. Once each of the Fourier coefficients $\{\vec{\tilde{\psi}}^{(m)} \}$ is known, we have successfully solved for $\vec{\psi}(t)$ using \eqref{eq:FT}.\footnote{\textbf{EXERCISE}: Find the complex impedance of the lumped edge in figure \ref{fig:impedance}. Then, solve for the node potentials in the network shown in figure \ref{fig:circuit} assuming that the current source is now an AC current source, the battery is now an AC voltage (at the same frequency), and the resistors are now impedances like that in figure \ref{fig:impedance}. You may assume that the node capacitance is zero and set all constants equal to 1. }

Solving dynamic networks in the frequency domain is simple whenever the boundary conditions are periodic. Depending on the problem, the choice of lumped elements and their arrangement with respect to each other (in series or parallel) may be different, in which case the operator in square brackets will be different, but the general technique of using Fourier series to find the potential for a periodic drive is the same. The method can be generalized to non-periodic driving, but one would need to perform a Fourier transform rather than a Fourier series, and thus evaluate a continuum of Fourier coefficients, and invert an uncountably infinite number of Laplacians. A better approach is to stick to the time domain.

\subsubsection{Method 2: Solving dynamic networks in the time domain}
In many cases, one is interested in transient or non-periodic boundary conditions, in which case the Fourier transform is impractical.\footnote{Transient effects can also be studied using a third method: the Laplace transform, which conveniently incorporates initial conditions, though may be impractical for general driving forces.} Here we will demonstrate the simplest and most intuitive approach to solving circuits in the time domain. For simplicity, assume that $L_\mu = 0$. Then, decoupling equations \eqref{Ohm_complex} and \eqref{eq:KCL_complex} gives\footnote{\textbf{EXERCISE}: The solution presented in this subsection fails when the capacitance at some of the interior nodes vanishes. Suppose only some interior nodes have capacitors while others have a known node current (either zero or a current source). Generalize the method in this section to account for these combined node boundary conditions. [Hint: Partition the space $\mathcal{I}$ into two spaces, and generalize the method from section III.B.] }
\begin{eqnarray}
    \frac{d}{dt} \vec{\psi} = - \hat{C}^{-1} \hat{\Delta}^T \hat{R}^{-1} \hat{\Delta} \vec{\psi} - \hat{C}^{-1}\vec{i}.
\end{eqnarray}
An initial condition $\vec{\psi}(0)$ must be specified. 

Numerically, one could solve this using a forward difference scheme: $\vec{\psi}(t+\delta t) = \vec{\psi}(t) + \frac{d\vec{\psi}(t)}{dt} \delta t$. Ensuring that the numerical scheme converges to the physical solution requires taking an appropriately small $\delta t$ and may work better for certain schemes over others. See \cite{Press1992NumericalRecipesC} for details. However, our linear problem can actually be solved without discretizing time. First, project onto the interior nodes (the potential at the boundary nodes is known for all times after all): 
\begin{eqnarray}
    \frac{d}{dt} \vec{\psi}_{\mathcal{I}} = - [\hat{C}^{-1} \hat{\Delta}^T \hat{R}^{-1} \hat{\Delta}]_{\mathcal{II}} \vec{\psi}_{\mathcal{I}} - [\hat{C}^{-1} \hat{\Delta}^T \hat{R}^{-1} \hat{\Delta}]_{\mathcal{IB}} \vec{\psi}_{\mathcal{B}} - \hat{C}^{-1}\vec{i}_{\mathcal{I}}.
\end{eqnarray}
The subscripts $\mathcal{II}$ and $\mathcal{IB}$ are defined using the $\hat{\mathbb{P}}_{\mathcal{I}}$ and $\hat{\mathbb{P}}_{\mathcal{B}}$ operators from section III.B. The above equation is now just a system of first-order linear differential equations \cite{Strogatz2015}. Given an equation of the form 
\begin{eqnarray}
    \frac{d}{dt}\vec{x} = \hat{A} \vec{x} + \vec{b}, \hspace{1cm} \vec{x}(0) = \vec{x}_0,
\end{eqnarray}
the solution is\footnote{The exponential of a matrix is defined by its power series $e^{\hat{A}t} = \sum_{k=0}^\infty \frac{(At)^k}{k!}$. If the matrix is diagonalizable with left eigenvectors $\{\vec{\xi}_L\}$, right eigenvectors $\{\vec{\xi}_R\}$, and eigenvalues $\{\lambda_i\}$, then this reduces to $e^{\hat{A}t} = \sum_{i} e^{\lambda_i t} \vec{\xi}_{R,i} \vec{\xi}_{L,i}^T$, where we now only have to sum over the finite number of eigenvectors of $\hat{A}$.} 
\begin{eqnarray}
    \vec{x}(t) = e^{\hat{A}t} \vec{x}_0 + \int_0^t e^{\hat{A}(t-s)} \vec{b}(s) ds.
\end{eqnarray}
For our $RC$ circuit, this suggests that the reciprocal of the eigenvalues of the matrix $[\hat{C}^{-1} \hat{\Delta}^T \hat{R}^{-1} \hat{\Delta}]_{\mathcal{II}}$ describe the characteristic $RC$ time scales.

\newpage
\section{Fluid flow networks}
So far, we have studied some general properties of linear flow networks and their application to electrical circuits. Here, we will show that there is a direct mapping between resistive electronic circuits and viscous fluid flow networks. However, unlike circuits, where the voltage drop across an edge occurs over discrete elements, fluid flow networks have a continuous pressure drop along each tube. It is important to develop an understanding of fluids in three spatial dimensions before proceeding to fluids in networks. First, we will review the fundamental equations of fluid mechanics and derive KCL. Second, we will show how steady, fully-developed flow gives rise to the analogue of Ohm's Law, with viscosity playing the role of resistance. Third, we will study flow in compliant vessels where the flow is no longer uniform across an edge. A review of fluid flow networks was written by Basu and Alim \cite{BasuAlim2025}.

\subsection{A brief review of fluid dynamics}
Before studying flow on networks, let's review the basic equations of fluid dynamics. A compact introduction can be found in the book by Childress \cite{Childress2009}. 

A fluid is described by a velocity field $\boldsymbol{v}(\boldsymbol{r})$, a stress field $\boldsymbol{\sigma}(\boldsymbol{r})$\footnote{This is the Cauchy stress tensor, a symmetric tensor describing the stress of a deformed configuration. $\sigma_{ij}$ describes the $i$-th component of the stress acting on a surface whose normal is in the $j$-th direction. }, and a density field $\rho(\boldsymbol{r})$. The continuity equation derived from mass conservation is 
\begin{eqnarray}
    -\div (\rho \boldsymbol{v}) = \frac{\partial \rho}{\partial t}. \label{eq:rhof_continuity}
\end{eqnarray}
We could immediately integrate this equation at a node, like we did in section III.A, to obtain KCL for fluid flow networks.\footnote{\textbf{EXERCISE}: Perform this calculation. Show that the conservation of fluid mass gives rise to a mass flow rate conserved at the nodes. The mass flow rate could be used to study the flow of a compressible gas through pipes.} In that case, the edge current would be the mass flow rate. However, this is almost never the quantity studied in liquid flows. We will consider the edge current more commonly used for liquid flow networks: the volumetric flow rate.  

When studying fluids, it is often necessary to consider the rate of change of a property of a fluid parcel as it is being carried by the flow. Mathematically, this is described by the material derivative:
\begin{equation}
    \frac{D}{Dt} \equiv \frac{\partial}{\partial t}+ \boldsymbol{v} \cdot \boldsymbol{\nabla} .
\end{equation}
An \textbf{incompressible fluid} has a constant density 
\begin{eqnarray}
    \frac{D\rho}{Dt} = 0  \label{eq:DrhoDt}.
\end{eqnarray}
Combining equations \eqref{eq:rhof_continuity} and \eqref{eq:DrhoDt} gives that the divergence of an incompressible fluid is zero:
\begin{equation}
    \boldsymbol{\nabla} \cdot \boldsymbol{v}  = 0 \label{eq:continuity}.
\end{equation}
This takes the form of a continuity equation, with conserved quantity volume V. Most liquids are very nearly incompressible, and equation \eqref{eq:continuity} is taken as a starting point. Define the volumetric flow rate, or \textbf{flow}: 
\begin{eqnarray}
    Q \equiv \int_S \boldsymbol{v} \cdot d\boldsymbol{A}. \label{eq:flow_def}
\end{eqnarray}
Then, integrating over a volume like that in figure \ref{fig:flux} gives rise to a KCL at the nodes:
\begin{eqnarray}
    (\hat{\Delta}^T \vec{Q})_j = \frac{d\text{V}_j}{dt} + q_j \label{eq:KCL_fluid}
\end{eqnarray}
where $\frac{d\text{V}_j}{dt}$ describes the accumulation of volume due to an expanding tube, and $q_{j}$ describes the volume flowing out of node $v_j$.

Fluids also satisfy momentum conservation:\footnote{We are neglecting body forces such as gravity.} 
\begin{equation}
    \rho \frac{D\boldsymbol{v}}{Dt} = \boldsymbol{\nabla}\cdot \boldsymbol{\sigma}. \label{eq:momentum}
\end{equation}
This equation states that the transport of momentum is caused by internal stresses. The stress tensor $\sigma_{ij}$ can be further decomposed into normal $-p\delta_{ij}$ and shear $\tau_{ij}$ components:
\begin{equation}
    \sigma_{ij} = -p \delta_{ij} + \tau_{ij}.\label{eq:stress}
\end{equation}
The scalar function $p$ is called the \textbf{pressure}. The pressure will be the potential function studied on the nodes of our network. Typically, we will concern ourselves with \textbf{Newtonian} fluids in which the shear stress $\tau_{ij}$ satisfies\footnote{This is the most general form for a linear relationship between $\tau_{ij}$ and $\partial_j v_i$ in an isotropic, incompressible fluid. If the fluid is compressible, then an additional term proportional to $\partial_k v_k \delta_{ij}$ is also allowed.}
\begin{equation}
    \tau_{ij} = \mu (\partial_j v_i + \partial_i v_j), \label{eq:Newtonian}
\end{equation}
where $\mu$ is the viscosity; in which case, equation \eqref{eq:momentum} becomes the \textbf{Navier-Stokes equations}, but for now, let's consider a general shear stress. Whereas the pressure describes the isotropic forces acting in the fluid, the shear stress can be thought of as frictional forces arising from fluid layers moving at different velocities. Thus, we expect $\tau_{ij}$ depends only on the gradients of the velocity. Combining equations \eqref{eq:momentum} and \eqref{eq:stress} gives
\begin{equation}
    \rho \frac{\partial \boldsymbol{v}}{\partial t} + \rho(\boldsymbol{v} \cdot  \boldsymbol{\nabla}) \boldsymbol{v} = -\boldsymbol{\nabla} p +\div \boldsymbol{\tau} .
\end{equation}
We are interested in fluid-carrying cylindrical pipes, so let us put our equations in cylindrical coordinates (and assume no $\phi$-dependence). Only one independent component of the shear is nonzero: $\tau \equiv \tau_{rx}=\tau_{xr}$. The mass continuity \eqref{eq:continuity} and momentum conservation \eqref{eq:momentum} equations become:
\begin{equation}
    \frac{\partial v_x}{\partial x} + \frac{1}{r}\frac{\partial(r v_r)}{\partial r}  = 0 \label{eq:continuity_cylindrical}
\end{equation}
\begin{align}
    \rho \left[ \frac{\partial v_x}{\partial t} + v_x \frac{\partial v_x}{\partial x} + v_r \frac{\partial v_x}{\partial r}\right] &= - \frac{\partial p}{\partial x} + \frac{1}{r} \frac{\partial (r \tau)}{\partial r}  \label{eq:momentum_x}\\
    \rho \left[ \frac{\partial v_r}{\partial t} + v_x \frac{\partial v_r}{\partial x} + v_r \frac{\partial v_r}{\partial r}\right] &= - \frac{\partial p}{\partial r} + \frac{\partial \tau}{\partial x} .\label{eq:momentum_r}
\end{align}

\subsection{Pipe flow}
\begin{figure}
    \centering
    \includegraphics[width=0.7\linewidth]{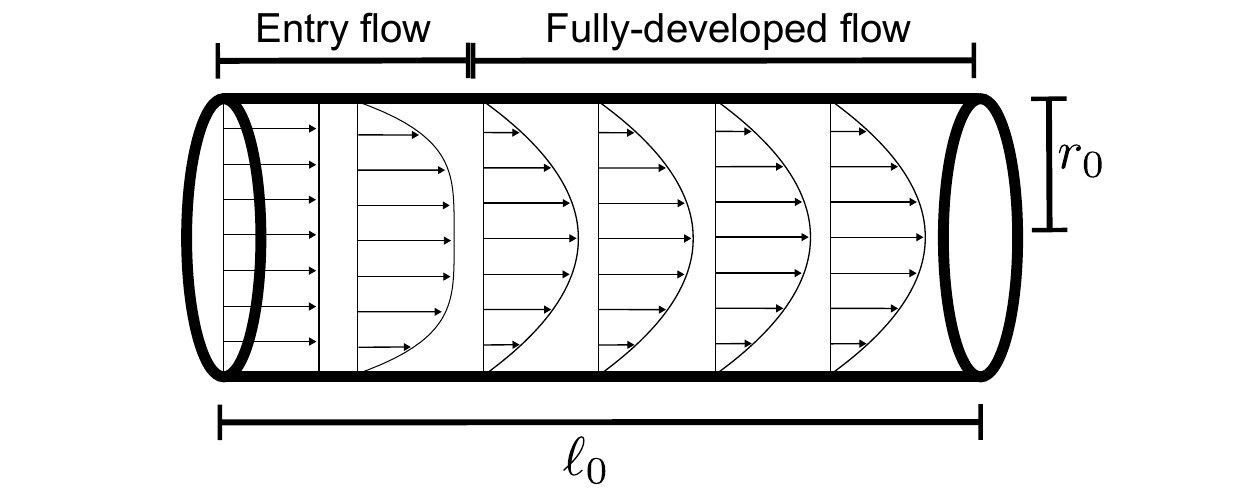}
    \caption{When a viscous fluid enters a pipe, it initially has a flat profile (entry flow), but as it travels down the pipe, it eventually develops a parabolic profile (fully-developed flow). If $\ell_0 \gg r_0$, then the flow is fully developed throughout almost all of the tube. }
    \label{fig:pipe_flow}
\end{figure}

Consider a cylindrical pipe whose length $\ell_0$ is much larger than its radius $r_0$. The flow is \textbf{steady}, meaning that the velocity and stress are independent of time, and \textbf{fully-developed}, meaning that the velocity profile is approximately independent of the position along the tube: $\boldsymbol{v}(r,x)=\boldsymbol{v}(r)$, and therefore the flow $Q$ is a constant across the tube. As a consequence of \eqref{eq:continuity_cylindrical}, we must have that $v_r(r)=0$, which also implies that $\tau(r,x)=\tau(r)$. The geometry is shown in figure \ref{fig:pipe_flow}. For steady, fully-developed flow, equations \eqref{eq:momentum_x} and \eqref{eq:momentum_r} simplify to 
\begin{equation}
    \frac{\partial p}{\partial x} = \frac{1}{r}\frac{\partial (r\tau)}{\partial r} 
\end{equation}
\begin{equation}
    \frac{\partial p}{\partial r} = 0. \label{eq:lubrication_approx}
\end{equation}
The second equation tells us that the $p(x,r) = p(x)$, which allows us to integrate the first equation to obtain:
\begin{equation}
    \frac{r}{2}\frac{\partial p}{\partial x} = \tau(r). \label{eq:tau(r)}
\end{equation}
Note that the right-hand side is not a function of $x$, so it must be that $p$ is a linear function of $x$: 
\begin{equation}
    \frac{\partial p}{\partial x} = \frac{p(\ell_0) - p(0)}{\ell_0}.
\end{equation}
Keeping in mind that $\tau$ only depends on $r$ through the gradients of the velocity, all we need to solve the pipe flow problem is a constitutive equation relating the shear stress to the relevant velocity gradient: $\tau(\partial_r v_x)$.\footnote{Some references prefer to define the \textbf{strain rate} $\dot{\gamma} \equiv  |\partial_r v_x|$, and specify the magnitude of the shear stress as a function of the strain rate: $|\tau|(\dot{\gamma})$. One can often write $|\tau|(\dot{\gamma}) = |\tau_y| + \mu_{\text{eff}}(\dot{\gamma}) \dot{\gamma}$ where $\tau_y$ is the yield stress and $\mu_{\text{eff}}$ defines an effective viscosity.}  This relationship is linear only for Newtonian fluids, equation \eqref{eq:Newtonian}. All other fluids are called non-Newtonian. Fluids that have an effective viscosity that increases with increasing strain rate are called shear thickening, while fluids that have an effective viscosity that decreases with increasing strain rate are called shear thinning. Fluids which have zero strain until a finite yield stress is applied are called plastic. Blood is an example of a Casson fluid, that is, one with both plastic and shear thinning properties. We will first explore pipe flow for a Newtonian fluid, while the Casson fluid is left as an exercise.

In any case, we apply \textbf{no-slip} boundary conditions by enforcing that the velocity vanishes on the boundary\footnote{In general, no-slip boundary conditions dictate that at a fluid-solid boundary, the fluid velocity matches the velocity of the solid.}:
\begin{equation}
    \boldsymbol{v}(r_0) = 0. \label{eq:no_slip}
\end{equation}

Once a constitutive equation for the shear stress has been specified, one can solve for the flow as a function of pressure. Let's assume the fluid is Newtonian: 
\begin{equation}
    \tau = \mu \frac{\partial v_x}{\partial r}.
\end{equation}
The coefficient $\mu$ is called the \textbf{viscosity}. With this simple form, integrate equation \eqref{eq:tau(r)} and apply the no-slip condition \eqref{eq:no_slip} to obtain the velocity: 
\begin{equation}
    v_x(r) = -\frac{r_0^2}{4\mu}\left(1 - \frac{r^2}{r_0^2} \right) \frac{\partial p}{\partial x}. \label{eq:parabolic_profile}
\end{equation}
The velocity as a function of $r$ is parabolic for steady, fully-developed flow of a Newtonian fluid. Integrating equation \eqref{eq:parabolic_profile} and applying the definition of the flow \eqref{eq:flow_def}, we find that the flow is related to the pressure drop across the pipe via
\begin{equation}
    Q = -\frac{\pi r_0^4}{8\mu \ell_0} [p(\ell_0)-p(0)].
\end{equation}
This is the most commonly used model of pipe flow, commonly known as the \textbf{Hagen-Poiseuille Law}. Notice that this equation is exactly of the form \eqref{eq:Ohm} provided that we define the conductance as 
\begin{equation}
    \Conductance = \frac{\pi r_0^4}{8\mu \ell_0}.
\end{equation}
Fluid viscosity behaves mathematically like an electrical resistor, with $Q$ playing the role of electric current and $\Delta p$ playing the role of a voltage drop.\footnote{\textbf{EXERCISE}: One can study the steady, fully-developed flow for non-Newtonian fluids as well. Consider blood as an example. The presence of red blood cells gives blood plastic and shear thinning properties. Blood can be modeled as a Casson fluid: \begin{equation*}
    \sqrt{ \Big|\mu \frac{\partial v_x}{\partial r} \Big|} = \begin{cases}
        0, & \abs{\tau} \leq  \abs{\tau_y} \\
        \sqrt{\abs{\tau}} - \sqrt{\abs{\tau_y}} , & \abs{\tau} > \abs{\tau_y}
    \end{cases} ,
\end{equation*} where $\tau_y$ is the yield stress. For stresses $|\tau|$ below the yield stress $|\tau_y|$, the fluid moves as a rigid body. Above the yield stress, the fluid is shear thinning, approaching a viscosity of $\mu$ for large strain rates. The magnitude of the shear stress is smaller near the center of the tube so define $r_y$ as the radius below which $|\tau|<|\tau_y|$. Find the velocity profile $v_x(r)$ for a Casson fluid, then integrate your expression to obtain the flow as a function of pressure. You will find that $Q(\Delta p)$ is a nonlinear function, as will be the case for any non-Newtonian fluid.} Solving for the pressures in the network amounts to solving the equation 
\begin{eqnarray}
    \hat{\Delta}^T \hat{\Conductance} \hat{\Delta} \vec{p} = -\vec{q}
\end{eqnarray}
where $\vec{q}$ is the vector of node flows. Just as with circuits, we will need to separate the knowns and unknowns and invert the reduced Laplacian to obtain the unknown pressures. Once the pressures at each node are found, the flows are obtained using Ohm's law:
\begin{eqnarray}
    \vec{Q} = -\hat{\Conductance} \hat{\Delta} \vec{p}.
\end{eqnarray}

The actual network equations are as simple as solving a resistive circuit. Appreciate the fact that we started from equations in three spatial dimensions (the Navier-Stokes equations), and obtained equations in zero spatial dimensions. Since we carefully applied a series of approximations, we can recover additional spatial information: the shear stress on the walls, the velocity field in the tubes, etc. The simplification from 3D $(r, \phi, x)$ to 1D $(x)$ was essentially obtained by integrating over the cross-section of the pipe. Though the math may get more involved, one can always define the flow as in \eqref{eq:flow_def}, and consider the cross-sectionally averaged pressure.\footnote{For steady, fully-developed flow in an axisymmetric pipe, we had that $p(r, \phi, x)$ was only a function of $x$, so there was no need to integrate out the transverse coordinates. But much like the 1D flow $Q$ is an integral of the 3D velocity field, we can obtain a 1D pressure $p(x,t)$ by averaging over the radial direction.} The reduction from 1D to 0D is more challenging. For steady, fully developed flow, it just so happened that $Q$ was independent of the position along the pipe, and $p$ was a linear function of $x$. In the next section, we will demonstrate how to reduce a PDE in one spatial dimension to a network problem in a more complicated scenario where $Q$ is a function of $x$. 

\begin{figure}
    \centering
    \includegraphics[width=0.8\linewidth]{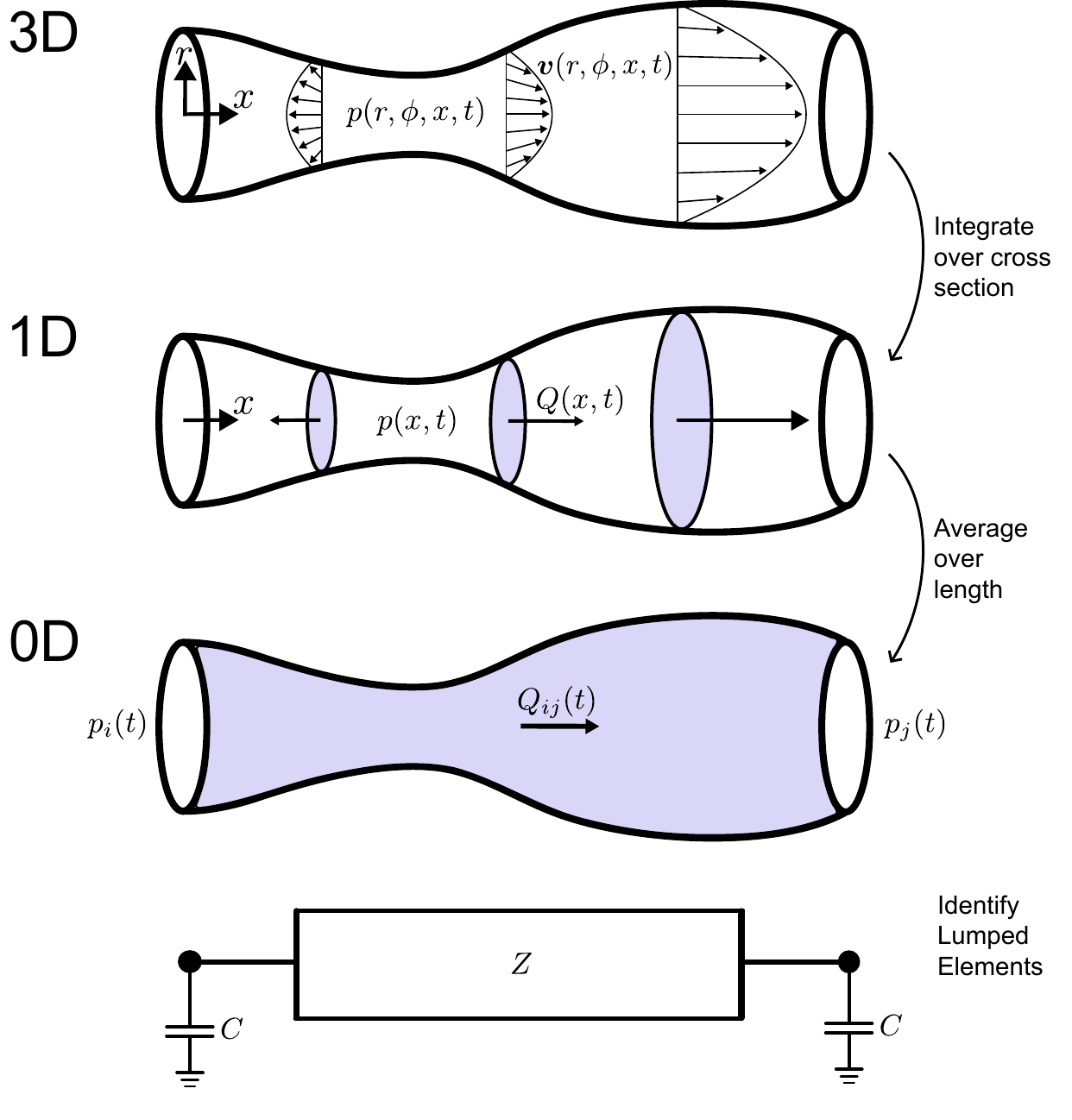}
    \caption{Pulsatile flow in three different descriptions. Developing a network model from fundamental physics often involves integrating over dimensions. For a fluid flow network, one begins with 3D equations for mass continuity and momentum conservation in a single tube, integrates over the cross section to obtain 1D equations for the pressure and flow, and finally integrates over the length of the tube to obtain 0D equations for edge flow $Q_{ij}$ and node pressures $p_i, p_j$. For the case of pulsatile flow, the dynamics of the 3D fluid and tube are characterized by the viscosity, density, and vessel stiffness. Integrating out the spatial dynamics introduces lumped elements: the impedance (which accounts for viscous and inertial effects) and the node capacitance (which accounts for vessel compliance). }
    \label{fig:fluid_flow_3Dto0D}
\end{figure}

\subsection{Solving PDEs in Networks (Pulsatile Flow Through Compliant Vessels)}

Fluid flow networks are commonly applied to vascular systems in biology. So far we have only studied steady flow, but the flow induced by a pumping heart is certainly not steady! Furthermore, biological vessels can expand to accommodate more volume. In order to take these effects into account, we will first revisit the continuity and momentum equations. Related calculations can be found in \cite{Rubenstein_Yin_Frame_2015, Fung1997, Zamir2000TheFlow}.

Let the tube radius change with space and time. The flow can still be defined by \eqref{eq:flow_def} provided that we make the replacement $r_0\rightarrow r_0 +  u_r(x,t)$, where $r_0$ is the rest radius of the tube, and $u_r(x,t)$ is the radial displacement. For the case of moving boundaries, the no-slip condition enforces that the fluid velocity equals the wall velocity. We'll assume that the wall only moves in the radial direction so the length remains $\ell_0$. The no-slip condition reads:
\begin{equation}
    v_x|_{r=r_0+u_r(x,t)} = 0,
    \hspace{1cm}v_r|_{r=r_0+u_r(x,t)} = \frac{\partial u_r(x,t)}{\partial t}.\hspace{1cm}
\end{equation}
The flow is now a function of space and time. Multiplying equation \eqref{eq:continuity_cylindrical} by $2\pi r$, integrating from 0 to $r_0 + u_r(x,t)$, and applying no-slip boundary conditions gives
\begin{equation}
    \frac{\partial Q}{\partial x} +  \frac{\partial }{\partial t} \left(\pi(r_0+u_r)^2\right)= 0.
\end{equation}
For an elastic pipe, the radius will increase as the pressure of the fluid in the pipe increases. For small deformations, this will typically look like Laplace's law \cite{Fung1997Circulation}:
\begin{equation}
    p(x,t) = \frac{Eh}{r_0^2}u_r(x,t), \label{eq:Laplace}
\end{equation}
where $E$ is the Young's modulus of the tube, $h$ is the tube thickness, and $r_0$ is the radius of the stress-free tube. Combining these equations and keeping only terms linear in the deformation, we have
\begin{equation}
    \frac{\partial Q}{\partial x} + \frac{2\pi r_0^3}{Eh}  \frac{\partial p}{\partial t} = 0. \label{eq:pulsatile_continuity}
\end{equation}
Next, we'd like a simple form of the momentum equation \eqref{eq:momentum_x}. Assume that the rest radius of an edge $r_0$ remains much less than the edge length $\ell_0$ and the wavelength of the deforming pipe $\lambda$. When the transverse length scale is much less than the axial length scale, we can apply lubrication theory: A careful scaling analysis for $r_0/\ell_0 \ll 1$ (see \cite{Tavakol2017ExtendedGeometry}) reveals that even though the flow is not steady, the pressure still only depends on $x$, and the gradient $\partial_r v_x$ dominates the shear terms. Under the lubrication approximation (and assuming a Newtonian fluid), we have 
\begin{equation}
    \rho \left[ \frac{\partial v_x}{\partial t} + v_x \frac{\partial v_x}{\partial x} + v_r \frac{\partial v_x}{\partial r}\right] = - \frac{\partial p}{\partial x} + \mu \frac{1}{r} \frac{\partial }{\partial r}  \left(r \frac{\partial v_x}{\partial r} \right) \label{eq:momentum_x_lube}
\end{equation}
\begin{equation}
   0 = \frac{\partial p}{\partial r} .\label{eq:momentum_r_lube}
\end{equation}
Finally, we will linearize our solutions about $Q=0$, $u_r= 0$. Under these approximations, the integrated momentum equation becomes 
\begin{eqnarray}
    \frac{\partial p}{\partial x} + \frac{\rho}{\pi r_0^2} \frac{\partial Q}{\partial t} + \frac{8\mu}{\pi r_0^4}Q = 0. \label{eq:momentum_1D}
\end{eqnarray}
These equations resemble the equations for transmission lines, with vessel compliance $2\pi r_0^3/Eh$ playing the role of capacitance per length, inertia $\rho/\pi r_0^2$ playing the role of inductance per length, and viscosity $8\mu / \pi r_0^4$ playing the role of resistance per length. We wish to solve equations \eqref{eq:pulsatile_continuity} and \eqref{eq:momentum_1D} in a network. There are two ways to proceed, as will be discussed in the next two subsections. 

\subsubsection{Method 1: Solving Network PDEs by Discretizing Each Edge}
One simple way of solving PDEs on networks is to further discretize the network such that each edge in the original network is decomposed into many smaller edges satisfying simpler equations. This is, in practice, how general PDEs are solved computationally. Here, we will be taking full advantage of the correspondence between the incidence matrix and the discrete finite difference operator. Throughout this subsection, an edge will be understood to be an infinitesimal edge in the discretized graph.

On each edge, we can define the edge-averaged flow:
\begin{equation}
    \bar{Q}_\mu (t) \equiv \frac{1}{\ell_{\mu}} \int_{0}^{\ell_{\mu}} Q_\mu (x,t) dx. \label{eq:edge_flow}
\end{equation}
Furthermore, we can associate the spatial pressure at an endpoint of an edge with the pressure at a node. In particular, 
\begin{equation}
    p_{\mu}(\ell_\mu,t) - p_{\mu}(0,t) = (\hat{\Delta} \vec{p}(t))_{\mu}. 
\end{equation}
Integrating equation \eqref{eq:momentum_1D} over an edge and applying the previous two equations gives
\begin{equation}
    -(\hat{\Delta} \vec{p})_{\mu} = \frac{\rho \ell_{\mu} }{\pi r_{0,\mu}^2} \frac{\partial \bar{Q}_\mu }{\partial t} + \frac{8\mu \ell_{\mu}}{\pi r_{0,\mu}^4} \bar{Q}_\mu . \label{eq:Ohm_fluid_spatial}
\end{equation}
Next, we want an integrated continuity equation. Equation \eqref{eq:KCL_fluid} still holds, though we have to be careful what we mean. Because the flow now depends on $x$, KCL will depend on how we draw the Gaussian surface! Here's where our approximation comes in. We can place the discs of the Gaussian surface (see figure \ref{fig:flux}) at the midpoint of the edges. In general, the flow at the midpoint need not equal the edge-averaged flow. However, if the edges are infinitesimally small, then the flow at the midpoint of the tube is equal to the edge-averaged flow (with corrections that scale with the infinitesimal edge length). Thus, equation \eqref{eq:KCL_fluid} holds with $\vec{Q}$ representing the edge-averaged flows $\vec{\bar{Q}}$. Note that, by our convention of enclosing half of each edge adjacent to node $j$ in our Gaussian surface, $\text{V}_j$ is equal to half the sum of all edge volumes adjacent to node $j$:
\begin{equation}
    \text{V}_j = \frac{1}{2} \sum_\mu |\Delta^T|_{j\mu} \text{V}_\mu.
\end{equation}
Recall that the unoriented incidence matrix $|\hat{\Delta}^T|$ sums over all edge-valued quantities adjacent to a node. Lastly, we use Laplace's law \eqref{eq:Laplace} to relate the change in volume to the pressure. Although this relation holds at each $x$, we can replace $p_\mu (x,t)$ with $p_j(t)$ since any corrections will be higher order in the infinitesimal edge length:
\begin{equation}
    \text{V}_j-\text{V}_{0,j} = \frac{1}{2} \sum_\mu |\Delta^T|_{j\mu} \frac{2\pi r_{0,\mu}^3 \ell_\mu }{Eh_\mu} p_j.
\end{equation}
Combining this result with equation \eqref{eq:KCL_fluid} gives
\begin{equation}
    (\hat{\Delta}^T \vec{\bar{Q}})_j = \frac{1}{2} \sum_\mu |\Delta^T|_{j\mu} \frac{2\pi r_{0,\mu}^3 \ell_\mu }{Eh_\mu} \frac{dp_j}{dt} + q_j.
\end{equation}
Defining
\begin{equation}
    R_\mu = \frac{8\mu \ell_\mu}{\pi r_{0,\mu}^4}  , \hspace{1cm} L_\mu = \frac{\rho \ell_\mu}{\pi r_{0,\mu}^2}, \hspace{1cm} C_j = \frac{1}{2} \sum_\mu |\Delta^T|_{j\mu} \frac{2\pi r_{0,\mu}^3 \ell_\mu }{Eh_\mu}, 
\end{equation}
our equations reduce to the equations for RLC circuits, equations \eqref{Ohm_complex} and \eqref{eq:KCL_complex}. This simple correspondence only holds on infinitesimal edges. Although equation \eqref{eq:Ohm_fluid_spatial} is exact for any edge length (given our convention that the edge flow is the edge-averaged flow), the continuity equation will, in general, be more complicated. 

This method can be generalized to other PDEs by finding the equations to leading-order in small edge length. This is a practical method in cases where analytic solutions are unnecessary and runtime is not a concern. However, one must check that the edge lengths are sufficiently small that the finite difference scheme converges to the physical solution.

One perk of having intuition for physical networks is that it becomes clear which operators to use when constructing a finite difference scheme, even if just solving a PDE on a line subdivided into many edges. Anyone who has ever built a PDE solver from scratch has likely considered whether to approximate a derivative using a forward difference or a backward difference.\footnote{Recall figure \ref{fig:circle_graph}: $\hat{\Delta}$ resembles a forward finite difference operator, and $-\hat{\Delta}^T$ resembles a backward finite difference operator.} Often, the approach is to use a staggered grid, where some quantities are defined at integer nodes, and other quantities are defined at the half-integer midpoints. This is more than just a mathematical tool to get convergence to the correct answer. It derives from an understanding of conservation laws and the distinction between 0-cochain (node-valued quantities like pressure) and 1-cochain (edge-valued quantities like flow). 

\subsubsection{Method 2: Solving Network PDEs Explicitly }
For some simple 1D PDEs, we can explicitly solve for $p_\mu(x,t)$ and $Q_\mu(x,t)$. We will still define the 0D edge flow $\bar{Q}_\mu(t)$ as in equation \eqref{eq:edge_flow}, but $Q_\mu(x,t)$ may have some complicated spatial structure. For simplicity, we will assume that the viscous losses are negligible for the rest of this section (just to simplify our calculations). Further, assume that the system is driven sinusoidally at a single frequency $\omega$, so that we can Fourier transform the equations and look for the steady state. Tildes will denote functions in the frequency domain. Equation \eqref{eq:Ohm_fluid_spatial} still holds, which with our current simplifications reads: 
\begin{eqnarray}
    \vec{\tilde{\bar{Q}}}(\omega) = -\hat{Z}(\omega)^{-1} \hat{\Delta} \vec{\tilde{p}}(\omega), \hspace{1cm} Z_\mu (\omega) \equiv \frac{i\omega \rho \ell_\mu}{\pi r_{0,\mu}^2} . \label{eq:Ohm_waves}
\end{eqnarray}

We can attempt to calculate KCL as we did in the previous subsection. Using a Gaussian surface with discs located at the midpoint of each edge results in a KCL relating flows at the midpoints of the edges $\tilde{Q}_\mu(\ell_\mu/2,\omega)$ incident to a node. However, since our edges are no longer infinitesimally small, we cannot equate $\tilde{Q}_\mu(\ell_\mu/2,\omega)$ with the edge-averaged flow $\tilde{\bar{Q}}_\mu(\omega)$. Along edge $e_\mu$, let $x=0$ be the tail node and $x=\ell_\mu$ be the head node. It is still true that fluid flow is conserved at a junction (assuming no flow sources/sinks):
\begin{equation}
    \sum_{\mu \rightarrow j} \tilde{Q}_\mu (\ell_\mu, \omega) - \sum_{\nu \leftarrow j} \tilde{Q}_\nu (0, \omega) = 0, \label{eq:local_KCL}
\end{equation}
but the edge-averaged flow is NOT conserved! We need to express $\tilde{Q}_\mu (\ell_\mu,\omega)$ and $\tilde{Q}_\nu (0,\omega)$ in terms of $\tilde{\bar{Q}}_\mu (\omega)$ and $\tilde{\bar{Q}}_\nu (\omega)$. Finding an appropriate KCL for $\tilde{\bar{Q}}$ will require us to solve for the pressure and flow along the edge. 

Let's forget about our network, for a moment, and solve our system of PDEs with Dirichlet boundary conditions. Decoupling equations \eqref{eq:pulsatile_continuity} and \eqref{eq:momentum_1D} and dropping the viscous term gives a wave equation with speed $c$:
\begin{eqnarray}
    \frac{\partial^2 p}{\partial x^2} -\frac{1}{ c^2} \frac{\partial^2 p}{\partial t^2} = 0, \hspace{1cm} c = \sqrt{\frac{Eh}{2 \rho r_0}}.
\end{eqnarray}
This wave speed, called the Moens-Korteweg formula, has a long history in biomechanics \cite{Fung1997Circulation}. The Fourier transformed pressure in a single tube satisfies 
\begin{eqnarray}
    \tilde{p}(x,\omega) = \tilde{p}(0,\omega) \frac{\sin \left( \frac{\omega}{c} (\ell_0-x)\right)}{\sin \left( \frac{\omega }{c}\ell_0\right)} + \tilde{p}(\ell_0,\omega) \frac{\sin \left( \frac{\omega}{c} x\right)}{\sin \left( \frac{\omega }{c}\ell_0\right)} \label{eq:p(x,omega)}.
\end{eqnarray}
This is a promising equation for applying to networks. It involves the pressure at the endpoints of the tube which for a network would be the node pressures. Next, calculate the flows, 
\begin{eqnarray}
    \tilde{Q}(x,\omega) = -\frac{\ell_0}{Z(\omega)} \frac{\partial \tilde{p}}{\partial x} =  \frac{\omega \ell_0}{c Z(\omega)}\left[ \tilde{p}(0,\omega) \frac{\cos \left( \frac{\omega}{c} (\ell_0-x)\right)}{\sin \left( \frac{\omega }{c}\ell_0\right)} - \tilde{p}(\ell_0,\omega) \frac{\cos \left( \frac{\omega}{c} x\right)}{\sin \left( \frac{\omega }{c}\ell_0\right)}  \right] \label{eq:Q(x,omega)}
\end{eqnarray}
Returning to equation \eqref{eq:local_KCL}, 

\begin{align}
    0 &= \sum_{\mu \rightarrow i} \tilde{Q}_\mu(\ell_\mu,\omega) - \sum_{\nu \leftarrow i} \tilde{Q}_\nu(0,\omega) \nonumber \\
    &= \sum_{\mu \rightarrow i} \frac{\omega \ell_\mu/c_\mu}{ \sin \left( \frac{\omega \ell_\mu}{c_\mu}\right)}\left[ \frac{\tilde{p}_\mu(0,\omega)}{Z_\mu(\omega)} - \frac{\tilde{p}_\mu(\ell_\mu,\omega)}{Z_\mu(\omega)} \cos \left( \frac{\omega \ell_\mu}{c_\mu}  \right) \right] - \sum_{\nu \leftarrow i}  \frac{\omega \ell_\nu/c_\nu}{ \sin \left( \frac{\omega \ell_\nu}{c_\nu}\right)}\left[ \frac{\tilde{p}_\nu(0,\omega)}{Z_\nu(\omega)} \cos \left( \frac{\omega \ell_\nu}{c_\nu}  \right) - \frac{\tilde{p}_\nu(\ell_\nu, \omega) }{Z_\nu(\omega)} \right] \nonumber \\
    &= \sum_{\mu \rightarrow i} \frac{\omega \ell_\mu/c_\mu}{\sin \left( \frac{\omega \ell_\mu}{c_\mu}\right)}\left[ \tilde{\bar{Q}}_\mu(\omega) + \frac{\tilde{p}_i(\omega)}{Z_\mu(\omega)}\left(1-  \cos \left( \frac{\omega \ell_\mu}{c_\mu}  \right) \right)\right] - \sum_{\nu \leftarrow i} \frac{\omega \ell_\nu/c_\nu}{  \sin \left( \frac{\omega \ell_\nu}{c_\nu}\right)} \left[ \tilde{\bar{Q}}_\nu(\omega) - \frac{\tilde{p}_i(\omega)}{Z_\nu(\omega)}\left(1- \cos \left( \frac{\omega \ell_\nu}{c_\nu} \right) \right) \right] \nonumber \\
    &= \sum_{\mu} \hat{\Delta}^T_{i \mu} \frac{\omega \ell_\mu/c_\mu}{\sin \left( \frac{\omega \ell_\mu}{c_\mu}\right)}\tilde{\bar{Q}}_\mu (\omega) + \sum_{\mu} |\hat{\Delta}|^T_{i \mu} \frac{\omega \ell_\mu/c_\mu}{Z_\mu(\omega) \sin \left( \frac{\omega \ell_\mu}{c_\mu}\right)} \left(1- \cos \left( \frac{\omega \ell_\mu}{c_\mu} \right) \right)\tilde{p}_i(\omega) \label{eq:KCL_complicated}
\end{align}
In the third line, we applied equation \eqref{eq:Ohm_waves} to express $\tilde{p}_\mu (0, \omega)$ in terms of $\tilde{\bar{Q}}_\mu(\omega)$ and $\tilde{p}_\mu (\ell_\mu, \omega)=\tilde{p}_i(\omega)$, and to express $\tilde{p}_\nu (\ell_\nu, \omega)$ in terms of $\tilde{\bar{Q}}_\nu(\omega)$ and  $\tilde{p}_\nu (0, \omega)=\tilde{p}_i(\omega)$. Equation \eqref{eq:KCL_complicated} is a continuity equation at the node purely in terms of the node pressures and edge-averaged flows. It is the appropriate generalization of KCL from the previous subsection to the case with finite edge lengths. The above equation, along with equation \eqref{eq:Ohm_waves}, define the network equations.\footnote{\textbf{EXERCISE}: By defining appropriate matrices on the edges and nodes, write this as a matrix equation. How would you solve for the pressures? Show that in the limit $\omega \ell_\mu/c_{\mu} \rightarrow 0$, your equation recovers the form from the previous subsection.} 

There are two advantages to solving the spatial dynamics of an edge by hand. One, it's computationally more efficient, at least in this case. By not having to employ finite differences, the Laplacian that we have to invert is much smaller. However, in some cases, the coefficients may involve complicated integrals with no analytic solution, and it's not clear that solving those integrals computationally saves runtime as compared with just employing finite differences. The second advantage is that we can probe the physics of a single edge. In this example, we notice something interesting. Whenever $\omega \ell_\mu/c_{\mu}$ is an integer multiple of $\pi$, there is a resonance in the pressure and flow. Although this result would have been observed computationally using the finite difference scheme, here we have an analytic form for the resonances. Furthermore, once the node pressures and edge-averaged flows are found, we can write down the full 1D dynamics of the edges using equations \eqref{eq:p(x,omega)} and \eqref{eq:Q(x,omega)}. The main drawback of this method is that even linear PDEs can be difficult to solve. Had we included the viscosity, the wave number would become complex, and one has to take care handling the square roots of complex numbers \cite{Fancher2022MechanicalNetworks}. Furthermore, dropping the lubrication assumption leads to flatter velocity profiles which can be expressed in terms of Bessel functions \cite{Zamir2000TheFlow}.

The bottom line is that if you can write down linear PDEs for the pressure and flow, there is probably a clever way to integrate the 1D equations into 0D network equations, using either of the two methods presented in this section. A visual summary is shown in figure \ref{fig:fluid_flow_3Dto0D}. One can apply this technique to other linear PDEs like the heat equation.

\newpage
\section{Mechanical Networks}
Mechanical networks come in many forms: balls attached by springs, lattice vibrations, filamentous networks, truss structures, etc. In any case, we wish to determine the displacement of the material elements when forces or displacements are applied. In the previous section, we saw that viscous fluid flow networks map directly onto resistive electric circuits. Mechanical networks possess a similar mapping, but with one key difference: each node possesses several degrees of freedom. Whereas in an electrical (fluid) network, each node is characterized by its electric potential (pressure), a mechanical network is characterized by its node position or node velocity, one number for each dimension of the embedding space. 

First, we will consider how momentum conservation gives rise to a KCL only if we use the Lagrangian coordinates of the edges. Next, we will see how static mechanical networks in one dimension map directly onto electronic circuits, but more interesting behavior can arise in higher dimensions. Lastly, we will show how dynamic networks can be mapped to RLC circuits. A complete discussion of mechanical networks and how they relate to general linear flow networks can be found in a new review paper \cite{OrtizTavarez2025UnifiedModel}, while their application to condensed matter physics is summarized in \cite{Lubensky2015Phonons}.

\subsection{Physical Networks are Lagrangian}
A continuum elastic solid is described by its displacement field $\boldsymbol{u}(\boldsymbol{R},t)$ from a rest configuration to the current configuration. In the context of elasticity theory, the coordinates $\boldsymbol{R}$ parameterizing the rest configuration are called material coordinates because they label a material element in the rest configuration. In general, coordinates that permanently label the particles of a material are called \textbf{Lagrangian}. This is to be contrasted with Eulerian coordinates $\boldsymbol{r}$ which label a particular point in space. For example, often when we describe fluids, we specify the velocity field $\boldsymbol{v}$ at a particular Eulerian point in space $\boldsymbol{r}$, not the velocity of individual parcels of fluid.

Physical networks are inherently Lagrangian. We measure quantities at a node or edge, not at a particular location in space. For a mechanical network, where the nodes are changing position with time, this is important to keep in mind.\footnote{It may seem that this problem goes away for a network like an electric circuit where we are measuring flows through the network rather than flows of the network itself, but actually, the problem arises even there. When we integrated the continuous current density $\boldsymbol{J}$ across the Eulerian volume $\Omega$ in figure \ref{fig:flux}, we secretly assumed that the circuit was not moving. But, there is no reason we can't analyze a circuit as it is being transported across a lab. In that case, our Gaussian surface should follow the circuit as it is being transported. We only care about current through the wires. We do not care that charge is being transported across the lab as we move the circuit. The relevant current $I$ is to be measured with respect to fixed Lagrangian coordinates of the wire, not at fixed coordinates in space. This may all sound silly and pedantic, but there are cases where this subtlety is quite important, such as the fluid flow through a network undergoing longitudinal stretching \cite{Winn2026GeometryFlows}.}
Ideally, the higher-dimensional models used to derive the network equations should also be expressed in Lagrangian coordinates. In terms of the Lagrangian coordinates of the solid, momentum conservation takes the form 
\begin{eqnarray}
    \div  \boldsymbol{\sigma} = \frac{\partial \boldsymbol{\rho}_{\textbf{p}}}{\partial t}, \label{eq:momentum_solid}
\end{eqnarray}
where $\boldsymbol{\rho}_{\textbf{p}}$ is the momentum density.\footnote{The stress that appears in equation \eqref{eq:momentum} is the Cauchy stress tensor relating area elements in the current configuration to forces in the current configuration. The stress appearing in equation \eqref{eq:momentum_solid} is the first Piola-Kirchhoff stress tensor relating area elements in the rest configuration to forces in the current configuration. To linear order in small displacements, which is all we will consider here, the Cauchy stress and first Piola-Kirchhoff stress tensors are identical.} Integrating this equation over a volume enclosing a junction at node $v_j$ in the rest configuration (fixed Lagrangian coordinates) gives rise to an integrated form of momentum conservation that functions as our KCL for a mechanical network. This is a fancy way of saying Newton's second law is satisfied at a node in a mechanical network. The ``edge current" in this case is the axial tension 
\begin{eqnarray}
    F_\mu \equiv \int_{S_\mu} \boldsymbol{n}_\mu \cdot \boldsymbol{\sigma} \cdot d\boldsymbol{A}.
\end{eqnarray}
Here, $S_\mu$ is a fixed cross-section in the rest configuration and $\boldsymbol{n}_\mu$ is the vector orientation of the edge.

\subsection{Static Mechanical Networks in One Dimension}
Before entering the uncharted territory of $d$-dimensional quantities on each node, let's formulate the problem in 1D, which maps directly to resistive circuits or viscous fluid flow networks. For simplicity, assume all edges are oriented to the right. 

One of the most famous equations in physics is Hooke's Law which states that the force in a spring is proportional to the change in length of the spring. Denote the rest length of the spring $l_0$, the current length $l$, the stiffness $K$, and the tension $F$. Then, Hooke's law on edge $e_\mu$ states
\begin{equation}
    F_\mu = - K_{\mu \nu} (l_\nu - l_{0,\nu}). \label{eq:Hooke_edge}
\end{equation}
In 1D, we can write the change in edge length as the difference in node displacements. If displacement of node $v_i$ is denoted $u_i$, then 
\begin{eqnarray}
    l_{ij} - l_{0,ij} = u_j - u_i.
\end{eqnarray}
Thus, we have our linear response law on the edges:\footnote{It may be surprising that Hooke's law is the analogue to Ohm's Law when both resistance and viscosity represent dissipative physics, but Hooke's law does not. Indeed, the analogy to circuits is most apparent when considering velocity, not position, as the node degree of freedom, in which case the dashpot (which is a dissipative element) plays the role of resistance. We will elaborate on this analogy in a later section.}
\begin{eqnarray}
    \vec{F} = -\hat{K} \hat{\Delta} \vec{u}.
\end{eqnarray}
KCL takes the form of force balance. Imagine that at the nodes, there is a net force $\vec{f}$ exerted by the node on the environment, or equivalently an external force $-\vec{f}$ imposed on the node. Then, the force balance on the nodes has contributions from the external force and the adjacent springs:
\begin{eqnarray}
    \hat{\Delta}^T \vec{F} = \vec{f}.
\end{eqnarray}
At this point, you know what to do: decouple force balance and Hooke's Law, fix the position of at least one node, and invert the reduced Laplacian $(\hat{\Delta}^T \hat{K}\hat{\Delta})_{\mathcal{II}}$ to obtain the unknown node displacements. The null space of $\hat{\Delta}$ is one dimensional, and it is spanned by the vector with constant entries. Much like how we could shift the electric potential of all nodes without introducing any currents, we can uniformly translate all nodes without altering any edge tensions. In higher dimensions, there are more ways in which we can displace nodes without changing the edge lengths.

\subsection{Static Mechanical Networks in \textit{d} Dimensions}
In higher dimensions, Hooke's law \eqref{eq:Hooke_edge} is still true. However, it is no longer so simple to relate the edge lengths to the displacements. If the rest position vector of node $v_i$ is denoted $\boldsymbol{r}_i$, and the displacement is denoted $\boldsymbol{u}_i$, then we can write\footnote{Recall that boldface denotes vectors with Cartesian components. An arrowhead is reserved for vectors whose elements correspond to a particular node or edge. Using a boldface and an arrowhead simultaneously indicates a node/edge-valued quantity with Cartesian components.}
\begin{equation}
    l_{0,ij} \equiv |\boldsymbol{r}_j - \boldsymbol{r}_i| =  \sqrt{(\boldsymbol{r}_j - \boldsymbol{r}_i)\cdot (\boldsymbol{r}_j - \boldsymbol{r}_i)}.
\end{equation}
\begin{equation}
    l_{ij} \equiv |(\boldsymbol{r}_j+\boldsymbol{u}_j) - (\boldsymbol{r}_i+\boldsymbol{u}_i)| =  \sqrt{(\boldsymbol{r}_j - \boldsymbol{r}_i + \boldsymbol{u}_j - \boldsymbol{u}_i)\cdot (\boldsymbol{r}_j - \boldsymbol{r}_i + \boldsymbol{u}_j - \boldsymbol{u}_i)}.
\end{equation}
Taylor expanding to linear order in $\boldsymbol{u}$, we get
\begin{align}
    l_{ij} - l_{0,ij} &= \sqrt{l_{0,ij}^2 + 2(\boldsymbol{r}_j-\boldsymbol{r}_i)\cdot (\boldsymbol{u}_j - \boldsymbol{u}_i) +(\boldsymbol{u}_j - \boldsymbol{u}_i)\cdot (\boldsymbol{u}_j - \boldsymbol{u}_i)} -l_{0,ij } \nonumber \\
    &\approx  \frac{(\boldsymbol{r}_j-\boldsymbol{r}_i)}{l_{0,ij}}\cdot (\boldsymbol{u}_j - \boldsymbol{u}_i).  
\end{align}
Thus, for small displacements, Hooke's law can be written as 
\begin{equation}
    F_{ij} = -K_{ij} \frac{(\boldsymbol{r}_j-\boldsymbol{r}_i)}{l_{0,ij}}\cdot (\boldsymbol{u}_j - \boldsymbol{u}_i) \label{eq:tauij} .
\end{equation}
The node displacements are now vector quantities! Recognize that 
\begin{equation}
    \boldsymbol{n}_{ij} = \frac{(\boldsymbol{r}_j-\boldsymbol{r}_i)}{l_{0,ij}} 
\end{equation}
is just the unit vector along edge $(v_i,v_j)$ at rest. As an edge vector, 
\begin{equation}
    \vec{\boldsymbol{n}} = \hat{l}_0^{-1} \hat{\Delta} \vec{\boldsymbol{r}}, \hspace{1cm} l_{0,\mu} = \sqrt{(\hat{\Delta} \vec{\boldsymbol{r}})_\mu \cdot (\hat{\Delta} \vec{\boldsymbol{r}})_\mu},
\end{equation}
where in the last expression, there is no summation over $\mu$. We will let $\hat{\boldsymbol{n}} = \text{diag} (\vec{\boldsymbol{n}})$. Returning to equation \eqref{eq:tauij}, we have 
\begin{equation}
    F_\mu = -K_{\mu \nu} \boldsymbol{n}_{\nu \rho} \cdot (\hat{\Delta}\boldsymbol{u})_\rho = -K_{\mu \nu} \boldsymbol{n}_{\nu \rho} \Delta_{\rho j}\cdot \boldsymbol{u}_j .
\end{equation} 
Thus, the relevant matrix describing the topology of our mechanical network is $\hat{\boldsymbol{\Delta}} \equiv \hat{\boldsymbol{n}} \hat{\Delta}$. Throughout the mechanical networks literature, this matrix is known as the \textbf{compatibility matrix}.\footnote{The transpose of the compatibility matrix $\hat{\boldsymbol{\Delta}}{}^T$ is known as the equilibrium matrix.} With this definition, we can finally write the desired network equations
\begin{equation}
     \vec{F} = - \hat{K} \hat{\boldsymbol{\Delta}} \cdot \vec{\boldsymbol{u}} \label{eq:Hooke}
\end{equation}
\begin{equation}
    \hat{\boldsymbol{\Delta}}{}^T \vec{F} = \vec{\boldsymbol{f}}. \label{eq:forceBalance}
\end{equation}
The sign convention is such that for $F_\mu > 0$, edge $e_\mu$ is in compression, and for $F_\mu < 0$, edge $e_\mu$ is in tension. 

By now, you may be annoyed that we are treating $\hat{\boldsymbol{\Delta}}$ as a matrix when it is really a three-index tensor, where one index runs over the $d$ spatial indices, but we will now see how it can be handled as a matrix. Rather than thinking of $\vec{\boldsymbol{u}}$ as a vector of vectors, let's flatten the nested vectors into one big vector of size $d|\VertexSpace|$. There are two convenient bases to work with: We may either write out all $d$ components of node 1 then all $d$ components of node 2..., or we may write the $x$-components of all $|\VertexSpace|$ nodes then all $y$-components of all $|\VertexSpace|$ nodes... We have found the latter to be more convenient. For example, in 2D, the compatibility matrix acting on the displacement vector can be simply written 
\begin{equation}
    \hat{\boldsymbol{\Delta}} \cdot \vec{\boldsymbol{u}} = \begin{bmatrix} \hat{n}_x \hat{\Delta} & \hat{n}_y \hat{\Delta} \end{bmatrix} \begin{bmatrix} \vec{u}_x \\ \vec{u}_y \end{bmatrix} .\label{eq:2DCompatibility}
\end{equation}
Unlike the mechanical 1D network, the null space of $\hat{\boldsymbol{\Delta}}$ in 2D is not one dimensional! Physically, we have, at least, uniform translations in $x$, uniform translations in $y$, and rotations through the origin that do not alter the edge lengths.\footnote{\textbf{EXERCISE}: Show that $\vec{\boldsymbol{u}} = \begin{bmatrix} \vec{1} \\ \vec{0} \end{bmatrix}$, $\begin{bmatrix} \vec{0} \\ \vec{1} \end{bmatrix}$, and $\begin{bmatrix} -\vec{y} \\ +\vec{x} \end{bmatrix}$ all satisfy $\hat{\boldsymbol{\Delta}}\cdot \vec{\boldsymbol{u}} = \vec{0}$.} These are the trivial rigid body motions; however, we may have additional displacements that leave our edge lengths invariant, something which was impossible in 1d. 

Let's now return to the general $d$-dimensional case. The compatibility matrix is a $|\EdgeSpace| \cross d|\VertexSpace|$ matrix. A \textbf{zero mode} is any solution $\vec{\boldsymbol{u}}$ to the equation\footnote{More precisely, these are the linear zero modes. Recall, the form of the edge lengths we are using is correct only to linear order in the displacements. It may be that the node displacements satisfy \eqref{eq:zeroMode}, but the edge lengths change at a higher order in $\vec{\boldsymbol{u}}$. Similarly, when we use the term ``rigid", we really mean ``first-order rigid".} 
\begin{equation}
    \hat{\boldsymbol{\Delta}}\cdot \vec{\boldsymbol{u}} = \vec{0} \label{eq:zeroMode}. 
\end{equation}
If a zero mode is not one of the $d$ rigid-body translations or $d(d-1)/2$ rigid-body rotations, then we call the zero mode a \textbf{floppy mode}. Systems which do not have any floppy modes are called rigid. While this analysis holds for crystal structures in equilibrium, crystals are always rigid, so working in the static limit doesn't say much; rather, one would see how the solid responds to perturbations by finding the normal modes (phonons) of the system. For ``soft" materials such as packed spheres, biopolymer networks, and glasses, the simple question of whether the system is rigid or not is an interesting one. 

Alongside the zero modes, we can also define a \textbf{state of self stress} as a set of nonzero tensions which exerts no net forces on the nodes, that is a solution $\vec{F}$ to the equation 
\begin{equation}
    \hat{\boldsymbol{\Delta}}{}^T \vec{F} = \vec{\boldsymbol{0}}.
\end{equation}
One can think of these states of self stress as cases where the system is over-constrained, so that even though the edges may be under tension, the excess bonds prevent the system from relaxing to a zero energy configuration. The most intuitive network that has a state of self stress is a 1D lattice with periodic boundary conditions. If each edge has an identical tension, then each node will be pulled equal amounts to the left and right such that the net force at each node is zero.

\begin{figure}
    \centering
    \includegraphics[width=.99\textwidth]{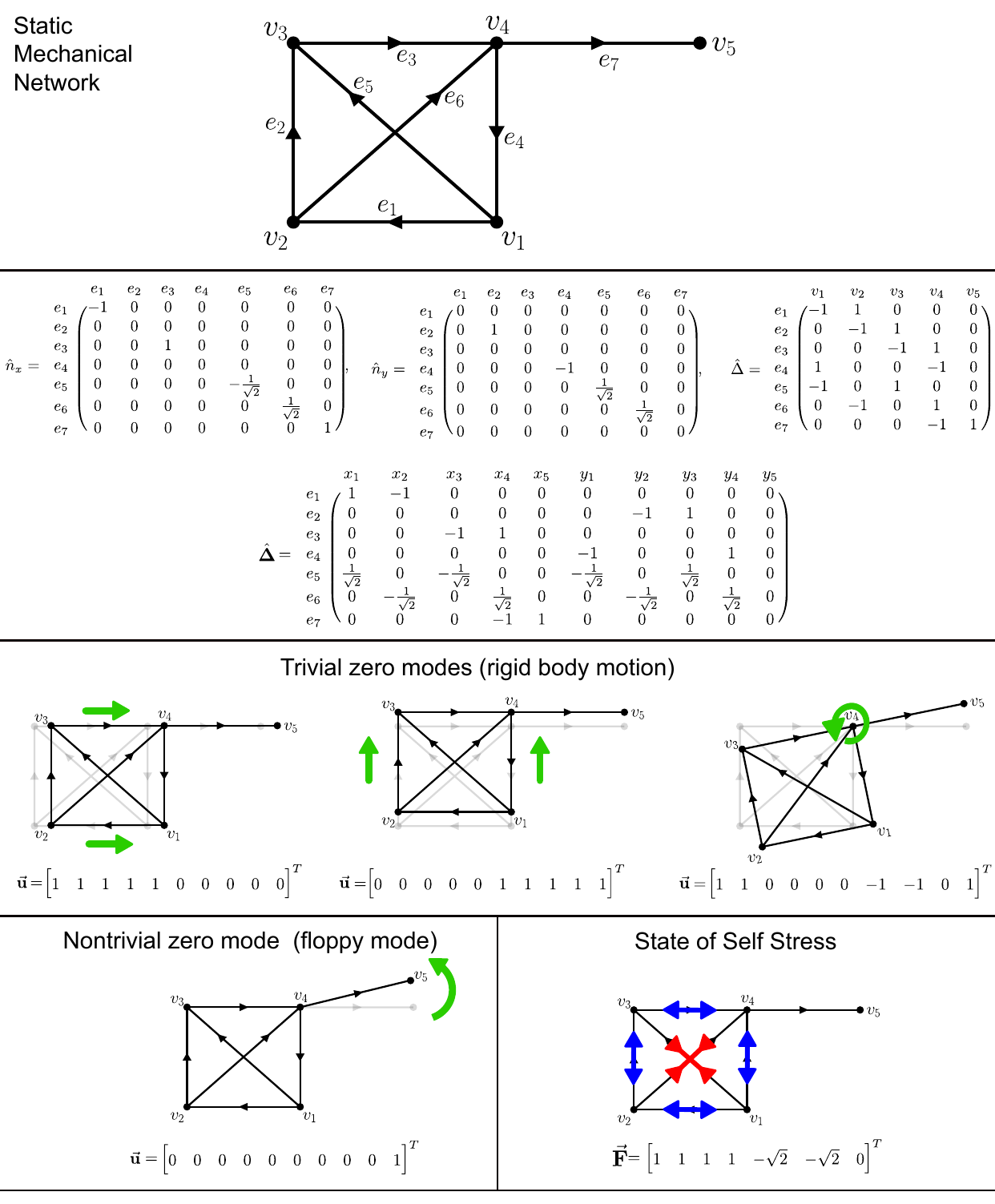}
    \caption{Mechanical network example. The compatibility matrix $\hat{\boldsymbol{\Delta}}$ is calculated from the incidence matrix $\hat{\Delta}$ along with the matrices $\hat{n}_x$ and $\hat{n}_y$ encoding edge directions: $[ \hat{n}_x \hat{\Delta} ,  \hat{n}_y \hat{\Delta} ]$. The zero modes satisfy $\hat{\boldsymbol{\Delta}}\cdot \vec{\boldsymbol{u}} = \vec{0}$ while the state of self stress satisfies $\hat{\boldsymbol{\Delta}}^T \vec{F} = \vec{\boldsymbol{0}}$. In two dimensions, there are three trivial zero modes: two translations and one rotation. The floppy mode corresponds to independently rotating node 5. The state of self stress occurs when the edges on the perimeter of the square are in compression while the crossbars are in tension. The arrows indicate the direction of the restoring forces that would relax each individual edge, but by geometric incompatibility, the entire network is stuck in a stressed configuration. Note that \eqref{eq:Maxwell_Calladine} is satisfied.}
    \label{fig:MechanicalNetwork}
\end{figure}

One can decouple equations \eqref{eq:Hooke} and \eqref{eq:forceBalance}:
\begin{equation}
    \hat{\boldsymbol{\LaplacianMatrix}} \cdot \vec{\boldsymbol{u}} \equiv \hat{\boldsymbol{\Delta}}{}^T \hat{K} \hat{\boldsymbol{\Delta}} \cdot \vec{\boldsymbol{u}} = - \vec{\boldsymbol{f}}.
\end{equation}
The operator $ \hat{\boldsymbol{\LaplacianMatrix}}$ that resembles the Laplacian is called the dynamical matrix.\footnote{The name makes more sense when we actually study dynamics.} Inverting this matrix is the key step to solving mechanical networks. One must apply enough constraints to eliminate the zero modes. For 2d, this means fixing at least 3 displacements (ex: the $x$ and $y$ component of one node, and the $y$ component of a different node), and constraining any floppy modes.

Much like our analysis of circuits, we can write down an energy function
\begin{equation}
    E = \frac{1}{2} \vec{\boldsymbol{u}}^T \cdot \hat{\boldsymbol{\LaplacianMatrix}} \cdot \vec{\boldsymbol{u}} + \vec{\boldsymbol{u}}^T \cdot \vec{\boldsymbol{f}}
\end{equation}
whose minimization gives the physical displacements. We can interpret the Laplacian for this problem as a Hessian of the elastic energy. Furthermore, we can now interpret our zero modes as either vectors in the null space of $\hat{\boldsymbol{\Delta}}$ or as displacements which cost zero energy. 

The zero modes and states of self stress can be counted using the Maxwell-Calladine index theorem \cite{Lubensky2015Phonons}. We can prove the theorem just by applying the rank-nullity theorem twice, once to $\hat{\boldsymbol{\Delta}}$ and once to $\hat{\boldsymbol{\Delta}}{}^T$:
\begin{equation}
    \text{rank}(\hat{\boldsymbol{\Delta}}) + \text{nullity} (\hat{\boldsymbol{\Delta}}) = d|\VertexSpace|
\end{equation}
\begin{equation}
    \text{rank}(\hat{\boldsymbol{\Delta}}{}^T) + \text{nullity} (\hat{\boldsymbol{\Delta}}{}^T) = |\EdgeSpace|.
\end{equation}
Recalling that the rank of a matrix is equal to that of its transpose, we can subtract these two equations to get 
\begin{equation}
    N_0 - N_{ss} = d|\VertexSpace| - |\EdgeSpace|, \label{eq:Maxwell_Calladine}
\end{equation}
where $N_0$ is the number of zero modes and $N_{ss}$ is the number of states of self stress. 

An example 2d static mechanical network is analyzed in figure \ref{fig:MechanicalNetwork}. The Python code to generate the compatibility matrix and find its null space is shown below. Note that the basis vectors of the null space that Python returns may not be the intuitive vectors shown in figure \ref{fig:MechanicalNetwork}.\footnote{\textbf{EXERCISE}: Calculate the zero modes for the left network shown in figure \ref{fig:circuit}, and classify them physically. Are there any states of self stress? }

\begin{lstlisting}[language=Python, caption={Analyzing a Mechanical Network}]
import networkx as nx
import numpy as np
import scipy.sparse as sp
from scipy.linalg import null_space

G = nx.DiGraph()
ordered_nodes = [1, 2, 3, 4, 5]
ordered_edges = [(1,2),(2,3),(3,4),(4,1),(1,3),(2,4),(4,5)]
G.add_nodes_from(ordered_nodes)
G.add_edges_from(ordered_edges)
restPos_dict = {1: (0,0), 2: (-1,0), 3: (-1,1), 4: (0,1), 5: (1,1)}
nx.set_node_attributes(G, restPos_dict, "restPos")
DeltaT = nx.incidence_matrix(G, oriented=True, 
                             nodelist=ordered_nodes, edgelist=ordered_edges)
Delta = DeltaT.T
restPos = np.array([restPos_dict[n] for n in G.nodes()])
edgeLengthVector = Delta @ restPos
DXVec = edgeLengthVector[:,0]
DYVec = edgeLengthVector[:,1]
l0Vec = np.sqrt(DXVec**2 + DYVec**2)
nXVec = DXVec/l0Vec
nYVec = DYVec/l0Vec
CompatibilityMatrix = sp.hstack([sp.diags(nXVec)@Delta, sp.diags(nYVec)@Delta])
print(CompatibilityMatrix.toarray())
print(null_space(CompatibilityMatrix.toarray()))
print(null_space(CompatibilityMatrix.toarray().T))
\end{lstlisting}

\begin{figure}
    \centering
    \includegraphics[width=.9\textwidth]{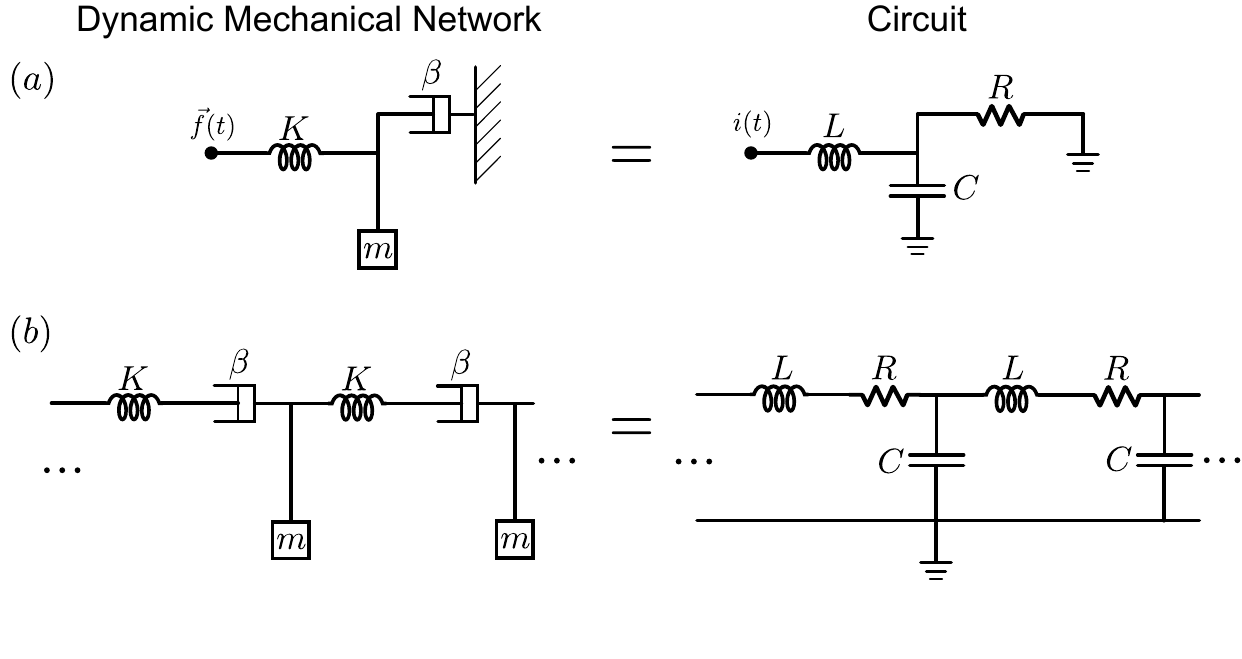}
    \caption{Demonstration of the correspondence between an electrical circuit and a dynamic mechanical network with two examples. In the mapping described by equations \eqref{eq:mechanical_R}, \eqref{eq:mechanical_L}, and \eqref{eq:mechanical_C}, the dashpot $\beta$ plays the role of the resistor $R$, the spring $K$ plays the role of the inductor $L$, and the mass plays the role of the capacitor $C$. Many other such mappings exist \cite{OrtizTavarez2025UnifiedModel}.}
    \label{fig:Mechanical_I}
\end{figure}

\subsection{Dynamic Mechanical Networks}

There are many ways in which we can map mechanical networks to circuits \cite{OrtizTavarez2025UnifiedModel}, though when dynamics are involved, the analogy is most apparent if we study the velocity of the nodes rather than the displacement:
\begin{align}
    \psi_i - \psi_j = R_{ij} I_{ij} &\iff v_i - v_j = \frac{1}{\beta_{ij}} F_{ij} \label{eq:mechanical_R}\\ 
    \psi_{i}-\psi_j = L_{ij} \frac{d I_{ij}}{d t} &\iff v_i - v_j = \frac{1}{K_{ij}} \frac{dF_{ij}}{dt} \label{eq:mechanical_L}\\
    i_j = C_{j} \frac{d \psi_j}{d t}  &\iff m_j \frac{d v_j}{d t} = f_j \label{eq:mechanical_C} 
\end{align}
The first equation describes frictional damping along an edge. The second equation is the time derivative of Hooke's law from the previous sections. The third equation describes Newton's second law at a node. Some example mappings are shown in figure \ref{fig:Mechanical_I}. Of course, this analogy only works in 1D where velocity is a scalar. Combining the techniques for RLC circuits introduced in section III.F with those to analyze vector-valued node quantities in the previous section, you can analyze dynamic mechanical networks. Furthermore, spatial dynamics such as wave propagation can be accounted for by using the technique described in section IV.C \cite{Fancher2025AnStructures}.

\newpage
\section{Random walks on graphs}

The theory of Markov chains and random walks is a thoroughly studied branch of mathematics that we will briefly review here. Some reviews on the topic, as applied to graphs, can be found in \cite{Aldous_Fill_2014, Masuda_Porter_Lambiotte_2017}. In particular, the connection to circuit theory was studied by Doyle and Snell in \cite{Doyle_Snell_1984}. 

A Markov chain is any process which can be described as a sequence of states such that the current state only depends on the previous state. We can represent each state with a node and each transition with a directed weighted edge. If the weight of edge $e_{ij}$ is $w_{ij}$, then the \textbf{transition probability} from $i$ to $j$ is 
\begin{equation}
    P_{ij} = \frac{w_{ij}}{\sum_k w_{ik}},
\end{equation}
where the denominator ensures that the probability to transition from $i$ to somewhere including $i$ is 1:
\begin{equation}
    \sum_j P_{ij} = 1.
\end{equation}
Thus, much like the Laplacian matrix, a probability matrix always has a $\vec{1}$ right eigenvector. Note that even if $\hat{W}$ is symmetric, $\hat{P}$ will typically not be symmetric, so the left and right eigenvectors are not the same. General Markov processes are studied on directed graphs. We will specifically consider the problem of random walkers, though the math can easily be adapted to other Markov processes. 

\subsection{Steady-state distribution of random walkers}
Consider a random walker that when released, moves randomly along the edges of a graph. Let $p_i(k)$ denote the probability that a random walker is present at node $i$ at time step $k$. We have a recursion relation relating $p_i(k)$ at different time steps:
\begin{equation}
    p_j(k+1) = \sum_i p_i(k) P_{ij}.
\end{equation}
In matrix form, 
\begin{equation}
    \vec{p}^T(k+1) = \vec{p}^T(k) \hat{D}_-^{-1} \hat{W} 
\end{equation}
where $\hat{D}_- = \mathrm{diag}\{\sum_j w_{ij} \}$ is the diagonal matrix of outdegrees and $\hat{W}$ is the adjacency matrix with elements $w_{ij}$.\footnote{It may look weird that we are left-multiplying our matrices, but this is essential to be consistent with how we defined $P_{ij}$. Some authors define the transition matrix to be the transpose of this which has the advantage that we can multiply probability vectors from the right, but it has the disadvantage that we need to read our expressions from right to left. Be sure to check which convention is being used when encountering such matrices!} Given some initial condition $\vec{p}(0)$, we can solve this equation iteratively to obtain:
\begin{equation}
    \Vec{p}^T(k) = \vec{p}^T(0) \left( \hat{D}_-^{-1}\hat{W}\right)^k.
\end{equation}
For each iteration, the operator $\hat{D}_-^{-1}\hat{W}$ causes the solution to spread out. Intuitively, the distribution of walkers diffuses until reaching a steady state $\vec{p}_*^T$ satisfying 
\begin{equation}
    \vec{p}_*^T \hat{P} = \vec{p}_*^T.
\end{equation}
For a strongly connected graph,\footnote{
A \textbf{strongly connected} directed graph has the property that every node can be reached by every other node along directed edges. A connected undirected graph is also strongly connected, but a connected directed graph may not be strongly connected.} there is a unique steady state $\vec{p}^T_{*}$,\footnote{\textbf{EXERCISE}: Show that for an undirected network, $p_{*,i} = d_i/\sum_i d_i$, where $d_i=\sum_{j} w_{ij}$ is the weighted degree of the node.} and the fraction of time one walker spends at node $v_i$ converges to $p_{*,i}$. Even if the time-averaged probability approaches a stationary distribution, the walker need not wander aimlessly. A circulating stationary flux is a unique feature of directed graphs: For electric circuits, it's impossible to have current around a loop without source terms, but for a directed graph, it is conceivable that a drunken sailor will walk in circles for eternity. If the Markov chain satisfies \textbf{detailed balance}: $p_{*, i} P_{ij} = p_{*, j} P_{ji}$, then no such cyclic dynamics can occur. Such Markov chains are also called reversible because, at equilibrium, the flux of random walkers from $i$ to $j$ is the same as that from $j$ to $i$.\footnote{\textbf{EXERCISE}: Show that for $P_{ij} = P_{ji}$, the equilibrium distribution of random walkers exactly maps onto solving an electric circuit. What is the additional constraint that uniquely determines $\vec{p}_{*}^T$?}

\subsection{Escape probability for a random walker}
What is the probability $x_i$ that a random walker (drunken sailor) starting at node $i$ reaches node $N$ (home) before it reaches node $1$ (bar)? It is easy to see that $x_N=1$, as the sailor starts the random walk already home, and $x_1=0$, as the sailor is already at the bar and has no chance to get home. 

As a self-consistency condition, 
\begin{equation}
x_i = \sum_j P_{ij} x_j 
\end{equation}
where the right-handside encodes all of the locations $j$ the walker could be at after one step, weighted by the probability of hopping from $i$ to $j$. 
In matrix form, this can be written 
\begin{equation}
    \vec{x} = \hat{P} \vec{x} = \hat{D}_-^{-1} \hat{W} \vec{x} \implies (\hat{D}_- - \hat{W})\vec{x} = \vec{0}.
\end{equation}
If our network is undirected, then this is equivalent to solving the circuit/flow problem with Dirichlet boundary conditions at 1 and $N$.

An absorbing state in a directed graph is a node whose outdegree is zero. Consider a graph which has the property that from any initial node, it is possible for our walker to reach an absorbing state.\footnote{An ergodic set consists of nodes for which our walker can completely explore all nodes in the set, but cannot leave the set \cite{Masuda_Porter_Lambiotte_2017}. Absorbing states are ergodic sets consisting of a single node. We are excluding graphs which have ergodic sets consisting of multiple nodes for simplicity.} Generalizing the idea of the drunken sailor, we see that absorbing states can be incorporated as boundary conditions to a Dirichlet problem. The interior nodes are all non-absorbing nodes. Using the method and notation introduced in section IIIB, we have that 
\begin{equation}
    \vec{x}_{\mathcal{I}} = - (\hat{\mathbb{P}}_\mathcal{I} (\hat{D}_- - \hat{W})\hat{\mathbb{P}}_\mathcal{I}^T)^{-1}\hat{\mathbb{P}}_\mathcal{I} (\hat{D}_- - \hat{W})\hat{\mathbb{P}}_\mathcal{B}^T \vec{x}_{\mathcal{B}}.
\end{equation}
Plugging in $x_{B,j} = \delta_{jk}$ gives the probability to reach node $k$ (home) before reaching the other absorbing boundaries (bars).\footnote{\textbf{EXERCISE}: Show that the operator in front of $\vec{x}_{\mathcal{B}}$ can be rewritten as $\sum_{n=0}^\infty (\hat{\mathbb{P}}_{\mathcal{I}} \hat{P} \hat{\mathbb{P}}_{\mathcal{I}}^T )^n \hat{\mathbb{P}}_{\mathcal{I}} \hat{P} \hat{\mathbb{P}}_{\mathcal{B}}^T$. Each operator in this sum can be interpreted as taking $n$ steps within the interior, then one step from the interior to the boundary.}
 
\subsection{Mean first-passage time for a random walker}
A related problem is to find the mean first-passage time. Define the mean first-passage time $h_i$ as the expected number of steps to go from node $i$ to node $1$. The time $h_i$ is equal to the sum of the neighbor's mean first-passage times weighted by the transition probability $P_{ij}$ plus 1 for the extra time step:
\begin{equation}
    h_i = \sum_j P_{ij} h_j +1 
\end{equation}
or in matrix form 
\begin{eqnarray}
    \vec{h} = \hat{D_-}^{-1} \hat{W} \vec{h} + \vec{1} \Rightarrow (\hat{D_-} -  \hat{W})\vec{h} = \vec{d}_-.
\end{eqnarray}
If the transition probabilities are symmetric, this is identical to a flow problem with distributed node currents and node $1$ grounded. Three other methods for calculating the mean first-passage times are summarized in \cite{Masuda_Porter_Lambiotte_2017} and extensively studied in \cite{Redner_2001}.

\newpage
\section{Nonlinear Flow Networks} \label{sec:nonlinear_networks}
Throughout this tutorial, we focused only on networks with a linear response. This fails for networks with diodes, valves, non-Newtonian fluids, non-Hookean solids, etc. Here, we consider networks with a nonlinear response. Using the notation of fluid flow networks, we wish to solve equations of the form:\footnote{Here, we are only considering static networks. In cases where the network equations have time dynamics (for fluid flow networks, inertia or compliance gives this, for instance), it's actually easier to handle. Given some known initial conditions, you can update the solution by some small amount each time step according to your nonlinear dynamics. For systems without a timescale (ex: Stokes flow), it's harder to treat nonlinearities, so that's what we'll focus on here. While it is incredibly computationally inefficient, it is sometimes possible to introduce a very small mass or large stiffness, run the dynamic nonlinear simulation, and check for convergence as the mass goes to zero or the stiffness goes to infinity. The problem is that as these timescales diverge, the simulations will take much longer to equilibrate, so avoid this whenever possible. However, in cases where nonlinearities have a cusp (ex: ideal valves), the methods in this section fail, so solving for the dynamics may be the best option. }
\begin{equation}
    \hat{\Delta}^T \vec{Q} = \vec{q}, \label{eq:nonlin_network1}
\end{equation}
\begin{equation}
    \vec{Q} = \vec{Q}(-\hat{\Delta}\vec{p}). \label{eq:nonlin_network2}
\end{equation}
The first equation is just the familiar KCL which is unchanged even when we have nonlinear edges. The second equation suggests that the flow can be written as a, possibly nonlinear, function of the pressure drop. Physically, we will usually deal with systems where the flow through an edge depends only on the pressure drop across that edge:
\begin{equation}
    Q_\mu = Q_{\mu } (-\Delta_{\mu j}p_j) .\label{eq:nonlinear}
\end{equation}
The linear case is recovered by taking $Q_\mu(-\Delta_{\mu j}p_j) = -\Conductance_{\mu \nu} \Delta_{\nu j} p_j$. 

Linear flow networks can always be studied using undirected graphs with arbitrary edge orientation. Nonlinear flow networks, on the other hand, may or may not be directed. If the function $\vec{Q}(-\hat{\Delta}\vec{p})$ is an odd function like a cubic, then the edge orientation is still arbitrary. However, in all other cases, like a valve or diode, we say the edge is non-reciprocal: $\vec{Q}(\hat{\Delta}\vec{p}) \neq -\vec{Q}(-\hat{\Delta}\vec{p}) $, and the edge orientation is not arbitrary. This doesn't actually pose any difficulties. Computationally, we have been using DiGraph() objects in Python all along, but now we just need to ensure that when adding an edge to a network, the oriented pair of nodes is consistent with the direction of the nonlinear element we wish to simulate.

Although there are many computational tools specifically designed to solve nonlinear systems, there are two techniques that relate nicely to the problems we have considered so far, which will be discussed in the following subsections.

\subsection{Solving networks with a small nonlinearity using perturbation theory} \label{sec:network_perturbation_theory}

Suppose we have a network where the pressure drops remain small. Furthermore, assume that \eqref{eq:nonlinear} is smooth near the origin. Then, we can Taylor expand \eqref{eq:nonlinear}:
\begin{equation}
    Q_{\mu } (-\Delta_{\mu j}p_j) \approx Q_{\mu}(0) + \partial_\nu Q_\mu (0) (-\Delta p)_\nu + \frac{1}{2} \partial_\nu \partial_\rho  Q_\mu (0) (-\Delta p)_\nu (-\Delta p)_\rho + ...
\end{equation}
where $\partial_\nu =  \partial/\partial(-\Delta p)_\nu$. For generality, we are allowing for an active flow term that is independent of the pressure. This equation can be combined with equation \eqref{eq:nonlin_network1} to obtain a single nonlinear equation in terms of $\vec{p}$ and $\vec{q}$. First, we will expand the knowns and unknowns in powers of a small parameter $\epsilon$. The precise scaling depends on the assumptions specific to the system we wish to solve, but a reasonable scaling for the knowns is:
\begin{equation}
    \vec{p}_\mathcal{B} = \epsilon \vec{p}_\mathcal{B}^{(1)}, \hspace{1cm} \vec{q}_\mathcal{I} = \vec{q}_\mathcal{I}^{(0)},
\end{equation}
such that the known interior flows are balanced by the active forcing, and the magnitude of the boundary pressures sets the scale for $\epsilon$. For the unknowns, 
\begin{equation}
    \vec{p}_\mathcal{I} = \epsilon \vec{p}_\mathcal{I}^{(1)} + \epsilon^2 \vec{p}_\mathcal{I}^{(2)} + ..., \hspace{1cm} \vec{q}_\mathcal{B} = \vec{q}_\mathcal{B}^{(0)} + \epsilon \vec{q}_\mathcal{B}^{(1)} + ... .
\end{equation}
The zeroth order equations reduce to 
\begin{equation}
    \Delta^T_{i\mu} Q_\mu (0) = q_i^{(0)}.
\end{equation}
On the interior, this gives us a consistency condition for known quantities. On the boundary, this tells us how the active edge flows dump flow to the boundaries. The first-order equations reduce to the typical linear problem:
\begin{equation}
    -\Delta^T_{i\mu} \partial_\nu Q_\mu(0) \Delta_{\nu j} p_j^{(1)} = q^{(1)}_i.
\end{equation}
The operator with matrix elements $\Delta^T_{i\mu} \partial_\mu Q_\nu(0) \Delta_{\nu j}$ is just the Laplacian for the linearized problem. Inverting this operator confined to the interior gives the unknown pressures $\vec{p}_{\mathcal{I}}^{(1)}$, and the unknown flows are then found using \eqref{eq:nonlin_network1}. The second-order equations are
\begin{eqnarray}
    -\Delta^T_{i\mu}  \partial_\nu Q_\mu (0) \Delta_{\nu j} p_j^{(2)} + \frac{1}{2} \Delta^T_{i\mu}  \partial_\nu \partial_\rho Q_\mu (0) (-\Delta p)_\nu^{(1)} (-\Delta p)_\rho^{(1)} = q_i^{(2)}.
\end{eqnarray}
The second term looks complicated, but it's completely determined by the known pressure drops from the first-order equations. Thus, we only have to invert the same Laplacian confined to the interior nodes to find the unknown pressures $\vec{p}_{\mathcal{I}}^{(2)}$. Although $\vec{p}_{\mathcal{B}}^{(2)} = 0$, we have an additional boundary term coming from the lower-order pressure contributions. For a weakly nonlinear system, or a nonlinear system with small pressure drops and smooth pressure-flow relation, this is the dominant source of nonlinearity. We can continue this series to higher orders, and, remarkably, we will only end up having to invert the same Laplacian at each order. This is a convenient technique when the main interest is to know how a small nonlinearity affects otherwise linear flow networks.

\subsection{Solving nonlinear networks using Newton's method} \label{sec:network_Newton}

The most common technique to solving nonlinear systems is Newton's method. Let's see how this works for nonlinear flow networks. Newton's method requires that the first derivative of $Q_\mu$ with respect to $(\Delta p)_\mu$ is defined and works best when there is an analytical form for $Q_{\mu } (-\hat{\Delta}\vec{p})$ and its first derivative. 

Decoupling equations \eqref{eq:nonlin_network1} and \eqref{eq:nonlin_network2} gives a nonlinear system of equations for the node pressure:
\begin{eqnarray}
    F_i(\vec{p}, \vec{q}) \equiv - \Delta^T_{i\mu} Q_{\mu} (-\Delta_{\mu j}p_j) + q_i = 0, \label{eq:F_i(p)}
\end{eqnarray}
The physical pressures and node flows satisfy $\vec{F}(\vec{p}, \vec{q}) = 0$. In the interior, $\vec{F}$ does not depend on the unknown boundary node flows, so we can write
\begin{equation}
    \vec{F}_{\mathcal{I}}(\vec{p}_{\mathcal{I}}, \vec{p}_{\mathcal{B}}, \vec{q}_{\mathcal{I}}) = 0
\end{equation}
where the only unknowns are $\vec{p}_{\mathcal{I}}$. Let's guess a solution to this equation. Call the guess $\vec{p}_\mathcal{I}^{(k)}$.\footnote{This is not to be confused with the $\vec{p}^{(k)}$ from the previous section! Here, $k$ indexes an iteration in an algorithm, while in the previous section, it indicated an order in a perturbative expansion.} If we're close, then we can approximate the error using a Taylor series. For $v_i \in \mathcal{I}$,
\begin{equation}
    F_i (\vec{p}_{\mathcal{I}}^{(k)}, \vec{p}_{\mathcal{B}}, \vec{q}_{\mathcal{I}}) \approx F_{i}(\vec{p}_{\mathcal{I}}, \vec{p}_{\mathcal{B}}, \vec{q}_{\mathcal{I}}) + \frac{\partial F_i}{\partial p_{j}}\Big{|}_{\vec{p}} (p^{(k)}_j - p_j) = \frac{\partial F_i}{\partial p_{j}}\Bigg{|}_{\vec{p}} (p^{(k)}_j - p_j) \approx \frac{\partial F_i}{\partial p_{j}}\Bigg{|}_{\vec{p}^{(k)}} (p^{(k)}_j - p_j) 
\end{equation}
Comparing with equation \eqref{eq:F_i(p)}, for $v_i, v_j \in \mathcal{I}$,
\begin{equation}
    J_{ij}^{(k)} \equiv \frac{\partial F_i}{\partial p_j}\Bigg{|}_{\vec{p}^{(k)}} =  \Delta^T_{i\mu} \frac{\partial Q_\mu}{\partial (-\Delta p)_{\nu}}  \Delta_{\nu j} \Bigg{|}_{\vec{p}^{(k)}} \label{eq:Newton_Jij}
\end{equation}
Thus, a closer approximation to the true solution is 
\begin{equation}
    \vec{p}_{\mathcal{I}}^{(k+1)} = \vec{p}_{\mathcal{I}}^{(k)} - \left(\hat{J}^{-1}_{\mathcal{I}\mathcal{I}}\right)^{(k)} \vec{F}_{\mathcal{I}} = \vec{p}_{\mathcal{I}}^{(k)} - \left(\hat{J}^{-1}_{\mathcal{I}\mathcal{I}}\right)^{(k)} [-\hat{\Delta}^T \vec{Q}(-\hat{\Delta} \vec{p}^{(k)}) + \vec{q}^{(k)}]_{\mathcal{I}}.
\end{equation}
This last equation summarizes Newton's method applied to flow networks. We iteratively solve for $\vec{p}_\mathcal{I}^{(k)}$ until the norm of $\vec{F}$ is sufficiently small. We need to provide an initial guess $\vec{p}^{(0)}_{\mathcal{I}}, \vec{q}^{(0)}_{\mathcal{B}}$, which could be, for example, the solution to the linearized problem. Note that $\hat{J}_{\mathcal{I}\mathcal{I}}^{(k)}$ given in equation \eqref{eq:Newton_Jij} is just a Laplacian at each iteration! The only difference from the linear case is that the conductance itself depends on the pressure, and so we need to update the conductance at each iteration to account for our new best guess of the pressure.

\newpage
\section{Concluding remarks}
In this tutorial, we showed how electric circuits, fluid flow networks, and mechanical networks are united by conservation laws and linear response. Whenever a transported quantity satisfies a continuity equation, we can write down a law analogous to Kirchhoff's Current Law. For most of the tutorial, we worked with the node analysis where the edge current is directly proportional to the gradient of a potential function on the nodes. This combines Kirchhoff's Voltage Law, which concerns the existence of such a potential function, with Ohm's Law, which encodes the physics of linear response (conductance, viscosity, damping, etc.). A summary of the connection between electrical networks, fluidic networks, and mechanical networks is given in table \ref{tab:summary}. Markov processes such as random walks are also closely related, but only when detailed balance is assumed. 

We showed that the node-edge incidence matrix $\hat{\Delta}{}^T$ is fundamental to understanding both the topology of oriented graphs and the physics defined on them. The incidence matrix functions both as a boundary operator (when acting on ``chains") and a derivative operator (when acting on ``cochains"). Physical networks have edge weights which encode the conductance of a linear response. From the incidence matrix and the conductance, one can build a Laplacian matrix $\hat{\LaplacianMatrix} = \hat{\Delta}^T \hat{\Conductance} \hat{\Delta}$. Solving linear flow networks usually amounts to inverting a Laplacian. In its simplest form, the Laplacian is not invertible, but once we apply constraints (boundary conditions) on the nodes, we can uniquely solve for the potential on all interior nodes. This technique can even be generalized to cases where vector quantities are defined on the nodes by using the equilibrium matrix $\hat{\boldsymbol{\Delta}}^T$ rather than the incidence matrix. For more information on how linear flow networks are united by linear algebra with particular focus on mechanical networks, we highly recommend a recent review paper by Ortiz-Tavárez, et al. \cite{OrtizTavarez2025UnifiedModel}. We hope our work supplements this and other existing work by providing explicit examples and Python code for students and researchers beginning to study physical networks.

We stressed that one should understand the underlying (often higher-dimensional) physics of the edges before solving the network equations. Begin with fundamental conservation laws, incorporate constitutive relations, integrate out spatial dimensions, and hope that you are left with something linear. In many cases, there is a nonlinear response between the current and potential difference across an edge, but we showed that Newton's method can be used to solve a nonlinear problem by iteratively solving linear problems, and that Kirchhoff's Laws remain unchanged. Thus, the techniques in this paper can even be used as a starting point for understanding how to solve nonlinear networks.

\begin{table}[h!]
\begin{center}
    \begin{tabular}{|  c | c | c | c |} 
      \hline
      \textbf{ } & \textbf{  Electrical  }& \textbf{\hspace{.9cm} Fluidic  \hspace{.9cm}  }& \hspace{.3cm}\textbf{Mechanical } \hspace{.3cm} \\
      \hline
     \textbf{Transported quantity}  &  Charge Q & Volume V & Momentum \textbf{p}\\
     \hline
     \textbf{Edge Current}  &  Current $I$ & Flow $Q$ & Tension $F$ \\
     \hline
     \textbf{Node Potential} &  Potential $\psi$  & Pressure $p$ & Node velocity $\boldsymbol{v}$ \\
     \hline
     \textbf{Edge Linear Response I}  &  Resistance $\sim \Delta \psi/I$ & Viscosity $\sim \Delta p/Q$ & Damping $\sim \Delta \boldsymbol{v}/F$ \\
     \hline
     \textbf{Edge Linear Response II}  &  Inductance $\sim \Delta \psi/\frac{dI}{dt}$ & Inertia $\sim \Delta p/\frac{dQ}{dt}$ & Elasticity $\sim \Delta \boldsymbol{v}/\frac{dF}{dt}$\\
     \hline
     \textbf{Node Linear Response}  &  Capacitance $\sim \text{Q}/\psi$ & Compliance $\sim \text{V}/p$ & Inertia $\sim \textbf{p}/v$  \\
     \hline 
    \end{tabular}
    \caption{Summary of physical quantities relevant for three types of networks studied in this tutorial. In the case of mechanical networks, only the dynamical networks considered in the last section are displayed.}
    \label{tab:summary}
\end{center} 
\end{table}

\begin{acknowledgments}
    We'd like to thank Phil Nelson, Felipe Martins, Niranjan Sarpangala, Adam Kline, Marcelo Guzman, Suman Kulkarni, and the PHYS 5570 Spring 2024 class for useful discussions and feedback on the draft. This is hopefully the first of multiple iterations of this tutorial, so we welcome comments.

\end{acknowledgments}

\newpage
\section{Appendix}
\subsection{Perron-Frobenius Theory}
The Perron-Frobenius Theorem gives us some useful information for the eigenvalues and eigenvectors of the sorts of matrices we encounter in graph theory. We will present the theorem in a way that is most intuitive for those studying networks \cite{Baez_Biamonte_2018}. A more thorough review of the spectral theory of graphs can be found in \cite{Van_Mieghem_2011}. 

First, note that we can associate any $N\cross N$ matrix with non-negative coefficients $\hat{W}$ with the adjacency matrix of a directed graph $G(\hat{W})$. 

The \textbf{Perron-Frobenius Theorem} states that for any matrix $\hat{W}$ with non-negative entries, if $G(\hat{W})$ is strongly connected, there exists a positive real eigenvalue $\lambda_1$ such that for any eigenvalue $\lambda_i$ of $\hat{W}$, $|\lambda_i| \leq \lambda_1$. The algebraic multiplicity of $\lambda_1$ is one, and the eigenvector $\vec{v}_1$ corresponding to the eigenvalue $\lambda_1$ can be made to have purely positive entries. Any other eigenvector which has purely non-negative entries is a scalar multiple of $\vec{v}_1$.

If in addition, $\hat{W}$ is symmetric (or equivalently, $G$ is undirected), we know that the eigenvalues are real, so we can order the eigenvalues $\lambda_1 > \lambda_2 \geq \lambda_3 ... \geq \lambda_N \geq -|\lambda_1|$. In particular, there is always a spectral gap $\lambda_1 - \lambda_2 > 0$. 

For a matrix whose only negative entries lie on the diagonal, we can simply add a sufficiently large positive number times the identity matrix, then apply Perron-Frobenius. For example, the negative of the Laplacian matrix for a connected undirected graph has only negative entries on its diagonal, so we can add the maximum degree $d_{max}$: $-\hat{\LaplacianMatrix} + d_{max} \hat{\mathbb{1}}  = \hat{W} + d_{max}\hat{\mathbb{1}} - \hat{D} \geq 0$. The unique positive eigenvector of $-\hat{\LaplacianMatrix} + d_{max} \hat{\mathbb{1}}$ is $\vec{1}$ with eigenvalue $d_{max}$ which is the largest eigenvalue of this matrix. Thus, the smallest eigenvalue of $\hat{\LaplacianMatrix}$ is zero with eigenvector $\vec{1}$, which is the only purely positive eigenvector. 

\subsection{Harmonic functions}

Suppose we wish to solve the linear flow network problem with $\current_j=0$ for each $v_j \in \mathcal{I}$:
\begin{equation}
    (\Op{\LaplacianMatrix}\NodeVec{\potential})_j = 0.
\end{equation}
This is the discrete analogue to the continuous Laplace's equation:
\begin{eqnarray}
    \nabla^2 \psi (\vec{x}) = 0.
\end{eqnarray}
Functions satisfying Laplace's equation are called \textbf{harmonic functions}. Harmonic functions have several  useful identities with direct analogues in the discrete case \cite{Grady_Polimeni_2010, Doyle_Snell_1984}. The \textbf{mean-value property} equates the value of a harmonic function averaged on the surface of a sphere to its value at the center of the sphere. Its discrete analogue can be written\footnote{\textbf{EXERCISE}: Prove the mean-value property for networks using the definition of $\Op{\LaplacianMatrix}$. Then, use it to prove the minimum (maximum) principle for networks.}
\begin{eqnarray}
    \psi_i = \frac{1}{d_i} \sum_{j\in \mathcal{V}} \Adjacency_{ij} \psi_j,
\end{eqnarray}
valid for any $\psi_i \in \mathcal{I}$. In words, the value of the potential at a node is equal to the weighted average of the potential at all of the adjacent nodes. Harmonic functions also satisfy the \textbf{minimum (maximum) principle} which states that $\psi$ takes its minimum (maximum) value on the boundary: For each $i \in \mathcal{I}$,
\begin{eqnarray}
    \min_{j\in \mathcal{B}} \psi_j \leq \psi_i \leq \max_{j\in \mathcal{B}} \psi_j.
\end{eqnarray}
Furthermore, equality in either of these expressions implies that $\psi_i$ is constant throughout $\mathcal{I}$.

\newpage

\bibliographystyle{unsrt}
\bibliography{refs}

\end{document}